\documentclass[aps,prd,prev,twocolumn,preprintnumbers,floatfix,nofootinbib]{revtex4-1}
\pdfoutput=1

\newcommand{\be}{\begin{equation}}
\newcommand{\ee}{\end{equation}}
\newcommand{\bea}{\begin{eqnarray}}
\newcommand{\eea}{\end{eqnarray}}

\newcommand{\tR}{\tilde{R}}

\usepackage{mathrsfs}
\usepackage{amsmath, amsthm, amssymb}
\usepackage{color}
\usepackage{epstopdf}
\usepackage[pdftex]{graphicx}
\usepackage{amssymb}
\usepackage{verbatim}
\usepackage{hyperref}
\usepackage{enumerate}

\begin{document}

\preprint{KCL-PH-TH/2017-33}

\title{Structure Formation and Microlensing with Axion Miniclusters}

\author{Malcolm Fairbairn$^{a}$}
\email{malcolm.fairbairn@kcl.ac.uk}
\author{David J. E. Marsh$^{a}$}
\email{david.marsh@kcl.ac.uk}
\author{J\'er\'emie Quevillon$^{a,b}$}
\email{jeremie.quevillon@lpsc.in2p3.fr}
\author{Simon Rozier$^{c,d}$}
\email{rozier@iap.fr}

\vspace{1cm}
\affiliation{
${}^a$ King's College London, Strand, London, WC2R 2LS, United Kingdom \\
${}^b$ Laboratoire de Physique Subatomique et de Cosmologie, Universit\'e Grenoble-Alpes, CNRS/IN2P3, 53 Avenue des Martyrs, F-38026 Grenoble, France \\
${}^c$ \'Ecole Polytechnique, Universit\'e Paris-Saclay, F-91128 Palaiseau, France \\
${}^d$ Institut d'Astrophysique de Paris --- UMR 7095 du CNRS, Universit\'e Pierre \& Marie Curie, 98\textsuperscript{bis} boulevard Arago, F-75014 Paris, France
}

\begin{abstract}

If the symmetry breaking responsible for axion dark matter production occurs during the radiation-dominated epoch in the early Universe, then this produces large amplitude perturbations that collapse into dense objects known as axion miniclusters.  The characteristic minicluster mass, $M_0$, is set by the mass inside the horizon when axion oscillations begin. For the QCD axion $M_0\sim 10^{-10}M_\odot$, however for an axion-like particle $M_0$ can approach $M_\odot$ or higher. Using the Press-Schechter formalism we compute the mass function of halos formed by hierarchical structure formation from these seeds. We compute the concentrations and collapse times of these halos and show that they can grow to be as massive as $10^6M_0$.  Within the halos, miniclusters likely remain tightly bound, and we compute their gravitational microlensing signal taking the fraction of axion dark matter collapsed into miniclusters, $f_{\rm MC}$, as a free parameter.  A large value of $f_{\rm MC}$ severely weakens constraints on axion scenarios from direct detection experiments. We take into account the non-Gaussian distribution of sizes of miniclusters and determine how this effects the number of microlensing events.  We develop the tools to consider microlensing by an extended mass function of non-point-like objects, and use microlensing data to place the first observational constraints on $f_{\rm MC}$.  This opens a new window for the potential discovery of the axion. 

\end{abstract}

\maketitle
\section{Introduction}

Models of particle dark matter (DM) can be broadly classified into two types: thermal and non-thermal. The prototypical thermal candidate is the Weakly Interacting Massive Particle~\cite{1996PhR...267..195J,2005PhR...405..279B}. The prototypical non-thermal candidate is the axion~\cite{2008LNP...741...19S,Marsh:2015xka}. 

In this paper, we explore astrophysical implications and constraints on axion DM in the so-called \emph{miniclusters} scenario~\cite{Hogan:1988mp}. In this scenario, dense lumps of axion DM form from the dynamics of symmetry breaking which leads to the axion production in the first place. Both the mass and the mass function of the miniclusters are determined by the axion particle mass, $m_a$, and in some cases miniclusters can be massive enough, dense enough, and abundant enough to impact astrophysical observables such as gravitational microlensing. Thus, searches for axion miniclusters are related to searches for non-particle DM candidates including MAssive Compact Halo Objects (MACHOs)~\cite{1986ApJ...304....1P,Griest:1990vu} and primordial black holes (PBHs)~\cite{1971MNRAS.152...75H,1974MNRAS.168..399C,1975ApJ...201....1C}.

The axion is a pseudo-Nambu-Goldstone boson of a spontaneously broken global $U(1)$ symmetry, known as a Peccei-Quinn (PQ) symmetry~\cite{pecceiquinn1977}. PQ symmetry breaking occurs when the temperature of the Universe drops below the symmetry breaking scale $f_a$. The cosmology of the axion, and consequent phenomenology of axion DM, is determined by the cosmic epoch during which symmetry breaking occurs. Miniclusters are formed if the PQ symmetry is broken after smooth cosmic initial conditions are established (we refer from now on specifically to inflation~\cite{2016A&A...594A..20P}, but the distinction is not important). The minicluster mass is determined by the axion Compton wavelength, with larger minicluster masses for lighter axions.

The cosmology and astrophysics of miniclusters comprised of the QCD axion~\cite{weinberg1978,wilczek1978,1979PhRvL..43..103K,1980NuPhB.166..493S,Zhitnitsky:1980tq,1981PhLB..104..199D} was explored in depth in a series of seminal papers by Kolb and Tkachev in the early 1990s~\cite{1993PhRvL..71.3051K,1994PhRvD..49.5040K,Kolb:1994fi,Kolb:1995bu}. At that time, microlensing observations were not yet precise enough to constrain QCD axion miniclusters. Modern observations are much improved, and here we perform the necessary improved theoretical calculations of the event rate to compare to the data. Furthermore, interest has been growing in the broader class of axion DM models, or ``axion-like particles'', inspired in part by the theoretical observation that light ($m_a\ll 1\,{\rm eV}$) axions are abundant in the string theory landscape~\cite{2006JHEP...06..051S,axiverse}. It is thus timely to reconsider the work of Kolb and Tkachev beyond the QCD axion. We take on this task, following also the work of Refs.~\cite{2007PhRvD..75d3511Z,2017JHEP...02..046H,Fairbairn:2017dmf}.

The fraction of axion DM bound up in miniclusters is $f_{\rm MC}$, and is not known \emph{a priori} from theoretical calculations. If $f_{\rm MC}\approx 1$ then axion DM direct detection is severely limited, as encounters with a minicluster in our own Galaxy will be exceedingly rare. This has profound implications for axion direct detection. If $f_{\rm MC}\approx 1$, then direct axion detection with ADMX~\cite{2010PhRvL.104d1301A}, MADMAX~\cite{2016arXiv161104549M,2016arXiv161105865T}, or the myriad other proposed experiements targeting the QCD axion in the $m_a\sim 1\,\mu{\rm eV}$ mass range would be much more difficult. Null results could erroneously be interpreted as excluding the axion, when in fact it was just ``hiding'' in miniclusters. On the other hand experiments like ARIADNE~\cite{2014PhRvL.113p1801A}, which detect axions via the forces they mediate~\cite{Moody:1984ba} could still detect axions even if $f_{\rm MC}=1$ or axions are not the DM. In practice even if $f_{\rm MC}\approx 1$ initially, tidal disruption will affect a few percent of miniclusters allowing for some prospect of DM direct detection, or even a measurement of $f_{\rm MC}$ in the laboratory~\cite{2016JCAP...01..035T,2017arXiv170103118O}. In the present study we take $f_{\rm MC}$ as a free parameter that, if sufficiently constrained by observation, could be used to rule out entire classes of models for axion production. Our method proposes to measure the minicluster fraction using gravitational microlensing.

In the course of considering this signal, we address issues of structure formation with miniclusters and present a series of possible models. A new consequence of this investigation is the computation of the mass function of minicluster halos (which we term ``MCHs''). Depending on the merging and tidal stripping of miniclusters, the MCH mass function may or may not be the appropriate mass function to consider for microlensing. There may be, however, other observational consequences of the existence of MCHs for which the mass function will be an important quantity. 

We begin in Section~\ref{sec:characteristic_mass}, where we present some introductory basics on miniclusters. There has been very little study in the literature on the subsequent gravitational evolution of axion miniclusters after their formation (although see \cite{2007PhRvD..75d3511Z}).  In Section~\ref{sec:mass_function} we therefore present a new computation of the MCH mass function following the Press-Schechter~\cite{Press:1973iz} formalism. The form of the mass function is seen to arise simply from basic physical principles, and can be easily parameterised. 

Miniclusters are extended objects and cannot be considered as point-like lenses (radius larger than the Einstein radius). We discuss minicluster density profiles in Section~\ref{sec:profiles}. Section~\ref{sec:microlensing} presents tools to compute the lensing signal from a mass function of non-point-like lenses. 

We apply our minicluster lensing methodology to the historical EROS survey~\cite{Tisserand:2006zx} and to the very recent Subaru Hyper Suprime Cam observations~\cite{2017arXiv170102151N} in Section~\ref{sec:results}. We conclude in Section~\ref{sec:conclusions}. 

Since miniclusters can only be produced if the PQ symmetry is broken after inflation, the axion parameters are constrained by the precise prediction for the axion DM relic density in this scenario. In particular, the relic density can be too large if the axion domain wall number $N_{\rm DW}>1$, due to the late decay of domain walls. Thus, in the standard QCD axion model, miniclusters are a generic prediction only of the ``KSVZ"-type models, with the scenario in the ``DFSZ"-type models being somewhat more complicated.\footnote{A comprehensive discussion of the variations on these general themes can be found in Ref.~\cite{2017PhRvL.118c1801D}.} In this paper we consider only models with $N_{\rm DW}=1$, where the minicluster scenario is simplest. Appendix~\ref{appendix:relic_density} collects results on the axion relic density, early time cosmology and thermal history, and determines the range of axion masses and decay constants for which the minicluster scenario can occur. 

Finding the relic density in general requires many complex computational steps: lattice QCD for the temperature dependence of the axion mass, and lattice field theory for the classical evolution of the axion field. These calculations have only been performed in the literature for the QCD axion. Our relic density computation uses analytic approximations to allow us to treat ALPs, and to account for uncertainties in the QCD calculation in a parameterised way. Our final constraints account for the uncertainties by allowing for a window in the relation between minicluster mass and axion mass for the QCD axion.

Appendix~\ref{appendix:modelling_mf} presents the theoretical modelling of the MCH mass function, and some analytic results. Appendix~\ref{appendix:non-gauss} discusses how non-Gaussianity of the axion initial conditions affects the MCH mass function. Appendix~\ref{appendix:axion_stars} discusses how formation of ``axion stars'' might modify our lensing results in the case of axion-like particles with temperature independent mass.

We use \emph{Planck} (2015)~\cite{planck_2015_params} cosmological parameters $h=0.67$, $\Omega_m=0.32$, $\Omega_ch^2=0.12$, $z_{\rm eq}=3402$, and for particle physics quantities we use natural units where $c=\hbar=1$. Throughout this work we consider a standard thermal history of the Universe after the inflationary epoch. 

\section{Axion Miniclusters}
\label{sec:characteristic_mass}

\begin{figure}
\vspace{-0.2em}\includegraphics[width=\columnwidth]{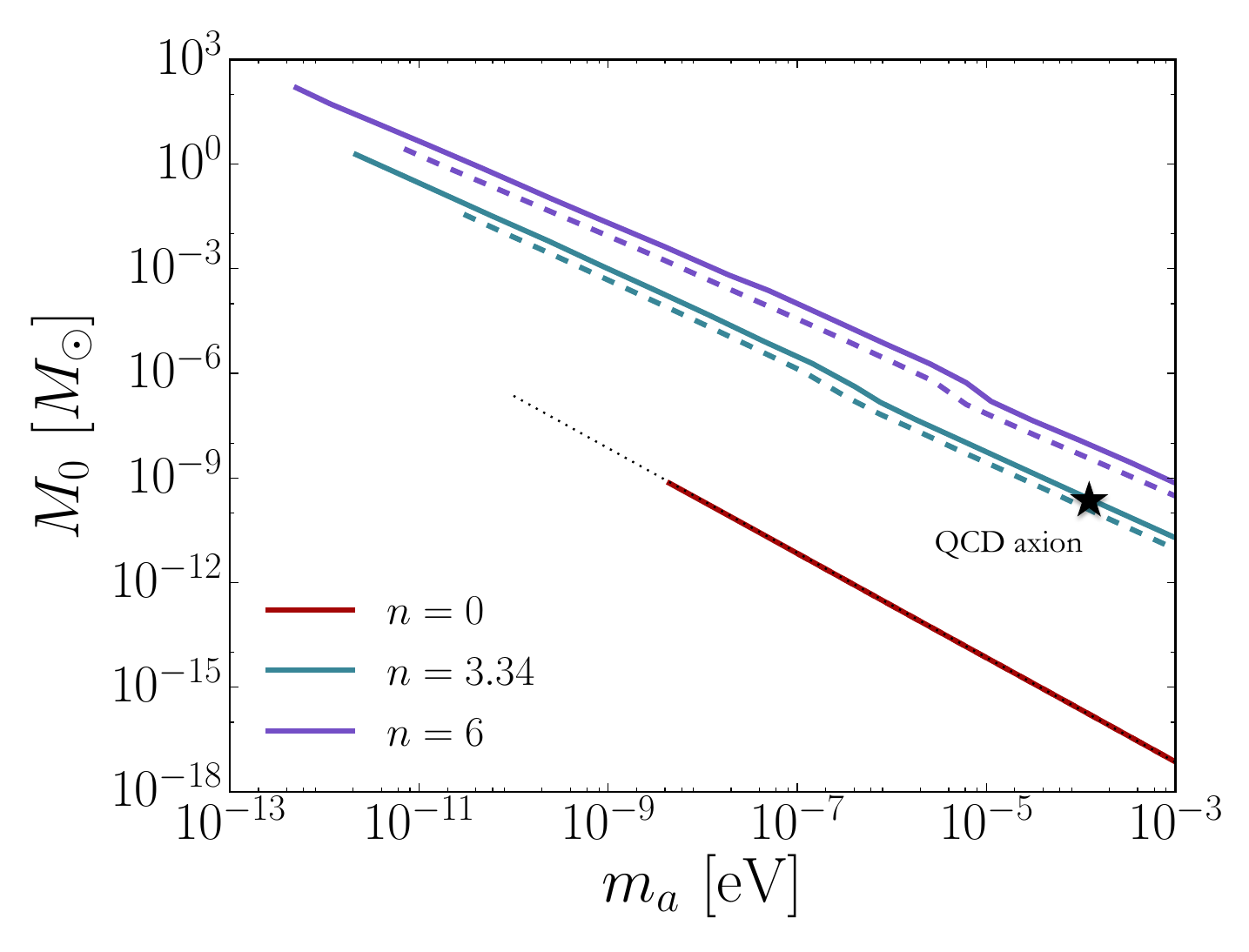}
\vspace{-2.5em}\caption{{\bf The Characteristic Minicluster Mass}: We plot $M_0$, as a function of the axion mass, $m_a$, for different temperature evolutions of the axion mass parameterised by index $n$. Solid lines show the most realistic assumptions about the relic density, while dashed lines relax those assumptions (see Appendix). When the axion mass is temperature independent ($n=0$), the two scenarios are equivalent for minicluster mass. The thin dotted line shows the approximation Eq.~\eqref{eqn:m0_n0_approx}. Lines terminate at a lower bound on $m_a$ set by the relic abundance.}  
\label{fig:M0_mass}
\end{figure}

This section gives the briefest outline of the minicluster scenario to establish some important language and physical scales. The main results are given in Fig.~\ref{fig:M0_mass} and Eq.~\eqref{eqn:m0_n0_approx}.

There are two energy scales that define the cosmological evolution of the axion field: the \emph{decay constant}, $f_a$, and the \emph{mass}, $m_a$. These two energy scales determine the two most important epochs in the life of a young axion.

The axion is the angular degree of freedom of a complex scalar field, $\varphi$, with a global $U(1)$ symmetry that is spontaneously broken by the potential:
\be
V(\varphi) = \frac{\lambda}{4!}\left( |\varphi|^2-\frac{f_a^2}{2} \right)^2 \, .
\ee
Spontaneous symmetry breaking (SSB) occurs when the temperature of the Universe cools to $T\lesssim f_a$.\footnote{The precise critical temperature for the phase transition from thermal field theory is calculated in e.g. Ref.~\cite{2016arXiv161001639B}.} After SSB, the complex field is given by $\varphi=(f_a/\sqrt{2})e^{i\phi/f_a}$, with $\phi$ the (real) axion field.

If PQ symmetry is broken before or during inflation then the axion field takes on a uniform value across the observed Universe, with the addition of small isocurvature perturbations from the finite temperature during inflation, and density perturbations inherited from the adiabatic perturbations in the hot Big Bang plasma. On the other hand, if PQ symmetry is broken after inflation then topological defects and large amplitude axion field fluctuations are present on scales of order the horizon size at symmetry breaking~\cite{Hogan:1988mp,Sikivie:1982qv,1987PhLB..195..361H}. In this case, since each horizon volume is causally disconnected, the axion field is uncorrelated across different horizon volumes and drawn from the distribution $\phi/f_a\in \mathcal{U}[-\pi,\pi]$. 

The Kibble mechanism~\cite{1976JPhA....9.1387K} smoothes the axion field on the horizon scale until such a time that the axion mass becomes cosmically relevant, which defines the second important epoch in the life of the axion. At this epoch, the topological defects decay, and the axion field is left with large amplitude isocurvature fluctuations on the horizon scale.\footnote{It is important to realise that these minicluster isocurvature fluctuations seeded by the PQ phase transition are unrelated to the ``axion isocurvature mode'' constrained by large scale CMB anisotropies~\cite{2009ApJS..180..330K}. The ``axion isocurvature mode'' is produced by vacuum fluctuations during inflation in the scenario when PQ symmetry is broken before inflation. The fluctuations are thus scale invariant, and contribute to large angle anisotropies. The minicluster isocurvature mode is not scale invariant: the power dies off rapidly on scales larger than the horizon size at the QCD phase transition ($k<k_0$) and does not contribute to the observed CMB anisotropies.} These fluctuations of order $f_a$ between horizon volumes provide the initial conditions for the axion field on small scales. It is these fluctuations that subsequently evolve into gravitationally bound miniclusters. Thus, as long as $f_a$ is low enough that SSB occurs  during the normal thermal evolution of the Universe, then miniclusters are a logical possibility. 

The time $t_0$ when the axion mass becomes significant is given by $3H(t_0)\approx m_a$, with $H$ the Hubble rate. The axion acquires its mass due to non-perturbative effects such as instantons~\cite{1981RvMP...53...43G}, which evolve with temperature as $m_a=m_{a,0}(T/\mu)^{-n}$. Therefore the critical time $t_0$ depends on the index $n$ giving the temperature evolution. From this time onwards the axion field oscillates about its own quadratic potential minimum, and the equation of state for the background axion energy density becomes the same as that of pressureless matter~\cite{1983PhLB..120..127P,1983PhLB..120..133A,1983PhLB..120..137D}. The epoch when axion oscillations begin thus determines the axion relic density. Appendix~\ref{appendix:relic_density} computes the range of $(m_a,f_a)$ for various $n$ for which axions provide the total DM relic density while having $f_a$ small enough for SSB to occur after inflation.

The initial axion fluctuations laid down by SSB remain smooth up to scales of order the horizon size at $t_0$. From this point on the density perturbations grow under gravity as usual, eventually collapsing into the gravitationally bound objects known as miniclusters. The total mass of axion DM contained within the horizon at time $t_0$ sets the characteristic minicluster mass, $M_0$, given by:
\be
M_0 = \bar{\rho}_a\frac{4}{3}\pi \left(\frac{\pi}{\mathcal{H}(t_0)}\right)^3\, ,
\label{eqn:M0_def}
\ee
where $\bar{\rho}_a$ is the energy density in axions today, and we have used the fact that the comoving wavenumber associated with the horizon size at this time is $k_0=\mathcal{H}(t_0)$ where $\mathcal{H}=aH$ is the conformal Hubble rate. 

In the case of a temperature independent axion mass, and approximating the number of relativistic degrees of freedom as constant in $T$, a reasonable approximation to $M_0$ can be obtained by using $(H/H_0)^2=\Omega_m (1+z_{\rm eq})^{-1} a^{-4}$ to find $a(t_0)$. A fudge factor of two leads to a good agreement with the full numerical calculation using $g_{\star}(T)$: 
\begin{align}
M_0(m_a,n=0)\approx& 2.3 \times 10^{-7}M_\odot \left(\frac{m_a}{10^{-10}\text{ eV}} \right)^{-3/2}\nonumber \\ &\left(\frac{\Omega_c h^2}{0.12} \right) \left(\frac{\Omega_m}{0.32} \right)^{-3/4} \left(\frac{1+z_{\rm eq}}{3403} \right)^{3/4}\, .
\label{eqn:m0_n0_approx}
\end{align}
This approximation is shown as the thin dotted line in Fig.~\ref{fig:M0_mass}.

In the general case, the characteristic minicluster mass $M_0$ is a function of the axion mass, the axion decay constant, and the index $n$ determining the temperature evolution of the mass. Fixing the decay constant as a function of the mass from the relic density, we show $M_0(m_a,n)$ for various $n$ in Fig.~\ref{fig:M0_mass}. The lines in Fig.~\ref{fig:M0_mass} for axion-like particles terminate at a lower bound on $m_a$. For masses below this bound axions cannot produce the correct relic density while maintaining a low enough $f_a$ to be of relevance for miniclusters. The lower bound on $m_a$ translates into an upper bound on $M_0$, which is around $10^3 M_\odot$ for $n=6$ temperature evolution.\footnote{Constraints from the Lyman-$\alpha$ forest flux power spectrum~\cite{2007PhRvD..75d3511Z} affect miniclusters with masses $M_0\gtrsim 10^4 M_\odot$, which can be achieved in certain exotic scenarios for symmetry breaking~\cite{2017JHEP...02..046H}.}

The QCD axion has a known temperature dependent mass with $n=3.34$ from the ``interacting instanton liquid'' model for the QCD topological susceptibility~\cite{Wantz:2009it}, which is consistent with the results from lattice simulations ($n\approx 3.55\pm 0.30$~\cite{2016PhLB..752..175B,Borsanyi:2016ksw}). The mass is given by $m_{a,{\rm QCD}}=6\,\mu {\rm eV} (10^{12}\text{ GeV}/f_a)$. For our modelling of the relic density the decay constant must be $f_a\approx 10^{10}\text{ GeV}$ leading to an axion mass $50\,\mu {\rm eV}\lesssim m_a\lesssim 200\,\mu {\rm eV}$. This implies a characteristic minicluster mass $M_0\approx 1.8\times 10^{-10}M_\odot$, in broad agreement with other estimates~\cite{1994PhRvD..49.5040K,Kolb:1994fi,Kolb:1995bu,2017JHEP...02..046H}. 

Our value of $M_0$ differs from the characteristic mass defined in e.g. Ref.~\cite{Kolb:1995bu}, who take the mass within a cubic volume of size the inverse horizon wavevector, rather than the spherical volume from the physical wavelength. This makes our definition of $M_0$ larger by a factor of $4\pi^4/3\approx 130$. We believe our definition captures the symmetries of minicluster formation better, and also better represents their likely formation history (see Section~\ref{sec:profiles}). In practice the constraints on miniclusters from microlensing cover a broad range of masses and even a factor of $10^2$ in $M_0$ does not have a large impact.

A far larger uncertainty is introduced if the axion string length parameter, $\xi$, rather than the horizon size, defines $M_0$. This factor is historically uncertain in simulations: it is currently favoured to be $\xi=1.0\pm 0.5$~\cite{2012PhRvD..85j5020H}, but could be much larger, with larger $\xi$ leading to smaller $M_0$. To an extent the effect of the uncertainty in $\xi$ on $M_0$ is captured by the uncertainty in the relic density (see Appendix~\ref{appendix:relic_density}).

\begin{figure}
\vspace{-0.2em}\includegraphics[width=\columnwidth]{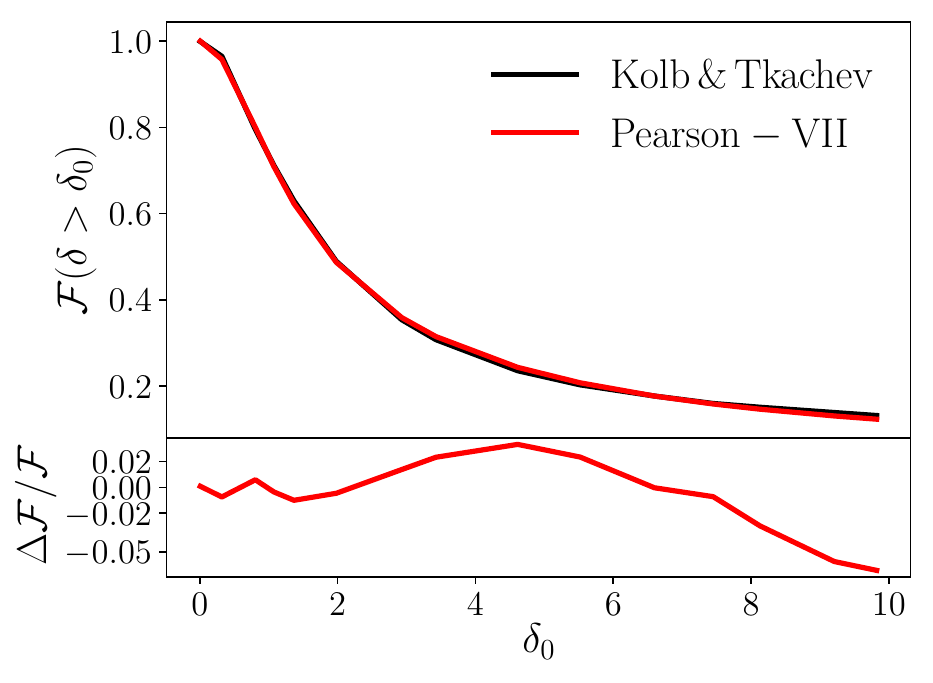}
\vspace{-2.5em}\caption{{\bf Minicluster Overdensity Distribution}: We show the cumulative mass fraction of miniclusters with overdensity parameter $\delta>\delta_0$. The black line shows the simulation results of Ref.~\cite{Kolb:1995bu}, which we have fit using a Pearson-VII distribution. The overdensity distribution determines the halo concentration parameter (i.e. compactness) of miniclusters.}  
\label{fig:fit_tkachev}
\end{figure}

The minicluster characteristic density, $\rho_{\rm MC}$, is another important quantity, since it sets the typical radius of a minicluster, and thus it's concentration. The characteristic density is given in terms of the initial overdensity parameter, $\delta$, by~\cite{Kolb:1994fi}:
\be
\rho_{\rm MC}=140\delta^3(1+\delta)\bar{\rho}_a(z_{\rm eq})\, ,
\label{eqn:minicluster_density_KT}
\ee
which can be derived using spherical collapse, and understood by considering how regions of large $\delta$ collapse when $z\gg z_{\rm eq}$. The initial condition simulations of Ref.~\cite{Kolb:1995bu} noted that while the mass of miniclusters was approximately fixed to $M_0$, the characteristic density showed a wide, non-Gaussian, variation due to the anharmonicities in the axion potential. 

Fig.~\ref{fig:fit_tkachev} shows the cumulative mass fraction, $\mathcal{F}(\delta>\delta_0)$, taken from digitising Fig.~2 of Ref.~\cite{Kolb:1995bu}. The non-Gaussian distribution is well fit by a Pearson-VII-type distribution:
\be
\mathcal{F}(\delta>\delta_0)=\frac{1}{[1+(\delta_0/a_1)]^{a2}}\, ,
\ee
with $a_1=1.023$, $a_2=0.462$ found by a non-linear least squares fit. 

We are not in possession of numerical simulations that would allow us to fully characterise the non-Gaussianity of the minicluster density field. In order to make some progress we assume that the non-Gaussianity in overdensity and minicluster size given by $\mathcal{F}$ is a small-scale phenomenon affecting only the stochastic distribution of minicluster sizes. 

This approach to model the non-Gaussian effects on the density profile can be partially justified by considering that the non-Gaussianities that affect the density profile are caused by the axion self-interactions. The large scale perturbations are of smaller amplitude (the dimensionless power spectrum falls as $k^3$ for small $k<k_0$). Small density perturbations mean small axion field perturbations. At small field values the axion potential is quadratic, giving free-field behaviour that should be close to Gaussian. Appendix~\ref{appendix:non-gauss} discusses non-Gaussian corrections to the large scale clustering of miniclusters and the minicluster halo mass function induced by the white noise fluctuations of $\theta$.

\section{Structure Formation with Miniclusters}
\label{sec:mass_function}

The discussion in the previous section asserts that the initial conditions for the axion field on small scales caused by SSB lead to the collapse of objects of mass $M_0$ around matter-radiation equality. In the following section we consider how miniclusters, once initially formed, go on to merge into larger bound structures, which we term ``minicluster halos'', or MCHs. This process has not been studied in great detail before. It deserves attention because the behaviour will be quite different from cold dark matter: the initial conditions are isocurvature, structure formation begins much earlier, and the power spectrum is truncated.  We address this situation by computing the standard halo mass function from linear growth of the minicluster initial conditions.  

MCHs are small scale structures; they are substructure within the larger-scale DM halos formed by the scale-invariant adiabatic initial conditions on large scales (see e.g. the combined power spectrum in Ref.~\cite{2007PhRvD..75d3511Z}).

The logic of computing the mass function for the small scale minicluster isocurvature initial conditions independently from the usual adiabatic large scale cosmology is the following. As we will see, miniclusters collapse very early, at $z\approx z_{\rm eq}$. Galactic halos like the Milky Way are formed from the dominant, but small amplitude, adiabatic perturbations on large scales and collapse at much lower redshifts. These galactic halos are, however, still formed of axion DM, and thus of the miniclusters formed early on. We treat these two periods of gravitational collapse independently, and assume that the minicluster mass function established early on provides the substructure mass function on small-scales within the larger galactic halos. In other words, the minicluster mass function is equivalent to the ``field'' mass function within the larger patches that collapse later on into galaxies. Since miniclusters collapse and ``freeze out'' from the expansion early on, they can be treated independently from the large scales. 

In the explicit examples in this section we use a Heaviside initial power spectrum to model the effects of the Kibble mechanism, and a Gaussian window function to define the mass variance. We consider other possibilities, and give analytic results, in Appendix~\ref{appendix:modelling_mf}. The examples also use the $M_0(m_a)$ relation for an axion-like particle with $T$-independent mass.

We work in conformal time $\tau$, with ${\rm d}t=a{\rm d}\tau$ and $a$ the scale factor of the Friedmann-Lema\^itre-Robertson-Walker metric. For simplicity of presentation, we begin our numerical calculations in the matter-dominated era once all $T$-dependence of the axion mass and $g_\star$ can be neglected. We comment briefly on the transfer function in the radiation era. 

\subsection{Evolution of Density Perturbations}
\subsubsection{Initial Conditions}
\begin{figure}
\vspace{-0.2em}\includegraphics[width=\columnwidth]{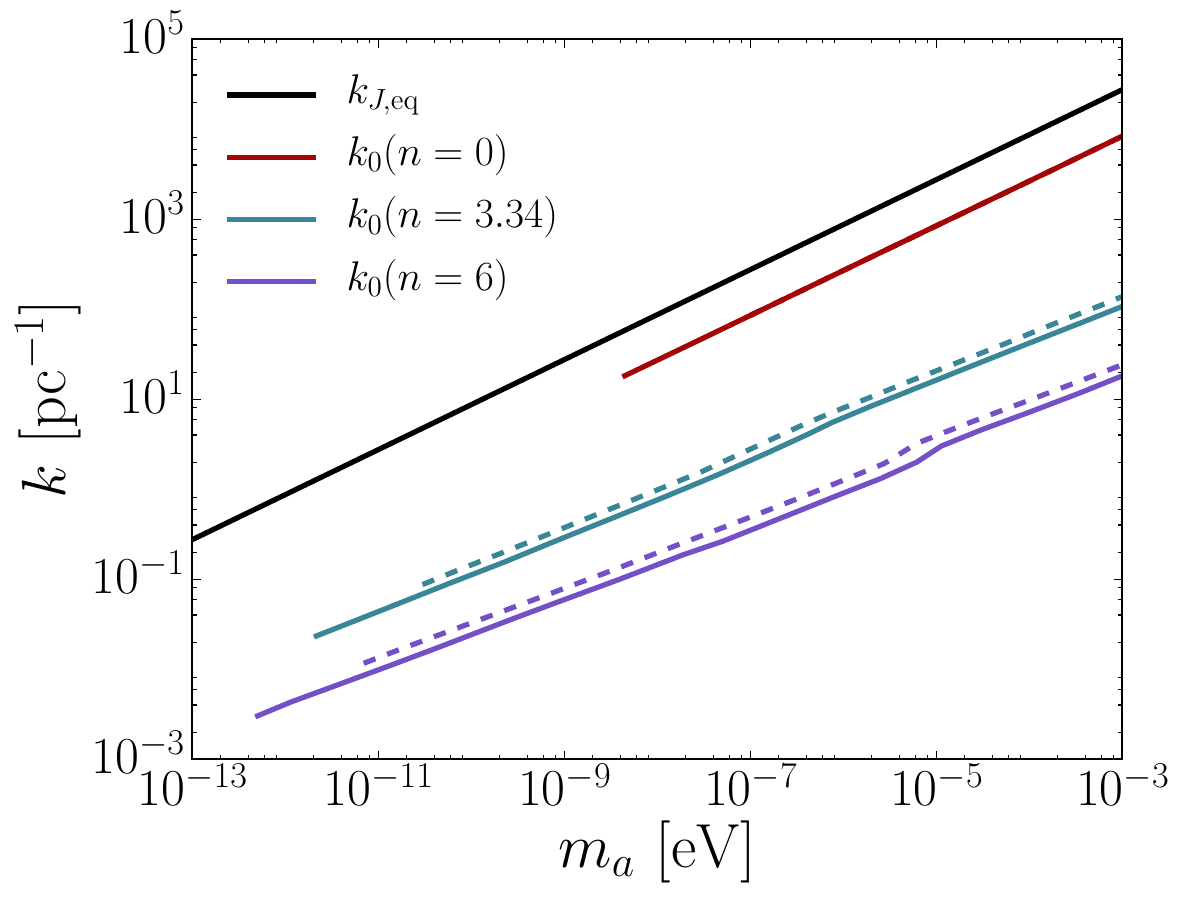}
\vspace{-2.5em}\caption{{\bf Cut-offs in the Power Spectrum}: We plot the wavenumbers that cut-off the axion power spectrum, $k_0$ and $k_{J,{\rm eq}}$ as a function of axion mass. For $k_0$, this depends on the temperature evolution of the mass as parameterised by $n$, and on details of the relic density computation. The Jeans wavenumber is fixed only by the zero-temperature mass. We notice that we always have $k_{J,{\rm eq}}>k_0$, which justifies that the Jeans scale can be ignored to a first approximation.}  
\label{fig:k0_plot}
\end{figure}

We define the axion density perturbation as $\delta_a(\textbf{r},\tau)=[\rho_a(\textbf{r},\tau)-\bar{\rho}_a(\tau)]/\bar{\rho}_a(\tau)$. The power spectrum $P_{\bf k}$ is defined as the Fourier transform of the two point correlation function:
\be
\xi({\bf r})=\langle \delta({\bf x})\delta ({\bf x+r})\rangle \, .
\label{eqn:2pt_corr}
\ee
We wish to specify the value of the power spectrum at the initial time of our study $\tau_0$: the time when the axion field begins oscillating. At $\tau_0$, the Kibble mechanism assures us that the axion field is fixed to a constant value over each causal horizon, and randomly distributed (white noise) over the different horizons. Hence, we can approximate the power spectrum by a sharp-$k$ function, cut at the typical (comoving) size of a horizon at $\tau_0$:
\begin{equation}
    P_k(\tau_0)=P_0\Theta(k_0-k) \, ,
\end{equation}
where $k_0=\mathcal{H}(\tau_0)$ and $\Theta (x)$ is the Heaviside function. 

We must now find the normalisation, $P_0$. From Eq.~\eqref{eqn:2pt_corr} it is easy to show that the integral of the power spectrum is (e.g. Ref.~\cite{Peacock:1999ye}):
\begin{equation}
    \displaystyle{\int}P_{\textbf{k}}(\tau)\frac{{\rm d}^3k}{(2\pi)^3}=\langle|\delta_a(\textbf{x},\tau)|^2 \rangle \, .
    \label{parsevalpowspec}
\end{equation}
From our ansatz for $P_k(\tau_0)$, the left hand side is equal to $\frac{4}{3}\pi k_0^3 P_0$. To find the value of $\int|\delta_a(\textbf{x},\tau_0)|^2d^3x$, we note that $\rho_a\propto \theta^2$, where $\theta\in [-\pi,\pi]$ is the uniformly distributed axion field. Thus the mean value of $|\delta_a(\textbf{x},\tau_0)|^2$ is $4/5$, and we find:
\be
P_0=\frac{24}{5}\pi^2k_0^{-3} \, .
\ee

This equation sets the initial condition on the modes of the perturbations. These initial conditions are of isocurvature type during the radiation-dominated epoch, and can be normalised to have $\delta_a(\tau_0)=1$ with all other perturbations absent.\footnote{This is sufficient accuracy for our purposes of computing the mass function in the matter era. For greater accuracy in the radiation era at $\tau>\tau_0$ for modes near the Jeans scale see the early-time expansion of Ref.~\cite{2016ApJ...830..155T}. For the complete treatment including $\tau<\tau_0$ see Refs.~\cite{marsh2013,Hlozek:2014lca}.}

\subsubsection{The Behaviour in the Radiation-Dominated Era}

It is well-known that sub-horizon isocurvature perturbations undergo only a very small amount of logarithmic growth in the radiation-dominated era, and that the transfer function up to matter-radiation equality is approximately $k$-independent for large $k$~\cite{1986MNRAS.218..103E}. The amplitude, i.e. the actual amount of growth in the raditation era, is somewhat cosmology-dependent. For our cosmological parameters using the analytic transfer function of Ref.~\cite{1986MNRAS.218..103E} we find: 
\be
\frac{\delta (z_{\rm eq})}{\delta_i} \approx 1 \, .
\ee

This approach is also taken in e.g. Refs.~\cite{Hogan:1988mp,2007PhRvD..75d3511Z,2017JHEP...02..046H} for the case of miniclusters. We have verified by full numerical solution of the Boltzmann equations that the transfer function indeed takes the form stated, and use this result in the following. 

\subsubsection{The Behaviour in the Matter-Dominated Era}

We consider the sub-horizon limit, assuming that most of the sizes we are interested in are sub-horizon-sized at matter-radiation equality. The equation of motion for the axion overdensity in an axion-dominated Universe on times $\tau\gg\tau_0$ is~\cite{Marsh:2015xka}: 
\be
\delta_a''+\frac{a'}{a}\delta_a'+(k^2c_s^2-4\pi Ga^2\bar{\rho}_a)\delta_a=0 \, ,
\label{growtheq_mat-dom}
\ee
where primes denote derivatives with respect to conformal time, and $c_{s}^2\approx k^2/4m_a^2a^2$ is the axion effective sound speed. The sound speed leads to a Jeans scale, which balances pressure and gravity~\cite{khlopov_scalar}:
\begin{align}
k_J&=(16\pi G a \rho_{a0})^{1/4}m_a^{1/2} \nonumber \, , \\
&=66.5\times10^6a^{1/4}\left(\displaystyle{\frac{\Omega_m h^2}{0.12}}\right)^{1/4}\left(\displaystyle{\frac{m_a}{10^{-10} \text{ eV}}}\right)^{1/2} \text{ Mpc}^{-1} \, . 
\label{eq:JeansScale}
\end{align}

The Jeans scale sorts the modes into three categories: 
\begin{itemize}
    \item $k<k_J(\tau_{eq})$: 
    these modes are already under the Jeans mode at matter-radiation equality, their behaviour is the usual growing/decaying as soon as matter-radiation equality is reached, and matches the transfer function used in Ref.~\cite{2007PhRvD..75d3511Z};
    \item$k_J(\tau_{eq})<k<k_J(\tau_{today})$:
    these modes are bigger (physically smaller) than the Jeans mode at matter-radiation equality, and as $k_J$ increases they cross the Jeans scale. The behaviour of these modes is to oscillate at the beginning of the matter-dominated era, then to follow the usual growing/decaying mode;
    \item$k>k_J(\tau_{today})$: 
    these modes are still today physically smaller (have larger $k$) than the Jeans mode, and still follow the oscillating behaviour. 
\end{itemize}
The Jeans scale will affect the growth of linear perturbations when $k_J<k_0$. However, as can be seen in Fig.~\ref{fig:k0_plot}, we notice that we always have $k_{J,{\rm eq}}>k_0$, which justifies that the Jeans scale can be ignored in our case to a first approximation when computing the variance and the evolution of the mass function. 
\begin{figure}[t!]
\includegraphics[width=\columnwidth]{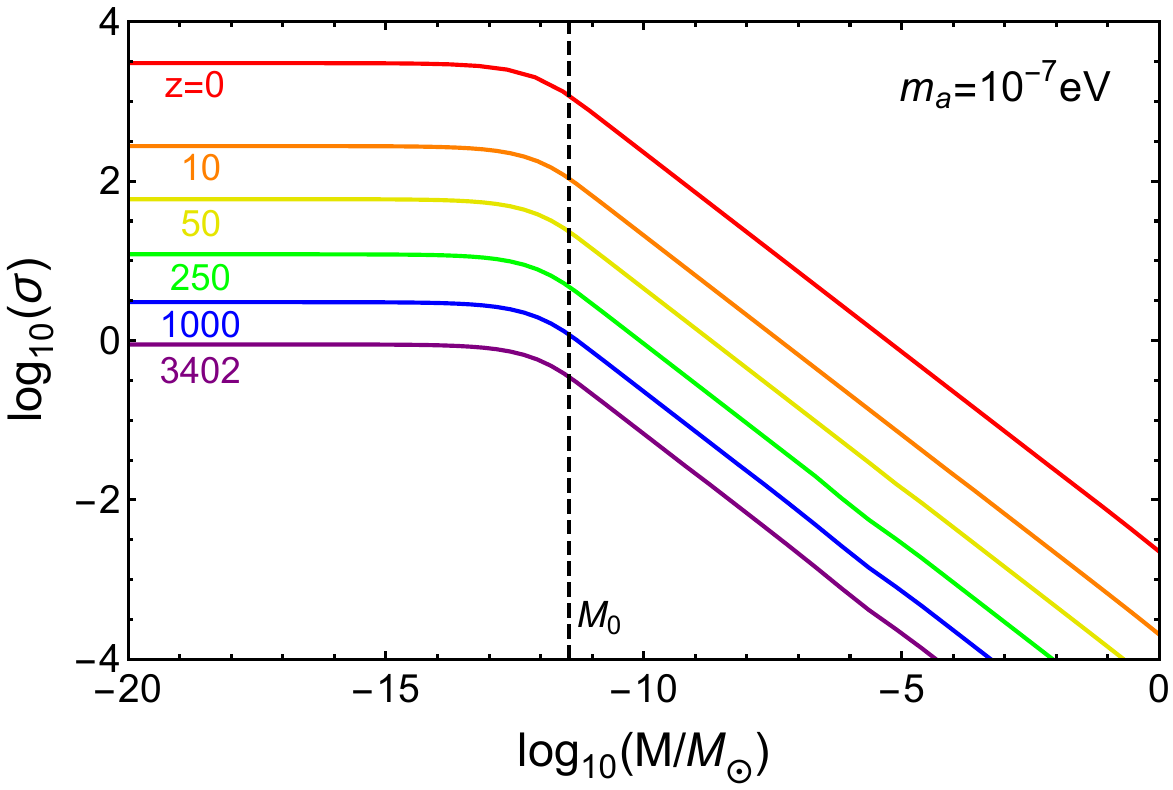}
  \caption{{\bf The RMS Density Fluctuation:} The curves show a $\sigma(M)\propto M^{-1/2}$ above $M_0$, which matches the RMS of white noise, as in Ref.~\cite{Hogan:1988mp}. The constant behaviour at low masses is due to the smoothed out axion field below the horizon at the phase transition, the effect of the Jeans scale cannot be seen in this diagram. The growth of the RMS mass fluctuation through time is due to the linear growing mode: $\delta\propto a$ above the Jeans scale. \label{sigmaMtimes}}
\end{figure}

\subsubsection{The RMS Mass Fluctuation}
\begin{figure*}
\begin{center}
$\begin{array}{@{\hspace{-0.1in}}l@{\hspace{-0.1in}}l}
\includegraphics[scale=0.45]{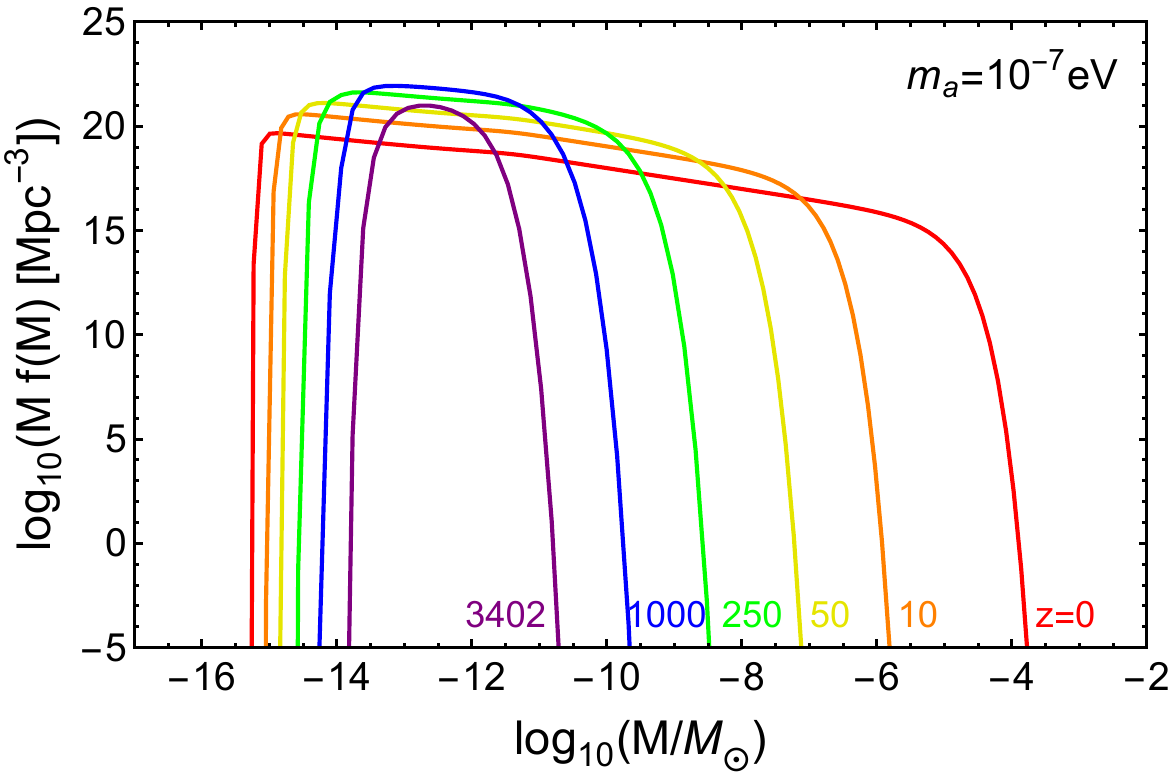} \hspace{0.8cm}
\includegraphics[scale=0.45]{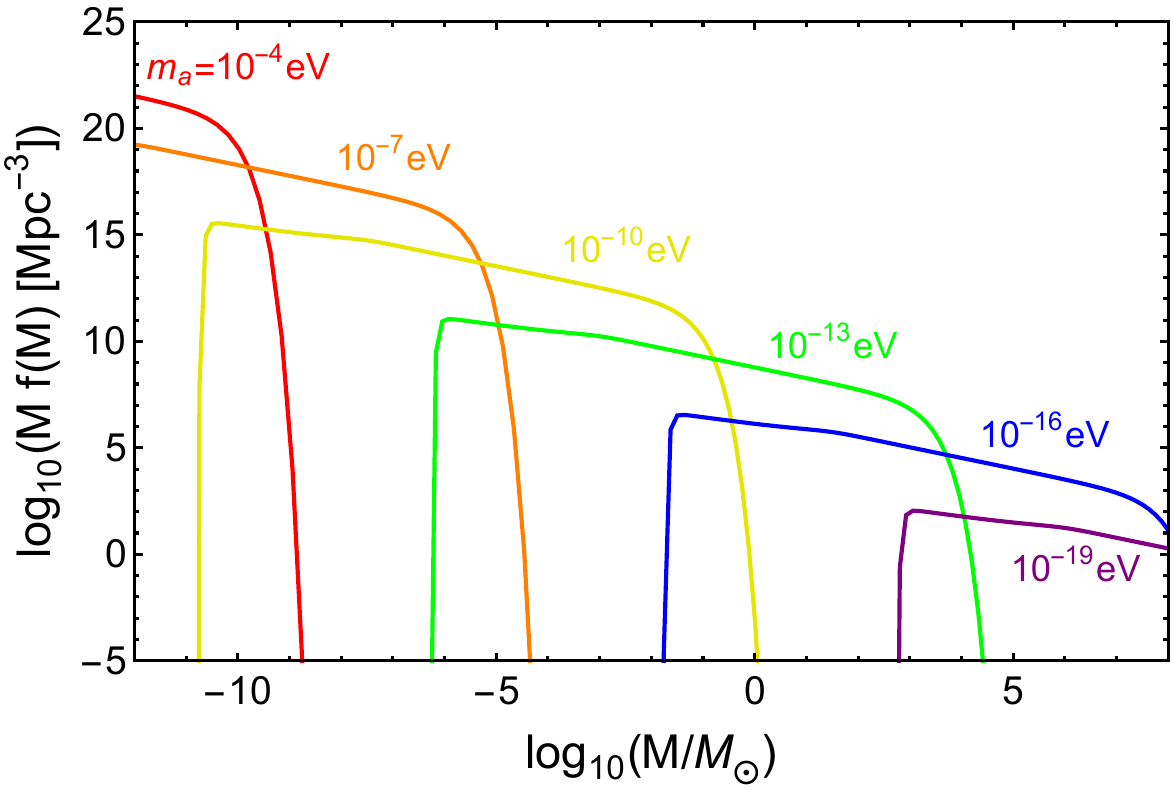}
 \end{array}$
 \end{center}
\caption{{\bf The MCH Mass Function:} \emph{Left Panel:} HMF as a function of time for fixed axion mass. The initial miniclusters at $M_0$ spread over time to form more massive objects due to hierarchical structure formation. Lighter objects are also formed as the late-time Jeans scale cutting off the mass function moves to scales smaller than $M_0$. \emph{Right Panel:} Minicluster mass function today for various axion masses. }
\label{fig:hmf_fig}
\end{figure*}

Let us define the variance of density fluctuations on a given scale by filtering the power spectrum with a window function of physical radius $R$:
\begin{align}
\sigma^2(R) &= \int \frac{{\rm d}k}{k} \Delta^2(k)|W(kR)|^2 \, , \nonumber\\ 
&= \int \frac{{\rm d}k}{k}\frac{k^3P(k)}{2\pi^2}|W(kR)|^2 \, .
\end{align}
For ease of numerical integrals, we take the window function to be a real-space Gaussian, which in Fourier space gives:
\begin{equation}
    W(kR)=e^{-k^2R^2/2}\, .
\end{equation}
We define the mass to be that contained within the comoving volume of the radius $R$ appropriate to the Gaussian window function, such that: 
\be
M=(2\pi)^{3/2} \bar{\rho}_{a,0}R^3\, .
\ee
Thus we have the mass variance, $(\delta M/M)^2\equiv \sigma^2(M)$ (see also Ref.~\cite{Kolb:1990vq}). The time dependence of $\sigma(M)$ comes from $\delta_a$, so that:
\begin{widetext}
\begin{equation}
    \sigma^2(M,\tau)=\int\frac{k^3P(k,\tau)}{2\pi^2}|W(k,M)|^2\frac{{\rm d}k}{k}=\int P(k,\tau_0)\mathcal{T}^2(k,\tau)|W(k,M)|^2\frac{k^2{\rm d}k}{2\pi^2}\, ,
\end{equation}
\end{widetext}
where we have defined the transfer function, $\mathcal{T}(k,\tau)$, to contain the evolution of the power spectrum. The above expression clearly separates the dependence of the mass fluctuation on the initial conditions, which depend on $k_0$, the dynamics, given by the transfer function, and the window function which we use to define the mass scale.

Our normalisation for the initial power spectrum sets the variance on the scale $R_0=\pi/k_0$ as:
\be
\sigma^2(R_0)=\frac{P_0}{2\pi^2}\int_0^{k_0}k^2 \exp \left[-\frac{\pi^2k^2}{k_0^2} \right] {\rm d} k\, ,
\ee
giving $\sigma(R_0)\approx 0.18$, which differs by a factor of two from the normalisation of Ref.~\cite{2007PhRvD..75d3511Z}, who consider quantum fluctuations of $\theta$ and find that the RMS density fluctuation on the scale $k_0$ is $\sigma\approx 1/(2\sqrt{2})\approx 0.35$. 

Fig.~\ref{sigmaMtimes} shows the behaviour of the RMS mass fluctuation at several moments in the matter-dominated era. It shows the white noise behaviour above $M_0$, with $\sigma(M)\propto M^{-1/2}$, and a smooth convergence under $M_0$, due to the sharp-k shape of the initial power spectrum. For this particular axion mass, the action of the Jeans scale is invisible, since the shape of $\sigma(M)$ is constant in time. 

\subsection{The MCH Mass Function}

We consider the formation of gravitationally bound structures from linear density perturbations using the analytic Press-Schechter formalism~\cite{Press:1973iz}. We use the original formalism, rather than modern updates such as the Sheth-Tormen~\cite{1999MNRAS.308..119S} as we are interested in only an approximate description of MCH formation and the mass function is subject to a number of theoretical uncertainties, which we discuss. 

The quantity $\delta_c$ is the critical overdensity  threshold for gravitational collapse, and plays a key role in the Press-Schechter formalism. In spherical collapse of cold dark matter it is given by $\delta_c\approx 1.686$, and is scale-independent. For every point in space, the probability to have $\delta > \delta_c $ using the filtered version of $\delta$ by $W_M$ is:
\begin{equation}
    p(\delta > \delta_c | W_M)=\displaystyle{\frac{1}{2}}\left(1-\text{erf}\left(\displaystyle{\frac{\delta_c}{\sqrt{2}\sigma(M)}}\right)\right)
\end{equation}
where $\sigma(M)$ is the RMS mass fluctuation we defined. We take the point of view where $\delta_c$ is time-independent, and put the time dependence onto $\sigma(M)$. Defining $f(M)$ such that $f(M)dM$ is the comoving number density of miniclusters in the range $dM$ around $M$, the Press-Schechter formalism gives:
\begin{equation}
    \displaystyle{\frac{M^2f(M)}{\bar{\rho}_{a0}}}=\left|\displaystyle{\frac{{\rm d} \ln \sigma}{{\rm d} \ln M}}\right|\sqrt{\displaystyle{\frac{2}{\pi}}}\displaystyle{\frac{\delta_c}{\sigma(M)}}e^{-\frac{1}{2}\left(\frac{\delta_c}{\sigma(M)}\right)^2}
\label{M2fM}    
\end{equation}
If $n(M)$ is the number density of structures of mass $M$, then the HMF, defined by ${\rm d}n/{\rm d}\ln M=Mf(M)$, is the comoving number density of structures of mass $M$ per logarithmic interval in masses. 

The low mass end of the HMF is subject to large theoretical uncertainty. The density field on scales $M<M_0$ (wavenumbers $k>k_0$) is non-Gaussian, and so the Press-Schechter formalism does not apply. The formalism can be applied perturbatively, or for certain special types of non-Gaussianity~\cite{2014MNRAS.439.3051M}, though such a calculation is beyond the scope of this work. The Press-Schechter mass function we present for $M<M_0$ applies only to ``the Gaussian part'' of the small-scale density field.

Even for a Gaussian field, on small scales there are theoretical uncertainties in the mass function in the case of a truncated power spectrum such as the axion power spectrum. With a real-space window function the formula Eq.~\eqref{M2fM} predicts that structures are formed on all scales below the non-linear scale even in the case of suppressed density perturbations. This is due to the asymptotic behaviour of the variance. However, a number of considerations modify this prediction and predict a cut-off in the mass function for $M\lesssim M_0$. These can be summarised as:
\begin{itemize}
\item {\bf ``Spurious Structure''}: Simulations with truncated power spectra display effects due to numerical discreteness on scales below the power spectrum cut-off~\cite{2007MNRAS.380...93W}.
\item {\bf Filter Dependence}: With a truncated spectrum the sharp-$k$ window function used to derive the Press-Schechter formula from the excursion set~\cite{1986ApJ...304...15B} predicts a cut-off in the mass function~\cite{2013MNRAS.433.1573S}.
\item {\bf Dynamical Effects}: The Jeans scale leads to a pressure on small scales that modifies the collapse barrier~\cite{2001ApJ...558..482B,Marsh:2013ywa}, the solution of the excursion set~\cite{2013MNRAS.428.1774B,2014ApJ...792...24P,2017MNRAS.465..941D}, and the formation of structure on small scales~\cite{Schive:2014dra,2016PhRvD..94l3523V}.
\end{itemize}

Appendix~\ref{appendix:modelling_mf} presents three different models for the mass function for $M<M_0$: a fit to remove structure below $M_0$, taken from the simulations of Ref.~\cite{2016ApJ...818...89S} (see also Ref.~\cite{2017PhRvD..95h3512C}); a sharp-$k$ filter, with mass normalised to the Gaussian filter on large scales~\cite{2013MNRAS.433.1573S}; a modified barrier at the Jeans scale, following the approximate implementation in Ref.~\cite{Marsh:2013ywa}. We show later that, for the range of MCH masses relevant for microlensing, and in particular for the particle mass range of the QCD axion, the cut-off dependence of the mass function for $M<M_0$ has negligible effect on the observables.

For the purposes of illustration, the MCH mass function computed with the modified barrier is shown in Fig.~\ref{fig:hmf_fig}. We show the mass function evolving over time for a fixed axion mass, and the mass function at $z=0$ for a range of axion masses. The $M_0(m_a)$ relation used in these examples is for a temperature-independent axion mass ($n=0$). The HMF is centred near $M_0$, and is cut off at high and low masses. The mass function spreads over time as structure formation progresses. For lower axion masses, the HMF is centred around larger MCH masses. 

\subsection{Parametrization of the Mass Function}
\label{sec:parameterize}
\begin{figure}
\center
\includegraphics[width=\columnwidth]{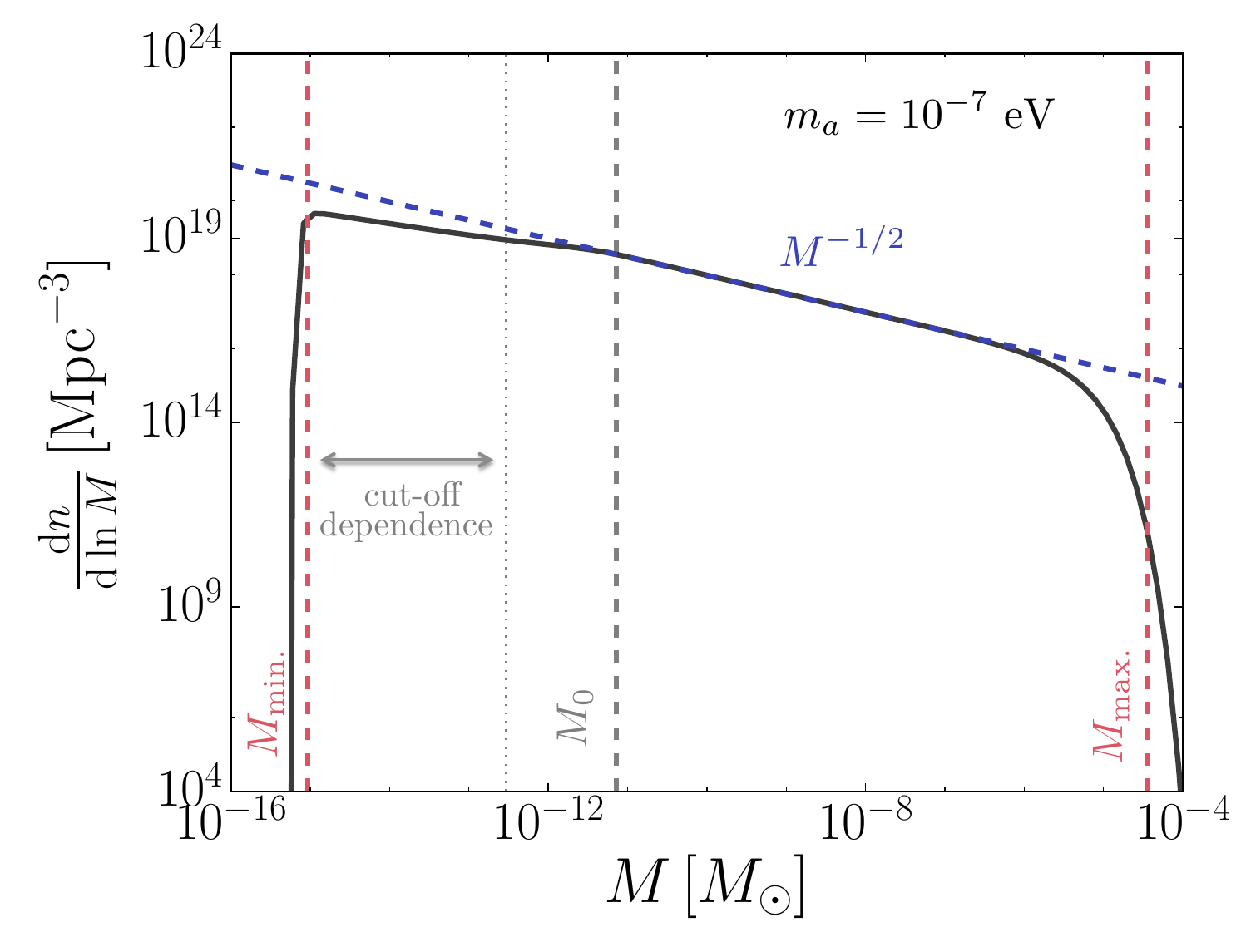}
  \caption{{\bf Parametrization of the MCH Mass Function}. The mass function can be well fit by two cut-offs and a single slope parameter, $M^{-1/2}$, derived form the white noise initial conditions. For the numerical calculation in the previous subsection (solid line), the normalization is fixed to be per unit volume. For the substructure mass function, we normalize by $f_{\rm MC}$ (see text). For $M\lesssim M_0/25$ where the variance becomes flat, there is cut-off dependence from the barrier, window function, and non-Gaussianities. For illustration we take $m_a=10^{-7}\text{ eV}$ and $n=0$ to use the analytic formula for $M_0$ and show the cut-off from the Jeans scale only.}
  \label{parametrize_HMF}
\end{figure}

The HMF for miniclusters that we have obtained can be parametrized by four quantities. We give an explicit parametrization of the HMF and show that (for arbitrary normalization) it matches our numerical calculations well. Since the parametric form is well understood, we can use this mass function to describe DM substructure within galactic halos, including the Milky Way. The substructure mass function is normalized by the host galaxy mass, and has units $[{\rm d}n/{\rm d}M]=[M]^{-1}$.

Our parameterized HMF is shown in Fig.~\ref{parametrize_HMF}, and compared to the numerical results from the previous subsection, for which the normalization per unit volume is fixed.

\subsubsection{Cut-offs in the HMF}

Cut-offs in the HMF are driven by the Gaussian term in Eq.~\eqref{M2fM}, the argument of which depends on $\delta_c(M)/\sigma(M,z)$, with the possible addition of a fit to remove spurious structure at low masses discussed above. We parameterize $\sigma(M)$ for the Heaviside initial power spectrum and Gaussian window function following Fig.~\ref{sigmaMtimes} as:
\begin{align}
\sigma(M<M_0/6.65)&=\sigma_0\approx 10^{3.4}\, , \\
\sigma(M>M_0/6.65)&=\sigma_0\left(\frac{25 M}{M_0}\right)^{-1/2}\, .
\label{eqn:sigma_param}
\end{align}

The high mass cut-off in the mass function depends on the form of the initial power spectrum. At large $M\gg M_0$, cut-off dependence vanishes, giving a Gaussian density field suppressed as $-\delta_c^2/2\sigma^2=-x^2/2$. Using $\delta_c=1.686$ we solve for $x(M_{\rm max})$ when the HMF has dropped by 0.01  for the Heaviside cut-off initial power spectrum:
\be
M_{\rm max}(z)\approx 4.9\times 10^6 M_0\times D(z)^2 \, ,
\ee
where $D(z)$ is the linear growth factor normalised to unity at $z=0$. The co-efficient of the Gaussian cut-off in the mass function is modified slightly when the non-zero bispectrum of the minicluster initial conditions is accounted for, leading to a slight increase in $M_{\rm max}$. This is a small effect, and depends on the model for non-Gaussian effects on the mass function. We do not account for it here. It is discussed in detail in Appendix~\ref{appendix:non-gauss}.

The maximum minicluster mass is fixed by $M_0$ and so depends strongly on the temperature evolution of the axion mass (see Fig.~\ref{fig:M0_mass}). MCHs of mass $M>M_0$ are formed by hierarchical structure formation from the seeds of mass $M\approx M_0$ present at matter-radiation equality. We discuss the mergers and density profiles of MCHs in Section~\ref{sec:profiles}.

We derive the low mass cut-off for the modified barrier (for the fit and Heaviside filtering cases the corresponding derivation is trivial). At small $M\ll M_0$, $\sigma(M)=\sigma_0$, and we use the asymptotic behaviour of $\delta_c(M<M_J)\propto\exp [1.8 (M/M_J)^{-1/2}]$, where $M_J$ is the Jeans mass, to find when the Gaussian argument $x(M_{\rm min})=1$:
\be
M_{\rm min}(z)\approx M_J \times \left[ \frac{1.8}{7.5+\log D(z)}\right]^2 \, .
\ee
There is only a logarithmic dependence of the minimum minicluster mass on the growth rate, and so the minimum mass varies very little over time. 

\subsubsection{Slope}

From the dependence of $\sigma$ on the mass scale $M$, Eq.~\eqref{eqn:sigma_param}, we can deduce the dependence of the HMF on $M$ far from either cut-off. Using Eq.~\eqref{M2fM} we have:
\begin{equation}
\frac{{\rm d}n}{{\rm d} \ln M}=Mf(M) \propto M^{-1/2}\, ; \quad \text{for } (M_{\rm min}<M<M_{\rm max})\, .
\end{equation}  
Our parameterization sets the HMF to zero outside these boundaries.

We see in Fig.~\ref{parametrize_HMF} that the slope of the mass function changes below $M_0/6.65$, where the variance becomes flat. The slope of the mass function in this regime is window function and cut-off dependent. 

\subsubsection{Normalization}

The normalization of the substructure mass function is simply fixed by the total mass of the host galaxy, $M_{\rm host}$. While we assume that all of the DM is composed of axions, we take the fraction collapsed into miniclusters, $f_{\rm MC}$, as a free parameter. The normalization condition is:
\be
f_{\rm MC}= \frac{1}{M_{\rm host}}\int \frac{{\rm d}n}{{\rm d}\ln M}{\rm d}M = \int \psi(M) {\rm d}M \, ,
\label{eqn:normalize}
\ee
where we have introduced the normalized mass function $\psi(M)$.

In principle, $f_{\rm MC}$ can be determined by numerical simulation of minicluster formation and the subsequent formation of galaxies from miniclusters. Such simulations will involve at least some modelling uncertainty. We take the alternate view that $f_{\rm MC}$ is a phenomenological quantity that can be used to constrain models of axion DM via astrophysical observations. Thus the mass function normalisation, $f_{\rm MC}$, is a free parameter in our constraints.


\section{Minicluster and MCH Density Profiles}
\label{sec:profiles}

\subsection{Hierarchical Structure Formation}
\begin{figure}[t!]
\vspace{-0.2em}\includegraphics[width=\columnwidth]{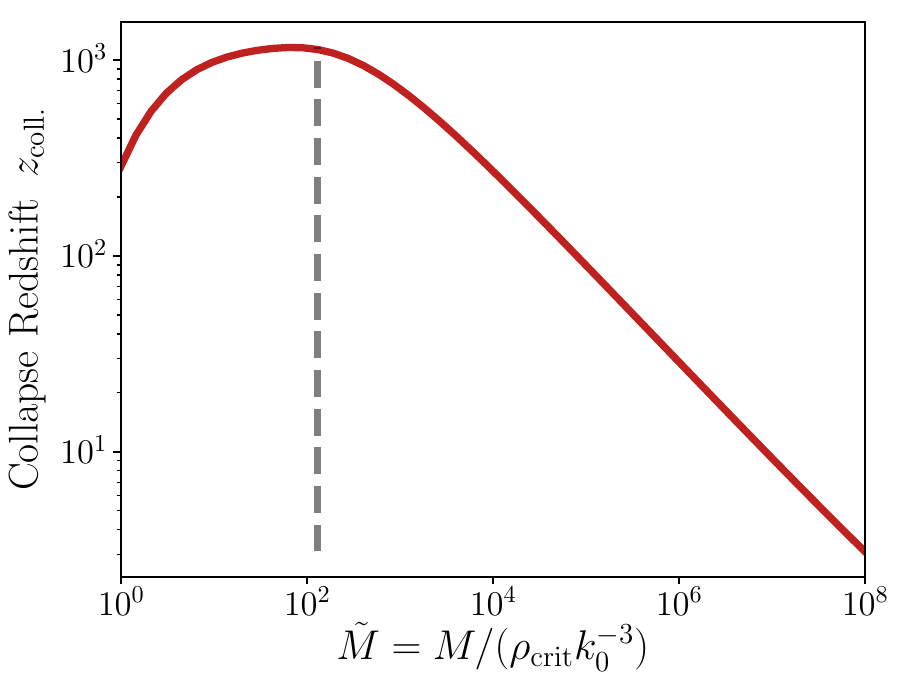}
\vspace{-2.5em}\caption{{\bf MCH Collapse Redshift}: The collapse redshift is computed following NFW (see main text). The earliest objects to collapse do so shortly after equality at $z_{\rm coll}\approx 10^3$. The vertical line shows $M_0=\rho_{\rm crit}(4\pi/3)(\pi/k_0)^3$. Miniclusters at the characteristic mass $M_0$ are the first bound structures to form.}  
\label{fig:collapse_plot}
\end{figure}
\begin{figure}[t!]
\vspace{-0.2em}\includegraphics[width=\columnwidth]{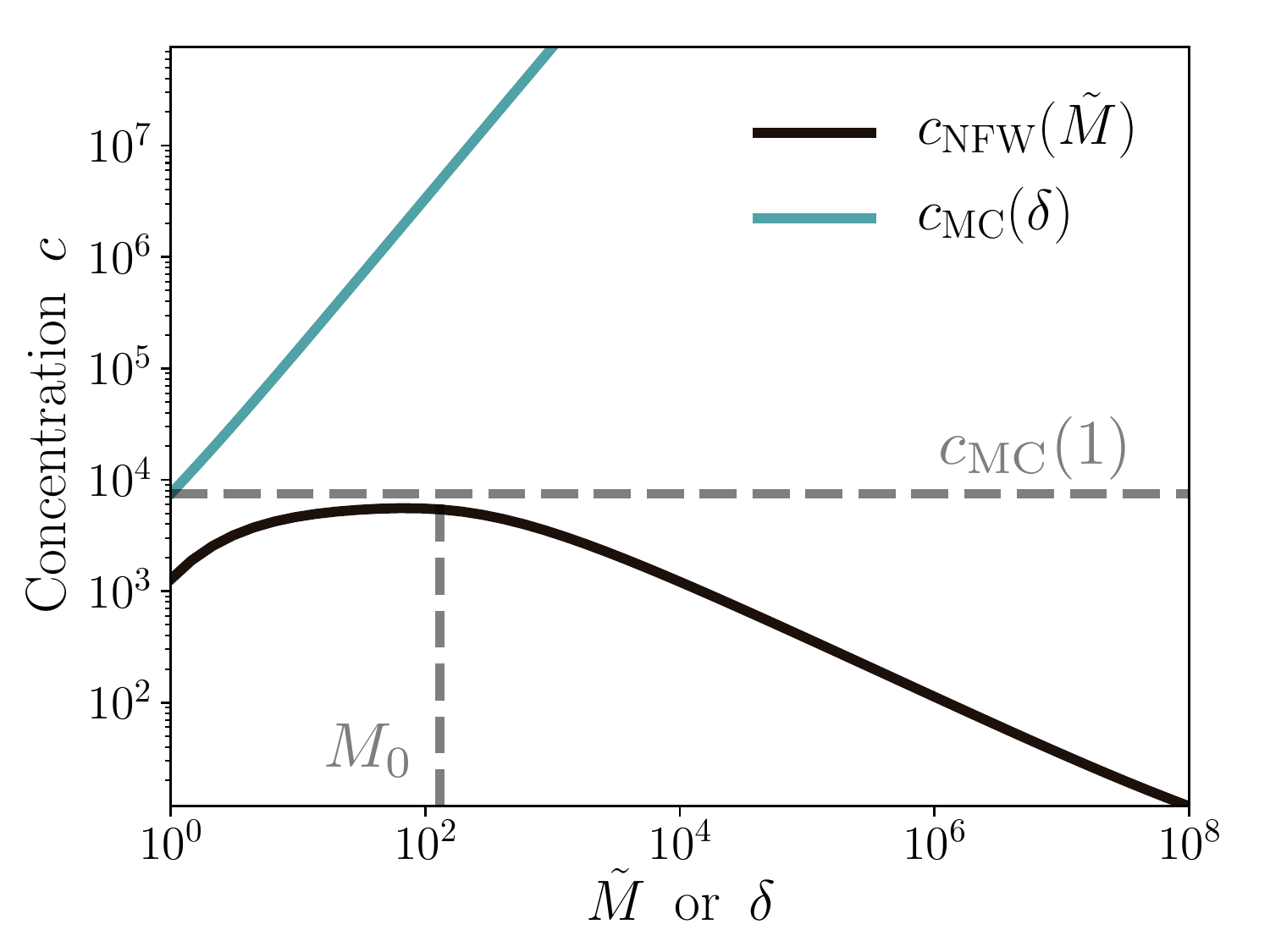}
\vspace{-2.5em}\caption{{\bf Concentration of Miniclusters and (Diffuse) MCHs}: For miniclusters we show $c_{\rm MC}(\delta)$ derived for an NFW profile with characteristic density fixed by Eq.~\ref{eqn:minicluster_density_KT}. For MCHs we show the concentration-mass relation assuming hierarchical structure formation. The vertical line shows the location of $M_0$, which almost maximises the concentration near $c_{\rm MC}(1)$.}  
\label{fig:concentration_plot}
\end{figure}

The scale of miniclusters is fixed by $k_0\ll k_J$ and so we expect density profiles to be well described by CDM. Density profiles formed by hierarchical structure formation of CDM are described by the famous Navarro-Frenk-White (NFW) profile~\cite{Navarroetal1997}:
\be
\rho_{\rm NFW}(r)=\frac{\rho_{\rm crit}\delta_{\rm char}}{r/r_s(1+r/r_s)^2} \, ,
\ee
where $r_s$ is the scale radius, which is specified in terms of the virial radius, $r_{200}$, and concentration, $c$, as $r_s=r_{200}/c$. The mass is defined as that contained within the virial radius, where the average density is 200 times the critical density. Thus the characteristic density is given in terms of the concentration as:
\be
\delta_{\rm char}=\frac{200}{3}\frac{c^3}{\ln (1+c)-c/(1+c)} \, .
\ee
 
Using our analytic result for the variance with a Heaviside filtering and Gaussian initial power spectrum, Eq.~\eqref{eqn:analytic_sigHG}, we first compute the collapse redshift, $z_{\rm coll}$. Following NFW, this is defined using the extended Press-Schechter formalism as the redshift when half the mass of the halo was first contained in progenitors more massive than one percent of the final mass. The result is plotted in Fig.~\ref{fig:collapse_plot}. The collapse redshift is maximised at $M\approx M_0$, indicating that objects of this characteristic mass are the first bound structures to form shortly after equality at $z\approx 10^3$. We note that this appearance of $M_0$ justifies our choice in Eq.~\eqref{eqn:M0_def} That is, the first objects to form are consistent with the spherical wavelength volume, and are somewhat heavier than the cubic volume estimate.

It is these first bound structures which are ``true'' axion miniclusters, that is MCHs with $M=M_0$. Due to the non-linear dynamics of axion interactions in the instanton potential, these miniclusters display a spectrum of sizes and characteristic densities given by Eq.~\eqref{eqn:minicluster_density_KT} and Fig.~\ref{fig:fit_tkachev}. The concentration is a stochastic function, which for the initially formed miniclusters has a large non-Gaussian tail.

We use Eq.~\eqref{eqn:minicluster_density_KT} to define the characteristic density of miniclusters, $\rho_{\rm MC}(\delta)=\rho_{\rm crit}\delta_{\rm char}(\delta)$, which defines a $\delta$-dependent concentration for miniclusters, $c_{\rm MC}(\delta)$. Due to the large spread in values of $\delta$, there are miniclusters with concentrations far exceeding that expected from ordinary hierarchical structure formation. The physical reason for this is, as discussed, the non-linear interactions in the axion potential, and the result is confirmed in numerical simulation.

The concentration-mass relation, $c(M)$, specifies the \emph{spatially averaged} density profile of an MCH. Assuming MCHs are formed hierarchically from the (assumedly) Gaussian large scale white-noise density fluctuations of miniclusters in mergers that are described entirely by CDM we can adopt the analytic model for $c(M)$ proposed by NFW (related approaches include e.g. Ref.~\cite{2001MNRAS.321..559B}): 
\be
\delta_{\rm char}=3000\, \, \Omega_m (1+z_{\rm coll})^3 \, .
\ee
The resulting concentration-mass relation, $c_{\rm NFW}(M)$, is plotted alongside $c_{\rm MC}(\delta)$ in Fig.~\ref{fig:concentration_plot}. We notice that miniclusters of mass $M_0$ maximise $c_{\rm NFW}(M)$ (since they have largest $z_{\rm coll}$) and that furthermore the NFW concentration agrees with the minicluster concentration for $\delta=1$: $c_{\rm NFW}(M_0)\approx c_{\rm MC}(1)$. This is a pleasing coincidence that further validates our use of $M_0$ as the characteristic mass. 

\subsection{Minicluster Mergers?}
\label{sec:mergers}

The NFW profile describes the average density profile of a halo. The Press-Schechter mass function (as we have used it) also simply groups all progenitors together into a single parent mass. Neither accounts for substructure. Using only the average density profile and ignoring the substructure is equivalent to the case where halo mergers completely disrupt the progenitors. Because observables, in particular microlensing, may be sensitive to the substructure of MCHs, and minicluster mergers may not be totally disrupting, we must be careful how the mass function and NFW density profile are used. 

We will satisfy ourselves with some estimates. We begin with considering the Hill sphere:
\be
r_{\rm Hill} = a (1-e)\left(\frac{M_{\rm sat}}{3M_{\rm host}}\right)^{1/3}\, ,
\ee
where $M_{\rm sat}$ is the satellite mass, $M_{\rm host}$ is the host mass, and the satellite has an elliptical orbit of semi-major axis $a$ and eccentricity $e$. 

Consider a minicluster satellite of mass $M_0$ inside a MCH of mass $M_{\rm host}$. The minicluster will be disrupted by tidal forces if the minicluster radius [calculated from $c_{\rm MC}(\delta)$] is larger than the Hill radius. In other words stripping is significant if:
\be
\frac{r_{\rm Hill}}{r_{\rm MC}}>1 \, ,
\ee
where we consider the minicluster radius to be given by the scale radius, $r_{\rm MC}(\delta,M_0)=r_{200}(M_0)/c_{\rm MC}(\delta)$.

\begin{figure}[t!]
\vspace{-0.2em}\includegraphics[width=\columnwidth]{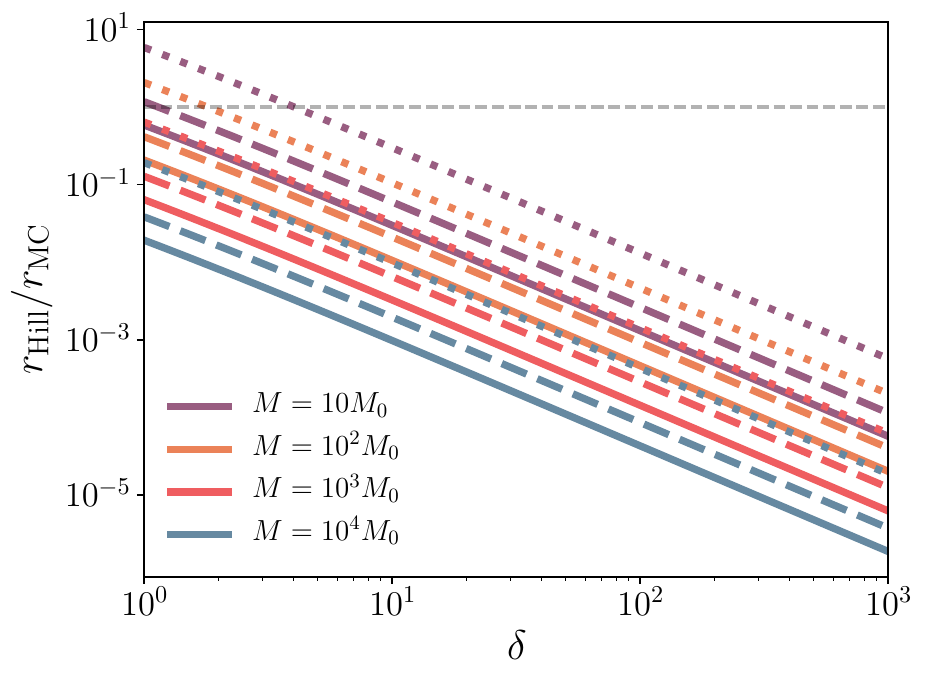}
\vspace{-2.5em}\caption{{\bf The Hill Radius}: When the ratio of the Hill radius to the minicluster radius is larger than unity (horizontal dashed line), tidal stripping is significant. We consider minicluster satellites of mass $M_0$ in hosts of mass $M$ located at the scale radius of the host. Solid, dashed, and dotted lines are for orbits of eccentricity $e=0,0.5,0.9$ respectively.}  
\label{fig:hill_ratio}
\end{figure}

For simplicity we consider the minicluster to be located at the scale radius of the host, where the majority of the mass is concentrated, with the concentration of the host given by $c_{\rm NFW}(M)$. Fig.~\ref{fig:hill_ratio} demonstrates that only the least concentrated miniclusters with the most eccentric orbits in relatively light MCHs are likely to undergo any significant tidal stripping as given by the Hill criterion.

A more complete analysis of stripping uses the tidal radius, $r_t$ for a satellite of mass $M_{\rm sat}$ orbiting at radius $r_{\rm sat}$ in a host halo of mass $M_{\rm host}$~\cite{1962AJ.....67..471K}:
\be
r_t= \left(\frac{GM_{\rm sat}(<r_t)}{\omega^2-\Phi''(r_{\rm sat)}}\right)^{1/3}
\ee
where $\Phi''$ is the second derivative with respect to radius of the gravitational potential in the host halo and $\omega$ is the angular velocity of the satellite. A full semi-analytic merger tree calculation including tidal stripping (e.g. with \textsc{Galacticus}~\cite{2012NewA...17..175B}) is beyond the scope of the present work. For simplicity we consider the satellite to be an initial minicluster of mass $M_0$ located at the scale radius of the host (where most of the mass is) and take both satellite and host to have concentration-mass relation given by $c_{\rm NFW}(M)$ (equivalent to only the least dense minicluster satellites with $\delta=1$). We find, similarly to the case with the Hill radius, that the tidal radius is always small compared to the minicluster radius, indicating that miniclusters do not undergo significant stripping within MCHs. We discuss this further in Section~\ref{sec:meaning_of_mergers}.

\section{Microlensing with Miniclusters}
\label{sec:microlensing}
\begin{figure}[t!]
\center
\includegraphics[width=1\columnwidth]{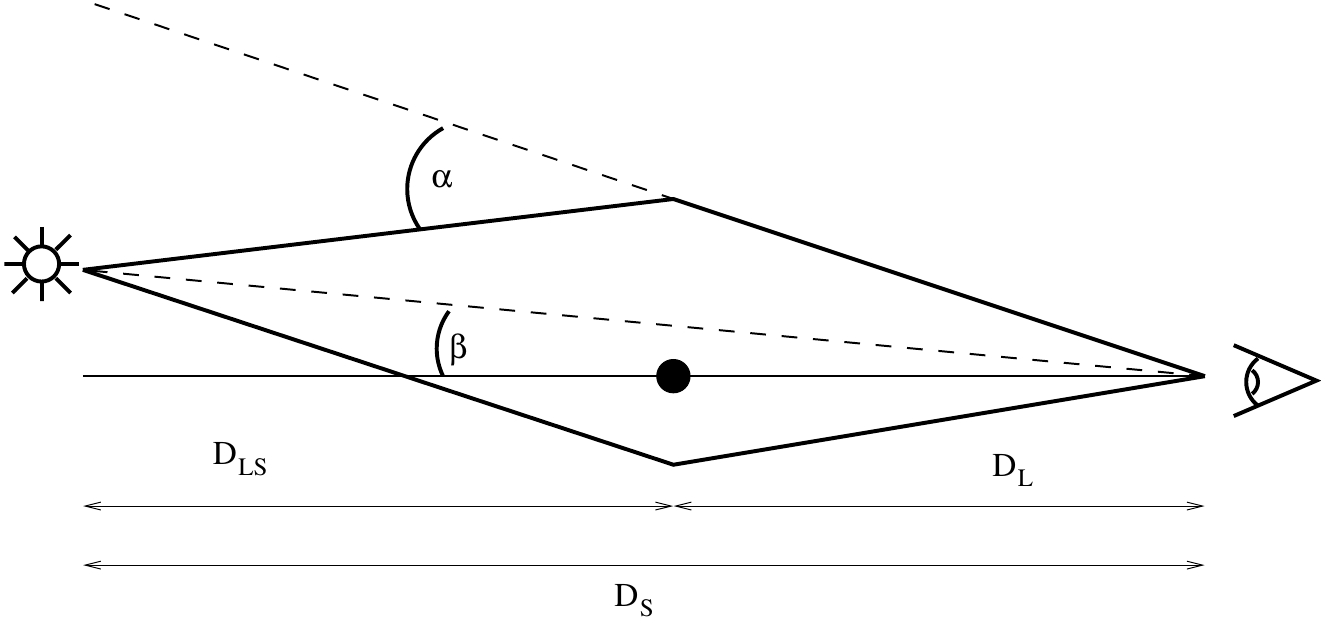}
  \caption{The Lensing of a source by a point mass at the origin.  $D_{S}$ is the distance from the observer (us) to the source, $D_{L}$ is the distance to the lens and $D_{LS}=D_{S}-D_{L}$.}
  \label{lensingfig}
\end{figure}

\subsection{Microlensing basics}

Microlensing is the fugitive amplification of a background star which occurs when a compact object passes close to the line of sight to that star~\cite{Paczynski:1985jf}.

The magnitude of the effect of microlensing by point objects is relatively simple to calculate and uses the normal equations for gravitational lensing \cite{Griest:1990vu}. The miniclusters that we are considering are not however point sources, and while they are very small, the deflection of light which they give rise to is also very small, so the extended nature of these objects is important to take into account.

For the configuration where observer, lens and source lie on the same line, the Einstein radius corresponds to the radius of closest approach of photons to the point mass lens as they pass by it.
\begin{equation}
R_E =\sqrt{\frac{4GM}{c^2}\frac{D_LD_{LS}}{D_S}}= 4 \times 10^{13} \sqrt{\frac{M}{M_{\odot}}\frac{D_{S}}{\text{kpc}}} \quad \text{cm}
\end{equation}
where the enumeration assumes that $D_{S}=2D_{L}=2D_{LS}$ and the quantities $D_{L}$ and $D_S$ etc. are shown in Fig.~\ref{lensingfig}. Getting numbers, for $10^{-8}M_{\odot}$ mass halos at the distance of Andromeda, the Einstein Radius will correspond to tens of nanoparsecs.\footnote{This expression differs from that in Ref.~\cite{Kolb:1995bu} by a factor of 100 due to considering local rather than cosmological sources, and using the modern values of the cosmological densities. This leads us to find that larger values of $\delta$ are necessary for lensing.}

A microlensing event occurs when a compact object passes through the microlensing `tube', which has a radius of $u_{T} R_{E}$ where $u_{T} \approx 1$ is the minimum impact parameter for which the amplification of the background star is above
the required threshold and $R_{E}$ is the Einstein radius:
\begin{equation}
R_{E}(x)= 2 \left[ \frac{ G M x (1-x)D_S}{c^2 } \right]^{1/2} \,,
\end{equation}
where $M$ is the lens mass, $D_S=L$ is the distance to the source,  and $x=D_L/D_S$ is the distance of the lens from the observer~\cite{Paczynski:1985jf}. In our case, the distance to the sources (Large Magellanic Cloud (LMC) or Andromeda (M31)) is much greater than its line of sight depth, so all of the source stars can be assumed to be at the same distance ($\sim 50$ kpc for LMC and $\sim 770$ kpc for M31) and the angular distribution of sources ignored. 

The differential event rate for lenses of mass $M$ in a halo with projected density profile $\rho(x)$ along the line of sight is given by~\cite{Griest:1990vu}~\footnote{This expression assumes a spherical halo with an isotropic
velocity distribution and ignores the transverse velocity of the microlensing tube, which has a small effect on the differential event rate~\cite{Griest:1990vu}.}:
\begin{equation}
\label{dGdt1}
\frac{{\rm d} \Gamma}{{\rm d} \hat{t}} =  \frac{32 L u_{T}^4}{{\hat{t}}^4{v_{c}}^2}
              \frac{1}{M}  
              \int^{x_{h}}_{0} \rho(x) R^{4}_{E}(x)
              e^{-Q(x)}  {\rm d} x  \,, 
\end{equation}
where $\hat{t}$ is the time taken by the lens to cross the Einstein diameter, $x_{h}$ is the extent of the halo and $Q(x)= 4 R^{2}_{E}(x) u_{T}^2 / (\hat{t}^{2} v_{c}^2)$, where $v_{c}  =220 \, {\rm km \, s}^{-1}$ is the local circular speed. The factor $u_T\approx 1$ defines the critical magnification for the lensing survey, which we take as $\mu=1.34$ (see below). The factor $e^{-Q}$ emerges by approximating the Bessel function in the lensing integral~\cite{Griest:1990vu,1996ApJ...461...84A}.

The expected number of events, $N_{\text{exp}}$, is given by
\begin{equation}
N_{{\text{exp}}} = E \int_{0}^{\infty} \frac{{\rm d} \Gamma}{{\rm d} \hat{t}}
           \,  \epsilon(\hat{t}) \, {\rm d} \hat{t} \,,
\end{equation}
where $E$ is the exposure in star years and $\epsilon(\hat{t})$ is the probability that a microlensing event with duration $\hat{t}$ is detected (detection efficiency). 

\subsection{Miniclusters as non-Point-Like Objects}
\subsubsection{Lensing Tube for non-Point-Like Objects}
\label{sec:malcolm_lensing_method}

We are interested in the case where there is a magnification of 1.34, which is the threshold adopted by the EROS and HSC surveys.  This magnification is not arbitrary, it corresponds precisely (in the point like mass case) to the outer ray passing the lens at a radius of $1.618\times R_E$ where $1.618$ is the Golden Ratio and $R_E$ is the Einstein radius.  The second image passes on the other side at a distance $0.618\times R_E$, i.e. the inverse Golden Ratio.  The magnification of the two images is given in terms of the distance at which the light rays pass the lens $x=r/R_E$ and the magnifications $\mu$ are given by
\begin{equation}
\mu_{\pm}=\left[1-\left(\frac{1}{x_\pm}\right)^4\right]^{-1}
\end{equation}
The sum of the magnification of the two images is 1.34, the majority of which comes from the outer image which gives a magnification of 1.17.  

We need to repeat the calculation for situations where the lens is potentially diffuse relative to the scales of interest, such that the enclosed mass is not a constant when $\delta$ is small.

We start by taking the characteristic density given by Eq.~\eqref{eqn:minicluster_density_KT}. We consider two functional forms for the minicluster density profiles. The NFW profile is a universal feature of CDM simulations emerging from hierarchical structure formation. However, miniclusters at the characteristic mass do not form from hierarchical structure formation, but probably from a more direct collapse. It has been suggested~\cite{2017arXiv170103118O} that a more suitable profile for the initial seed miniclusters be given by that due to self-similar infall~\cite{1985ApJS...58...39B}:
\begin{equation}
\rho(r)=\rho_s\left(\frac{r_s}{r}\right)^{9/4}\, .
\label{ssprof}
\end{equation}
Such a power-law profile also appears in the minicluster $N$-body simulations of Ref.~\cite{2007PhRvD..75d3511Z}.

For both of the NFW and self-similar profiles we need to make an identification with the density $\rho_{\rm MC}$ and the characteristic density.  For the NFW profile we simply say that $\rho_{\rm MC}=\rho_{\rm crit}\delta_{\rm char}$ as above, and rescale $r_s$ until we obtain the correct mass of the halo at $r_{\rm max}$. If we were to calculate the virial radius for these objects, they would be hugely larger than the scale radius by many orders of magnitude, but since the halos are within a galactic halo, we make the approximation that the NFW profile is cut off at a radius $r_{\rm max}=100r_s$. This simplifies the numerical lensing calculation by reducing the dynamic range.
\begin{figure}[t!]
\center
\includegraphics[width=0.9\columnwidth]{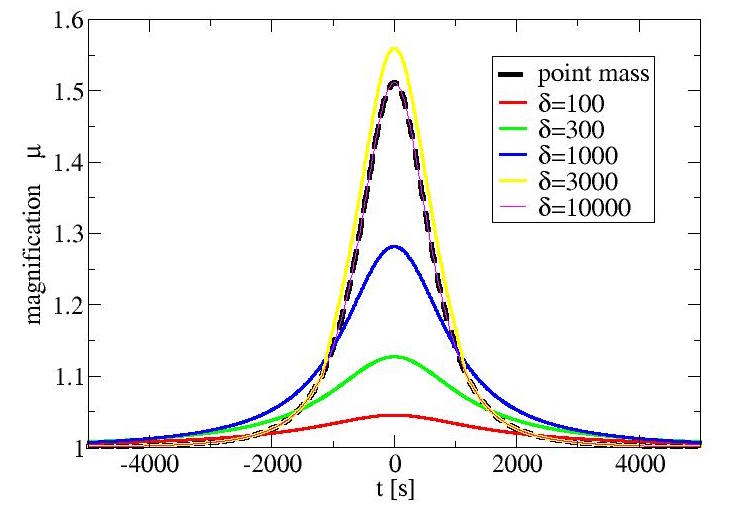}
\caption{{\bf Microlensing Lightcurves}: Magnification of source star as a minicluster passes through the lensing tube for various values of $\delta$. The minicluster has an NFW profile, a mass of $10^{-9}M_\odot$, a tangential velocity of 200 km s$^{-1}$ and an impact parameter with the line to the source of 1.6 milli-AU.  The source star is assumed to be in the Andromeda Galaxy (M31), with $D_{{\rm LS}}=D_{\rm S}/2$. As we increase the value of $\delta$ the lensing curve approaches the value for a point mass, as described in the text.}
\label{lightcurves}
\end{figure}
\begin{figure*}[t!]
\center
\includegraphics[width=2\columnwidth]{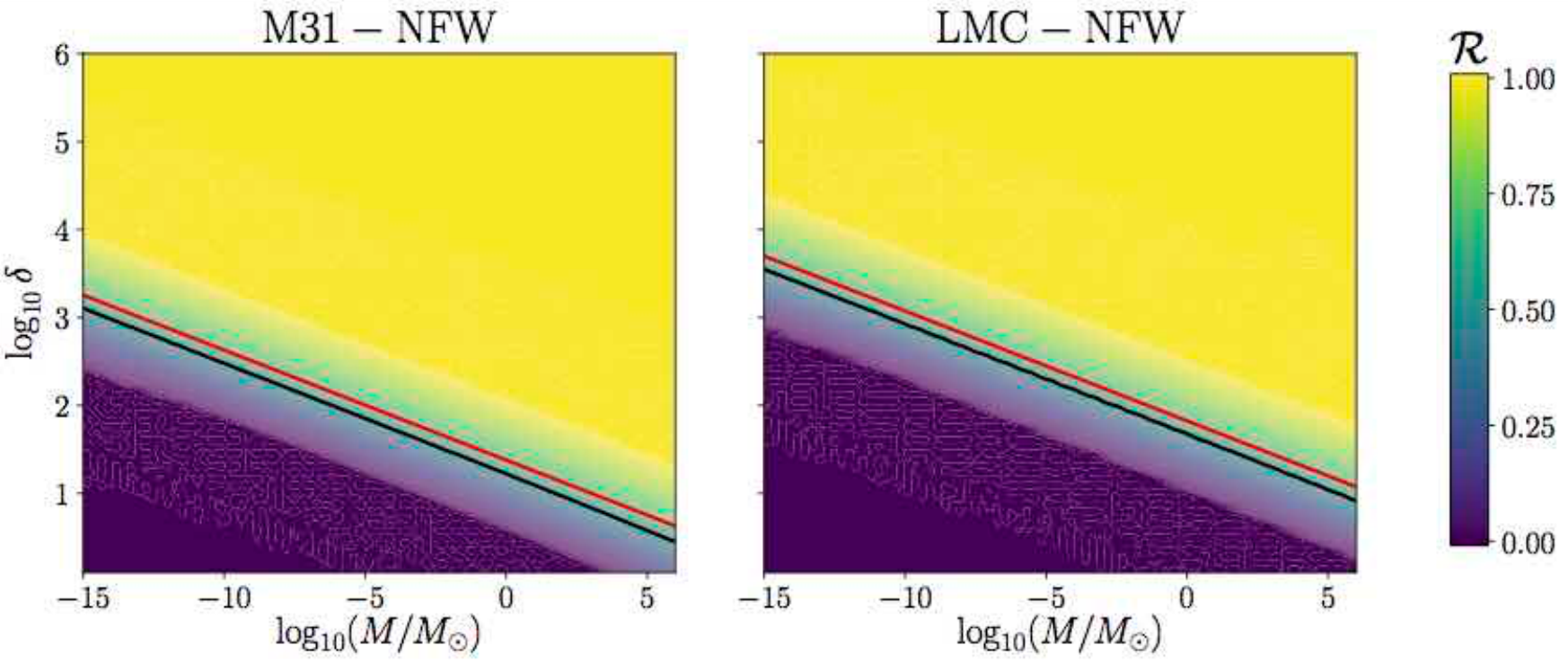}
  \caption{{\bf Minicluster Microlensing Tube}: Rescaling factor, $\mathcal{R}$, for a minicluster of mass $M$ and overdensity $\delta$ with NFW radial profile. The black contour shows the location of $\mathcal{R}=0.5$, while the red line shows the analytic result for $\delta_{\rm lens}(M)$ defined by equating the point mass Einstein radius to the hard-sphere minicluster radius.}
  \label{plotR}
\end{figure*}

The situation is slightly more complicated for the self-similar profile (Eq.~\ref{ssprof}) because $r_s$ is completely degenerate with $\rho_s$ due to the scale invariance.  The overall mass of such a halo which is truncated at a radius $r_{\rm max}$ is
\begin{equation}
M=\frac{16\pi}{3}\rho_sr_s^{9/4}r_{\rm max}^{3/4}
\end{equation}
and the average density of such a halo is
\begin{equation}
\rho_{\rm av}=4\rho_s\left(\frac{r_s}{r_{\rm max}}\right)^{9/4}
\end{equation}
or 
\begin{equation}
r_{\rm max}=\left(\frac{3M}{4\pi\rho_{\rm av}}\right)^{1/3}
\end{equation}
and then we make the identification $\rho_{\rm av}=\rho_{\rm MC}$.  

We then want to turn the three-dimensional density into a surface density by collapsing it onto the lensing plane.  We do this by integrating the three-dimensional density profile along the line of sight towards the centre of the halo.  In the situation where $r_{\rm max}$ is larger than the lensing radius we are probing, we only integrate the mass within the two cones defined by the radius of interest and the distance between the lens and the source and the lens and the image.

Once we have surface mass as a function of radius we can calculate the magnification using the expression for an axisymmetric mass distribution (this can be derived from the equations in Ref.~\cite{Narayan:1996ba})
\begin{align}
&\mu=\left[\left(1-B\right)\left(1+B-C\right)\right]^{-1} \\
C=\frac{1}{\Sigma_c \pi r}&\frac{dM(r)}{dr}\,\,; \,\, B=\frac{M(r)}{\Sigma_c \pi r^2}\,\, ; \,\, \Sigma_c=\frac{c^2D_S}{4\pi G D_L D_{LS}}
\end{align}
What we need to do is to define the ``lensing tube''~\cite{Griest:1990vu} as being the tube within which a lens would create a magnification of at least 1.34.  

We do this by starting at large r and zooming into the radius at which the magnification is the same as the outer image in the point mass case, i.e. $\mu=1.17$.  When a  halo is diffuse, this will occur at a radius much less than the radius at which a point mass would give rise to the same lensing.  The lightcurves for lensing computed as a function of $\delta$ for fixed minicluster mass are shown in Fig.~\ref{lightcurves}. An interesting feature occurs as one increases the parameter $\delta$ in that the magnification rises above that for a point mass before settling down to the same value as the point mass.  This is because there is an intermediate regime where there are multiple paths for photons from the same source to pass the lens at different radii and still arrive at the observer, adding to the overall magnification. This is captured by the gradient terms in the lens equations.

As $\delta$ increases and  the halo becomes more compact it behaves increasingly like a point mass until once $r_{\rm max}$ is well below any of the scales of interest, the lensing tube is the same size as for the point mass case. One can check that for large values of $\delta$ the radius for the distributed mass and the radius for the point mass coincide. 

For each value of $M$ and $\delta$ we average the ratio between the radius of magnification 1.17 for ten values of $x$ between 0 and 1 in order to obtain a numerical correction factor which we can then apply to the point mass lensing equation.  For small values of $\delta$ which give rise to diffuse halos, there is no radius at which the magnification reaches 1.17 and such halos cannot contribute to the lensing integral.

\begin{figure}[t!]
\center
\includegraphics[width=0.9\columnwidth]{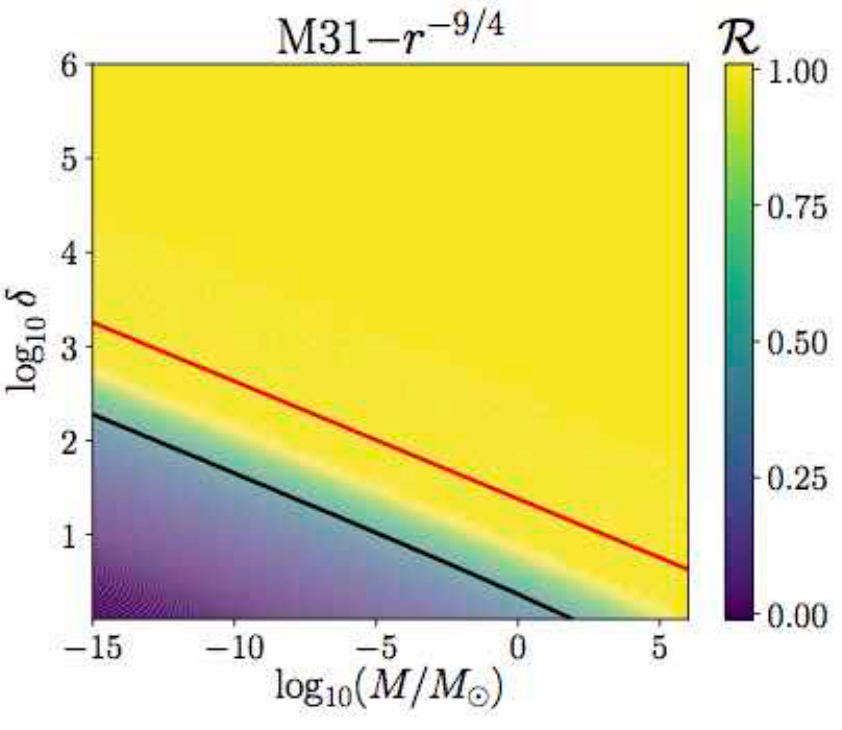}
\caption{As Fig.~\ref{plotR} for M31 in the case of the self-similar infall density profile, $\rho\propto r^{-9/4}$. This profile is more compact than the NFW profile (or a hard sphere) leading to a larger effective lensing tube at fixed $M,\delta$.}
\label{fig:Rplot_self-similar}
\end{figure}

We define the correction factor $\mathcal{R}$ which corresponds to the average radius of the lensing tube with the specific magnification we are looking for divided by the average radius for a point mass.  We plot values of $\mathcal{R}$ for the NFW and self similar profiles in Figs.~\ref{plotR} and \ref{fig:Rplot_self-similar}.  It is clear that for small values of $\delta$ the halos are diffuse and the corresponding lensing is supressed.  For larger values of $\delta$ the lensing increases and eventually when the vast majority of the mass is inside the lensing tube, the objects are indistinguishable from point-like objects. 

An estimate for the transition from $\mathcal{R}=0$ to $\mathcal{R}=1$ can be obtained using the hard-sphere minicluster radius:\footnote{This differs by approximately a factor of three from the scale radius obtained by setting $\rho_{\rm MC}=\rho_{\rm char}$ and solving numerically for $c_{\rm MC}(\delta)$.}
\begin{equation}
r_{\rm mc} = 4 \times 10^{16} \frac{1}{\delta((\delta+1)\Omega_{m})^{1/3}}\left(\frac{M_0}{M_{\odot}}\right)^{1/3} \quad \text{cm} \, .
\label{eqn:hard_sphere_mc}
\end{equation}
We equated this to the Einstein radius and solved for $\delta_{\rm lens} (M)$, the value of $\delta$ when microlensing will be large. The result is plotted in Figs.~\ref{plotR} and \ref{fig:Rplot_self-similar} as a red line. For the NFW case, the estimate is close to the contour for $\mathcal{R}=0.5$ from the full numerical lensing calculation (black line) and accurately estimates the transition in the microlensing behaviour.

It should be noted that we expect greater lensing if we adopt the self similar profile rather than the NFW profile.  The reason for this is that the majority of the mass contributing to the overall profile is greatest at the scale radius for the NFW profile, but reaches a maximum towards the centre of the scale invariant profile, so in the situation where only a fraction of the halo is within the normal lensing tube, the self similar profile will give rise to a bigger lensing effect than the NFW profile.  This can be seen in Fig.~\ref{fig:Rplot_self-similar}, where the transition in lensing behaviour occurs at a smaller value of $\delta$ than for the NFW case, and the estimate for $\delta_{\rm lens}$ from treating miniclusters as hard spheres is an overestimate.

\subsubsection{Microlensing Event Rate for non-Point-Like Objects}

From the previous numerical lensing calculations, we find that the shape of the microlensing tube is still reasonably well described by the Einstein radius, $R_{\rm E}(x,M)$, but with a rescaling factor, $\mathcal{R}$, that depends on $M$ and $\delta$, such that the minicluster lensing tube is given by: 
\be
R_{\rm MC}(x,M,\delta)=\mathcal{R}(\delta,M)R_E(x,M) \, .
\ee
When a  mincluster/MCH is diffuse, the tube is smaller. There is a minimum value of $\delta$ below which there is no existing value of impact parameter $\ell$ for which $A\geq 1.34$, i.e. $\mathcal{R}(\delta<\delta_{\rm min})=0$ with $\delta_{\rm min}=\delta_{\rm min}(M)$ given approximately by $r_{\rm mc}/R_E>1$. As we will see, this treatment reduces significantly the expected number of microlensing events for miniclusters compared to point masses (MACHOs, PBHs). For $\delta\gg \delta_{\rm min}$ the limiting behaviour is that of a point mass, $\mathcal{R}\rightarrow 1$.

Miniclusters of mass $M_0$ treated as non-point like objects have a rate of microlensing events of duration $\hat{t}$ given by:
\be
\hspace{-0.1in}\frac{{\rm d}\Gamma}{{\rm d}\hat{t}}= \frac{32 L u_{T}^4}{\hat{t}^4v_c^2}\frac{1}{M_0}\int_0^\infty \frac{{\rm d}n}{{\rm d}\delta}\left[\int_0^1\rho(x) R_{\rm E}(x)^4e^{-Q(R_{\rm E})}{\rm d }x\right]{\rm d}\delta \, ,
\ee
which is a modified version of Eq.~\ref{dGdt1} where the Einstein radius has been replaced by the minicluster lensing tube radius $R_{\rm MC}$. The partition function ${\rm d n}/{\rm d \delta}$ is deduced from the Fig.~2 of Ref.~\cite{Kolb:1995bu}, which we showed earlier in Fig.~\ref{fig:fit_tkachev} and the fit Eq.~\eqref{eqn:minicluster_density_KT}.

\subsection{Mergers and the Meaning of the Mass Function for Microlensing}
\label{sec:meaning_of_mergers}

In the case of an extended mass function, Eq.~\ref{dGdt1} should be replaced by \cite{Green:2016xgy}:
\be
\label{dGdt2HMF}
\hspace{-0.1in}\frac{{\rm d}\Gamma}{{\rm d}\hat{t}}= \frac{32 L u_{T}^4}{\hat{t}^4v_c^2}\int_0^\infty \frac{\psi(M)}{M}\left[\int_0^1\rho R_{\rm MC}^4e^{-Q(R_{\rm MC})}{\rm d }x\right]{\rm d}M \, ,
\ee
where we have used the normalized mass function of Eq.~\eqref{eqn:normalize}.

The assumption in Eq.~\ref{dGdt2HMF} is that an object of mass $M$ in the mass function contributes to the microlensing only on a single time scale $\hat{t}$. The integral over the mass function ignores the possibility of microlensing events caused by substructure within an object of mass $M$. This is appropriate for point-like objects such as PBHs and MACHOs~\cite{Green:2016xgy}. For MCHs, which can have substructure, such an integral over the mass function assumes a smoothing of the lensing signal which is equivalent to assuming that miniclusters merge completely upon formation of an MCH. 

The estimates in Section~\ref{sec:mergers} suggest that miniclusters are dense enough to remain tightly bound within MCHs, like ``plums in a pudding'', and do not undergo significant merging. In this case, there will always be lensing events with time scale $\hat{t}(M_0)$. A microlensing observation effectively smooths the lightcurve of a lensing event over a timescale given by the cadence [the lower limit of the efficiency $\epsilon (\hat{t})$]. If the microlensing observation in question is sensitive to $\hat{t}(M_0)$, and miniclusters do not merge when they form MCHs, then the effective mass function for microlensing is a Dirac-delta function, $\delta_{\rm D}(M-M_0)$. We call this the \emph{isolated miniclusters} scenario, and take it as our default model for the lensing constraints in Section~\ref{sec:results}.\footnote{In the isolated mincluster scenario MCHs can still play a role in modulating the arrival of miniclusters leading to correlation in the events of $\hat{t}(M_0)$ over the longer time scale $\hat{t}(M)$. Modeling such multi-time-scale microlensing would be challenging, but could be used in future to probe the MCH mass function and improve sensitivity to lower values of $M_0$.}

However, we may be wrong about minicluster mergers if scalar field dynamics plays an important part (i.e. if miniclusters cannot be considered as pure CDM). In such a case tunnelling through the tidal radius can cause additional effects that need to be accounted for~\cite{2017PhRvD..95d3541H}. Scalar field dynamics also plays a significant role in axion star formation and core mergers (see Appendix~\ref{appendix:axion_stars}). To account for our inability to fully model these processes we consider two additional \emph{ad hoc} possibilities for minicluster mergers into MCHs. Both scenarios assume complete merging of miniclusters (the ``plums'' dissolve), and use the Press-Scechter mass function to consider lensing by smooth MCHs.
\begin{itemize}
\item {\bf ``Dense MCHs''}: large non-Gaussian effects and interactions could cause MCHs to remain very dense. This scenario assumes that the concentration of an MCH is independent of mass and follows the same distribution with $\delta$ of miniclusters, $c_{\rm MC}(\delta)$.
\item {\bf ``Diffuse MCHs''}: miniclusters could become completely stripped when they form MCHs. This scenario assumes MCHs follow $c_{\rm NFW}(M)$ with no substructure.
\end{itemize}

In the dense MCH scenario, the MCHs themselves can still be dense enough to contribute to the microlensing signal, and so the lensing integral must account for both ${\rm d}n/{\rm d}\delta$ and the mass function $\psi(M)$, being:\footnote{For ease of notation in Ref.~\cite{Fairbairn:2017dmf} we used ${\rm d}n/{\rm d}M$ with implicit normalization by the host mass.}
\begin{widetext}
\be
\frac{{\rm d}\Gamma}{{\rm d}\hat{t}}= \frac{32 L u_{T}^4}{\hat{t}^4v_c^2}\int_0^\infty \left\{\frac{{\rm d}n}{{\rm d}\delta}\int_0^\infty\left[\frac{\psi(M)}{M}\int_0^1\rho R_{\rm MC}^4e^{-Q}{\rm d }x\right]{\rm d}M\right\}{\rm d}\delta \, ,
\ee
\end{widetext}
The spread of masses away from $M_0$ can lead to a significant amount of mass moving in or out of the efficiency range of a given microlensing survey. In the range of scales relevant to the HSC micolensing survey, this effect ends up reducing the signal for the QCD axion minicluster masses and so (despite the vastly increased density of MCHs) it is \emph{more pessimistic} than the isolated minicluster scenario (for lower minicluster masses than the QCD axion the microlensing event rate is increased and the dense MCH scenario is more optimistic).

In the diffuse MCH scenario we take only a small window of masses $M_0/10<M<10 M_0$ to be described by $c_{\rm MC}(\delta)$, i.e. those objects that have not undergone any mergers. We throw out all MCHs with $M>10 M_0$ from the mass function integral, since $c_{\rm NFW}(M>M_0)<c_{\rm MC}(\delta)$ for all $M$, and thus such diffuse MCHs will have $\mathcal{R}\ll 1$ and contribute negligibly to lensing. The diffuse MCH scenario is equivalent to a huge reduction in $f_{\rm MC}$ caused by structure formation and is the \emph{most pessimistic} scenario for microlensing.

\begin{figure*}[!htb]
\center
\includegraphics[scale=0.6]{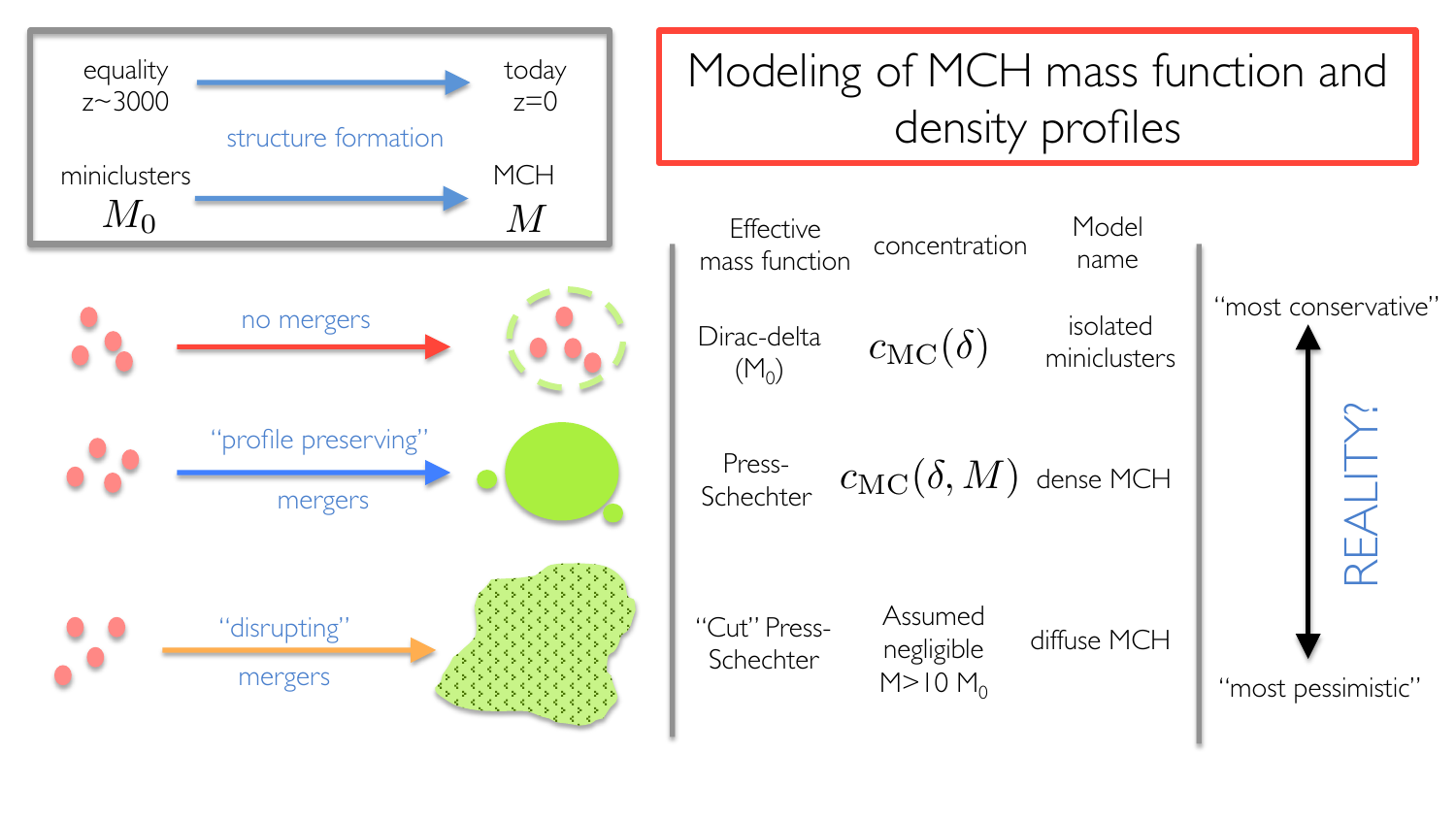}
  \caption{Cartoon showing the modelling of the mass function and density profiles applied to the computation of the expected number of lensing events.}
  \label{diagram1}
\end{figure*}

A cartoon depicting our three models for the microlensing signal of miniclusters and MCHs is shown in Fig.~\ref{diagram1}.

\section{Results: Microlensing Constraints on Axions}
\label{sec:results}

\subsection{The EROS Microlensing Survey}

The EROS survey observed the LMC, at a distance $d_{\rm LMC}=50\text{ kpc}$, considering only lensing events of LMC stars by DM in the Milky Way (MW). EROS models the MW as a cored isothermal sphere:
\be
\rho_{\rm MW,EROS}(r)=\rho_0\frac{R_c^2+R_\oplus^2}{R_c^2+r^2}\, ,
\ee
where $R_\oplus=8.5\text{ kpc}$ is the radius of the Earth from the centre of the MW and the MW halo parameters are $\rho_0=0.0079\, M_\odot\text{pc}^{-3}$ and $R_c=5\text{ kpc}$. A minicluster in the MW at distance $d$ from Earth on the line of sight to the LMC has radial coordinate in the MW halo $r_{\rm MW}^2(d)=R_\oplus^2-2R_\oplus d\cos l_{\rm LMC}\cos b_{\rm LMC}+d^2$, where $(l,b)=(280^o,-33^o)$ are the measured Galactic coordinates.

For the EROS-2 survey $E=3.68 \times 10^{7}$ star years. We extract the detection efficiency, in terms of Einstein radius crossing time, from the Fig.~11 of Ref.~\cite{Tisserand:2006zx}.\footnote{We follow Ref.~\cite{Tisserand:2006zx} by multiplying the efficiency by an extra factor of 0.9 to take into account lensing by binary lenses} The efficiency is large ($\epsilon\gtrsim 0.5$) for time periods of between one day and 1000 days.

\subsection{Microlensing Survey of Subaru HSC}
\label{sec:subaru}

The major limiting factor that prevents the EROS data from constraining MACHOs and miniclusters at low masses is the small-time efficiency, driven by the cadence of the observation. Very recently, observations in Ref.~\cite{2017arXiv170102151N} used data from the Subaru Hyper Suprime Cam (HSC) to place constraints on  low mass primordial black holes (PBHs) with $10^{-13}M_\odot<M_{\rm PBH}<10^{-6}M_\odot$. The HSC observations exclude a PBH fraction $f_{\rm PBH}\gtrsim \mathcal{O}({\rm few})\times 0.01$. This constraint was made by performing a microlensing survey with a cadence of two minutes over a single night of seven hours.

For the microlensing survey, HSC observed Andromeda (M31). A major difference to the EROS survey, apart from the target galaxy, is that for microlensing in M31, due to the high DM density, one must account not only for lensing by DM in the MW, but also for lensing by DM in M31 itself. Thus the differential event rate is given by ${\rm d}\Gamma={\rm d}\Gamma_{\rm MW}+{\rm d}\Gamma_{\rm M31}$.  

HSC adopt NFW radial density profiles for the DM in the MW and M31 with parameters from Ref.~\cite{2002ApJ...573..597K} of $r_s=21.5\text{ kpc}$, $\rho_c=4.88\times 10^6\,M_\odot{\rm kpc}^{-3}$ giving $M_{\rm vir}=10^{12}M_\odot$ for the MW and $r_s=25\text{ kpc}$, $\rho_c=4.96\times 10^6\,M_\odot{\rm kpc}^{-3}$ giving $M_{\rm vir}=1.6\times 10^{12}M_\odot$ for M31. The different host masses for the MW and M31 normalise the minicluster mass function differently in each case.

The DM density profile along the line of sight is $\rho_{\rm DM}(x)=\rho_{\rm MW}(x)+\rho_{\rm M31}(x)$, with $x=d/d_s$ for source distance $d_s=770 \text{ kpc}$. The line of sight distances are given in terms of the radial co-ordinate $r$ as
\begin{align}
r_{\rm MW}(d)&=\sqrt{R_\oplus^2-2R_\oplus \cos{l}\cos{b}+d^2}\, , \\
r_{\rm M31}(d)&=d_s-d \, ,
\end{align}
with $R_\oplus=8.5\text{ kpc}$ the radial co-ordinate of Earth from the MW centre and $(l,b)=(121.2^\circ,-21.6^\circ)$ the Galactic co-ordinates of M31.

HSC has a microlensing efficiency of $\epsilon\sim 0.1-0.8$ for time periods $2\,{\rm minutes}\lesssim\hat{t}\lesssim 7 \,{\rm hours}$ with a number of stars $N_s\sim 10^7-10^9$. The advanced treatment of the efficiency and candidate selection employed in Ref.~\cite{2017arXiv170102151N} is beyond the scope of the present work. In order to get a sense for the constraints that could be obtained with a dedicated analysis, we model the HSC microlensing efficiency as a step function with $\epsilon=0.5$ in the given time scale (see Fig.~14 of Ref.~\cite{2017arXiv170102151N}). To normalise the exposure we compute the expected number of events for PBHs, and rescale $E$ to approximately match the constraints in Fig.~29 of Ref.~\cite{2017arXiv170102151N} without accounting for finite source size and using the extrapolated number counts of source stars (this is the most optimistic constraint).  

It is the short cadence that gives HSC access to low PBH masses, and for our purposes will allow constraints on the minicluster fraction for the QCD axion. This is because low mass objects create lensing events on shorter timescales due to the smaller radius of the microlensing tube. Thus detecting microlensing events by such objects requires large efficiency on small times scales, i.e. a short cadence.

HSC use pixel lensing and image subtraction to select microlensing candidates. Using this technique, they identify a large number of variable stars, eclipses, and other transient events. They find a single event with a light curve consistent with a PBH microlensing event, though the genuine nature is not confirmed. Thus, the Poisson statistics 95\% C.L. upper limits on the expected number of microlensing events are
\begin{eqnarray}
N_{\rm exp}\leq \left\{ 
\begin{array}{ll} 
3 \quad \text{w/o. the PBH candidate} \\
4.74 \quad \text{w. the PBH candidate}
\end{array}
\right. .
\end{eqnarray}
We take $N_{\rm exp}\leq 3$ as the conservative limit on the minicluster lensing events. A dedicated analysis of the HSC data with the minicluster light curve would be required to be more precise, and this is beyond the scope of the present work.

\subsection{Results: Expected Number of Microlensing Events}

\begin{figure}
\includegraphics[width=\columnwidth]{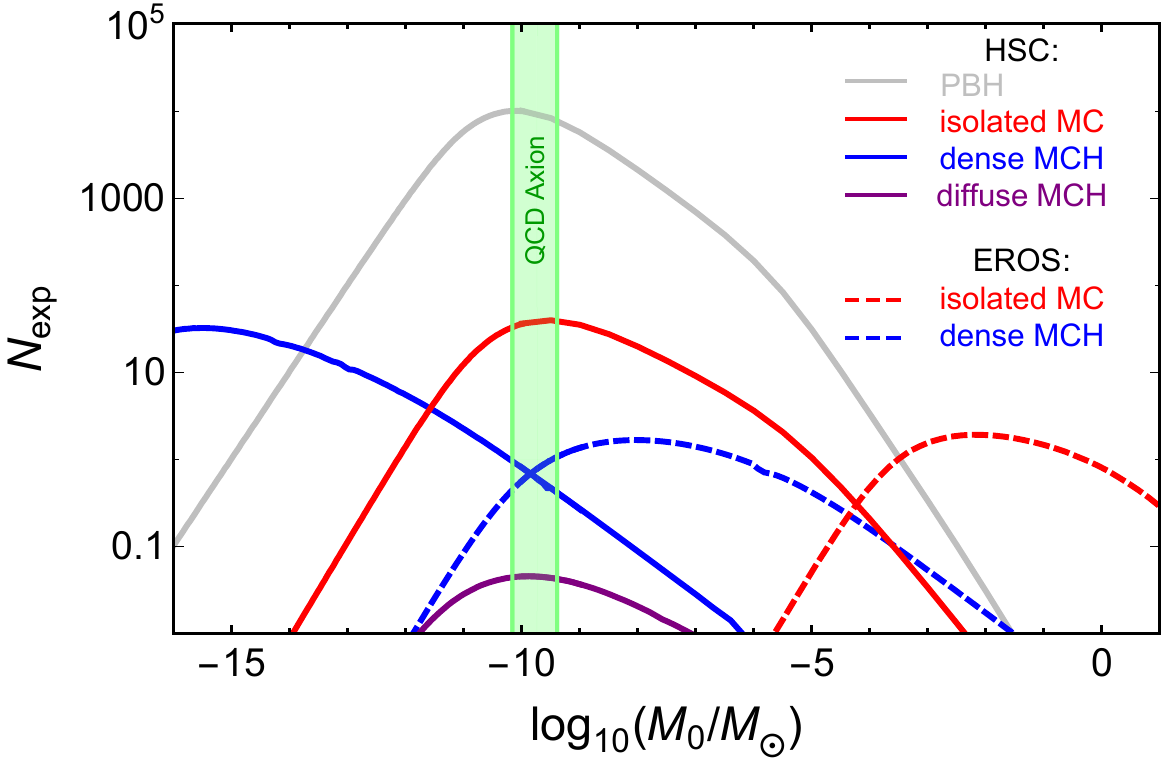}
\vspace{-2.5em}\caption{{\bf Expected Microlensing Events}: Here we assume that all the DM is composed of miniclusters on small scales. Lines show the effects of our modelling of the minicluster mass function and density profile for HSC and the EROS survey.}  
\label{fig:Nexp_M0}
\end{figure}

In Fig.~\ref{fig:Nexp_M0} we show the expected number of microlensing events in various minicluster scenarios as a function of $M_0$  for HSC and EROS assuming $f_{\rm MC}=1$. The number of events in HSC is generally far larger than for EROS due to the huge volume of DM between the Earth and M31 leading to a larger optical depth to microlensing for HSC~\cite{2017arXiv170102151N}. In order to understand the effects of our modelling we show four different calculations of $N_{\rm exp}$ for HSC. 

In the first (gray full-line in Fig.~\ref{fig:Nexp_M0}), we compute the event rate for point-like objects (i.e PBHs) of fixed mass $M_0$ (i.e. Dirac-delta-function mass distribution) to normalise the exposure and efficiency. 

We then compute the case of isolated miniclusters (Dirac-delta-function mass distribution but non-point like objects), with density profiles determined by ${\rm d}n/{\rm d}\delta$ extracted from Fig.~\ref{fig:fit_tkachev}. This corresponds to the red full/dashed line in Fig.~\ref{fig:Nexp_M0} for the HSC/EROS survey. This additional treatment reduces the number of events by a factor of $\mathcal{O}(10^2)$ due to the requirement of large $\delta$ such that $\mathcal{R}>0$. We consider this scenario as the most conservative: miniclusters are too dense to suffer much disruption on mergers, and MCHs are likely to be a ``plum pudding'' of objects of mass $M_0$. In this case, the modulating role of the MCH mass function is not relevant for the HSC cadence and QCD axion.
 
The dense MCH case includes in addition the effects of ${\rm d}n/{\rm d}M$ i.e an extended mass function. A microlensing survey is sensitive to objects of fixed mass $M$. The mass function spreads the MCHs to $M>M_0$ (with more total mass at larger $M$), shifting the central $M_0$ to smaller values. The density profiles of the dense MCHs are also computed using ${\rm d}n/{\rm d}\delta$ i.e. mergers forming MCHs are assumed to preserve the distribution of halo concentrations. This treatment corresponds to the blue full/dashed line in Fig.~\ref{fig:Nexp_M0} for the HSC/EROS survey. This scenario is more conservative for the HSC survey and the QCD axion since it reduces the number of events by moving mass out of the central region of sensitivity.

Finally, the diffuse minicluster case uses ${\rm d}n/{\rm d}M$, but assumes that all MCHs with $M$ outside the small window $M_{0}/10 \leq M \leq 10 M_0$ have too low density for microlensing. Mergers are assumed to disrupt the miniclusters and the MCHs with $M>10M_0$ are uniform with concentration $c_{\rm NFW}(M)$, far too low to lens. The cut in ${\rm d}n/{\rm d}M$ reduces significantly the number of events. This is the most pessimistic model, corresponding to an effective reduction in $f_{\rm MC}$ caused by mergers. This scenario corresponds to the purple line in Fig.~\ref{fig:Nexp_M0} for the HSC survey.

\begin{figure}
\includegraphics[width=\columnwidth]{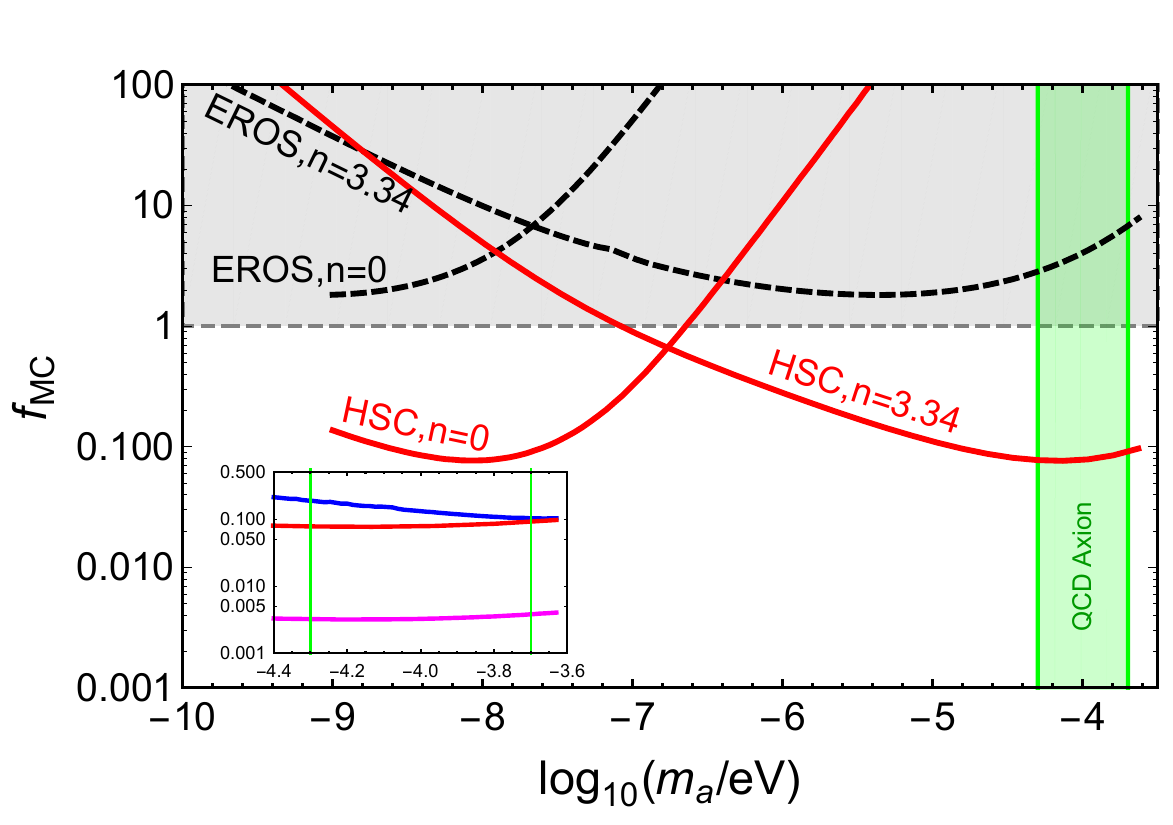}
\vspace{-2.5em}\caption{{\bf Limits on the Fraction of DM collapsed into Miniclusters}: The model adopted is for ``isolated miniclusters'', which we consider the most realistic. The shaded region shows the allowed mass for the QCD axion with miniclusters. Where the $n=3.34$ lines intersect this region, $f_{\rm MC}$ is constrained for the QCD axion. The inset shows a zoom-in. The magenta (blue) line in the inset shows a hypothetical improved observation by HSC ten nights with an efficiency $\epsilon\sim 1$ in the case of isolated miniclusters (dense MHCs).}  
\label{fig:fMC_ma}
\end{figure}

Using the Poisson statistics 95\% C.L. limit from above we find constraints on $f_{\rm MC}$ as a function of $m_a$, the axion mass, presented in Fig.~\ref{fig:fMC_ma} for the most realistic isolated miniclusters case. The dashed black lines correspond to the EROS limits in the $n=0$ or $n=3.34$ hypothesis. The full red lines correspond to the HSC limits in the $n=0$ or $n=3.34$ hypothesis.
We find that EROS does not place any bound on $f_{\rm MC}<1$ however HSC places very strong bounds on $f_{\rm MC}$ for an axion-like particle with $n=3.34$, reaching as low as $f_{\rm MC}\approx 8.0 \times 10^{-2}$ for $m_a\approx 50\,\mu\text{eV}$.

The shaded green band shows the allowed mass for the QCD axion fixed by $m_a=6.6\,\mu\text{eV}(10^{12}\text{ GeV}/f_a)$~\cite{weinberg1978,wilczek1978} and the relic density: $50\,\mu\text{eV}\lesssim m_a\lesssim 200\,\mu\text{eV}$~\cite{2016arXiv161001639B}. The $n=3.34$ lines represent temperature evolution of the axion mass: where these lines intersect the shaded band, $f_{\rm MC}$ is bounded for the QCD axion, and we find $f_{\rm MC}<0.083(m_a/100\,\mu\text{eV})^{0.12}$.

As shown in the inset of Fig.~\ref{fig:fMC_ma}, these results could be improved. Indeed the magenta line shows a hypothetical improved observation by HSC, extending the current one night to ten nights with an efficiency $\epsilon\sim 1$, leading to a forecast bound of $f_{\rm MC} \lesssim 0.004$ for the QCD axion in the isolated miniclusters case. The improved observation is also able to bound $f_{\rm MC} \lesssim 0.1$ in the more pessimistic (for the QCD axion) dense MCH scenario. We advocate a dedicated analysis of the HSC microlensing data to place more rigorous bounds on $f_{\rm MC}$ than we have approximated here, and for a longer microlensing survey in order to improve those bounds further.  We are confident that a more thorough analysis by the observing teams will show that HSC, and microlensing in general, is now a powerful tool to constrain the QCD axion and, more generally, axion-like particles.

\subsection{Theoretical Uncertainties attached to our Results}

We now discuss various theoretical uncertainties and modeling that can give small shifts in the constraints. As we already discussed, an additional uncertainty comes from our simplified modeling of the lensing efficiency. 

\begin{figure}
\includegraphics[width=\columnwidth]{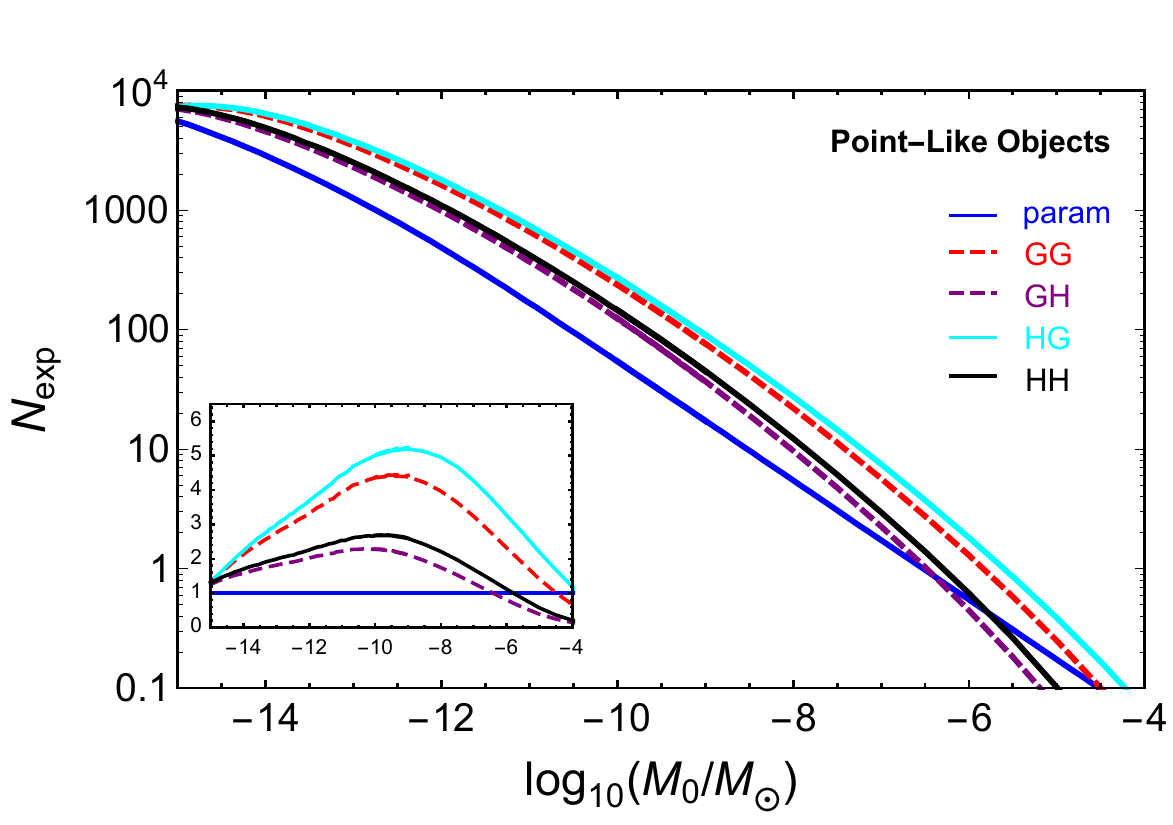}
\vspace{-2.5em}\caption{{\bf Theoretical Uncertainties in the Mass Function}: Lines show the effects of our modeling concerning the mass function for the HSC survey. In order to isolate this effect from the effect of the density profile we considered the MCHs as point-like objects (i.e. PBHs). The inset displays the various distributions normalised to our default simple parametric mass function (blue line).}  
\label{fig:Nexp_M0_HMFs}
\end{figure}

In Fig.~\ref{fig:Nexp_M0_HMFs}, we study the impact of the modelling of the mass function on the computation of the expected number of lensing events. We used the analytical formulations of Appendix~\ref{appendix:modelling_mf}. The full-blue line in Fig.~\ref{fig:Nexp_M0_HMFs} corresponds to our simple parametric mass function discussed earlier in section~\ref{sec:parameterize}. 

For all combinations of initial power spectrum (Gaussian or Heaviside) and window function (Gaussian or Heaviside), we inject the variance $\sigma^2_{\rm XY}$ of Appendix~\ref{appendix:modelling_mf} in the usual Press-Schechter mass function also given in Appendix~\ref{appendix:modelling_mf} in order to estimate the expected number of microlensing events. In the case of the Gaussian window function, we used the half-mode models for the low-mass cut-off. In order to isolate this source of theoretical uncertainty from other mass-dependent effect (such as the rescaling lensing tube factor) we considered in Fig.~\ref{fig:Nexp_M0_HMFs} the MCHs as point-like objects which is a quite unrealistic scenario. However we said previously that the additional treatment of considering MCHs as non-point like objects reduces the number of events by a factor of $\mathcal{O}(10^2)$ due to the requirement of large $\delta$ such that $\mathcal{R}>0$. Our result is that our simple description of an extended mass function is conservative since for some characteristic mass,
$M_0$, a different choice of mass function could increase the expected number of lensing events by up to a factor of five. 

The analytic results for the variance and associated MCH mass functions (``GG", ``GH", ``HG", ``HH") have generically a smaller upper limit compared to our simple parametric approximation (see Fig.~\ref{fig:hmf_modelling}). This small difference starts to matter when this upper limit approaches the critical value of the experiment. This critical value for HSC is around $M_{0}/M_{\odot}=5\times10^{-5}$ and corresponds to the typical PBH mass when the lensing experiment loses all sensitivity due to the finite observing time (grey line of Fig.~\ref{fig:Nexp_M0}). At a particular moment, the parametric HMF will have MCH masses outside the sensitivity region of HSC, however for the other HMF all the masses would remain in the sensitive region and would consequently predict more lensing events. When every single HMF have MCHs heavier than that critical value then the discrepancy between them starts to shrink. This feature explains why the lines associated to the more refined HMF are close to the parametric HMF for low and large values of $M_{0}$ and reach a maximum around $M_{0}/M_{\odot} \simeq 10^{-9}$. The inset in Fig.~\ref{fig:Nexp_M0_HMFs} displays the various distributions normalised to our default simple parametric mass function (blue line).

\begin{figure}
\includegraphics[width=\columnwidth]{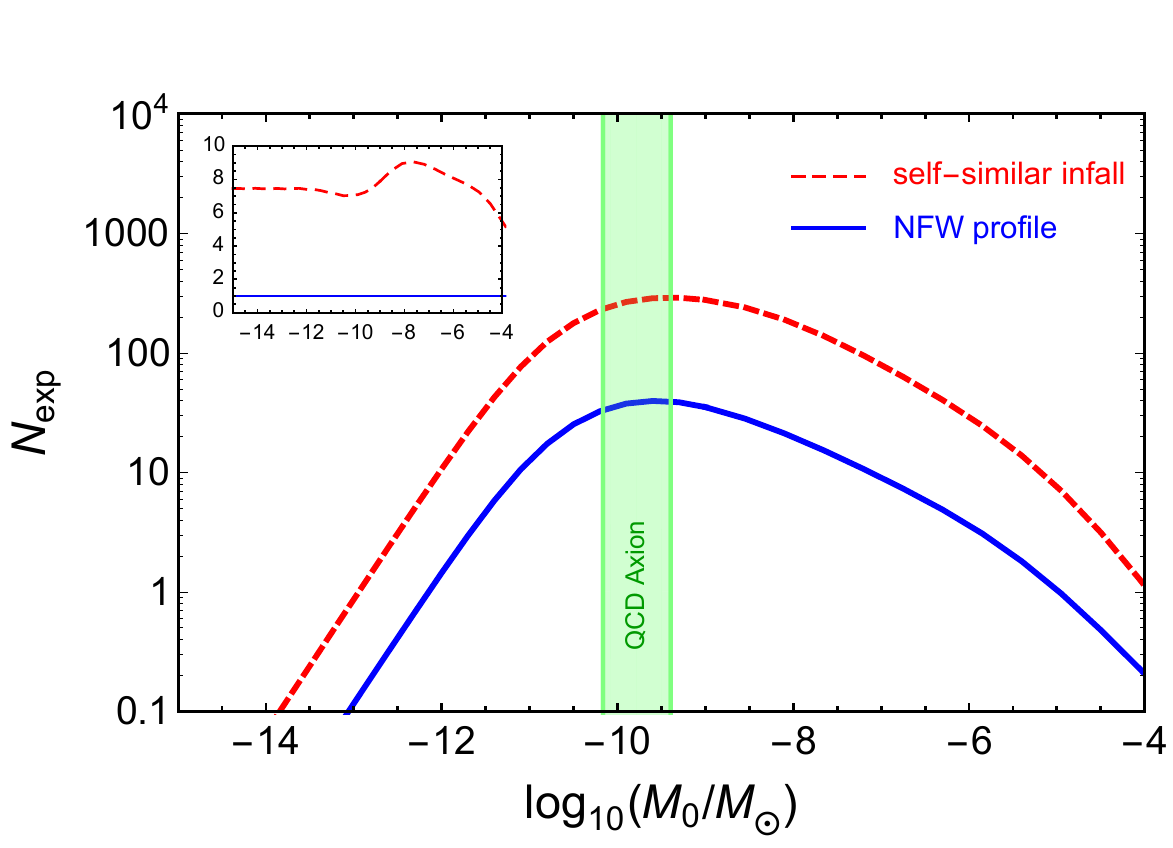}
\vspace{-2.5em}\caption{{\bf Theoretical Uncertainties in the Density Profile}: Here we assume that all the DM is composed of miniclusters on small scales and take the isolated minicluster scenario. Lines show the effects of our choice concerning the density profile for the HSC survey. The dashed-red line assumes a self-similar infall profile while the full-blue line assumes our default NFW profile. The inset displays the distributions normalised to the default NFW case (blue line).}  
\label{fig:Nexp_M0_DPs}
\end{figure}

In Fig.~\ref{fig:Nexp_M0_DPs}, we study the impact of the density profile on the computation of the expected number of lensing events for the HSC survey. The dashed-red line assumes a self-similar infall profile (cf. section \ref{sec:profiles}) while the full-blue line assumes our default NFW profile. The inset displays the two distributions normalised to the the default NFW line (blue line). As we can see, the self-similar infall profile predicts a higher expected number of lensing events by a factor up to eight for $M_{0}$ masses below $10^{-6}M_{\odot}$. This is caused by the self-similar profile being more compact, and thus requiring lower threshold values of $\delta$ for microlensing (see Fig.~\ref{fig:Rplot_self-similar}).

\section{Summary and Conclusions}
\label{sec:conclusions}

The QCD axion remains one of the best motivated dark matter candidates some 40 years after it was originally proposed. Unlike the case of thermal WIMPs, the QCD axion parameter space remains wide-open in the face of direct constraints due to the extraordinarily weak interactions between axions and the standard model. All of that is about to change, with a wide range of proposed experiments set to probe a large part of the parameter space in the coming decades. It is therefore timely to study more subtle aspects of axion DM that may affect direct detection signals.

If PQ symmetry breaking occurs during the radiation-dominated phase in the early Universe (or indeed during a putative matter-dominated phase) then the axion DM model is severely constrained and makes rather precise predictions. One such prediction is the existence of miniclusters: gravitationally bound lumps of axions with masses of the order $M_0\sim 10^{-10}M_\odot$. There is no theoretical prediction for the fraction of DM in miniclusters, $f_{\rm MC}$, but naively we expect it to be of order unity. If $f_{\rm MC}$ is large, the direct detection signal for axions is reduced by $1-f_{\rm MC}$ (assuming that the probability of an encounter between the Earth and a minicluster over the course of an experiment is essentially zero).

Despite in some sense representing ``half'' of axion parameter space, the minicluster scenario has attracted relatively little attention over the years. In the present work we have attempted to revive interest in it, and attempted to observationally bound $f_{\rm MC}$. 

We began by computing the mass function of miniclusters. We predict the slope of the mass function should be $M^{-1/2}$ arising from the white noise initial conditions. This is the mass function on large scales, where we have assumed Gaussianity of fluctuations. The minicluster halos (MCHs) that we predict are composed of many individual miniclusters (up to $10^6$) and would not be seen in simulations such as Ref.~\cite{2007PhRvD..75d3511Z} except in very large boxes with total mass many times larger than $M_0$. We outlined uncertainties in the MCH mass function due to the initial power spectrum and window functions.

Our modelling of minicluster and MCH formation used the simplest possible implementation of Press-Schechter in an analytic form. More advanced semi-analytic formulations of extended Press-Schechter can be used to answer deeper questions about the merger history and tidal effects and will provide more accuracte mass functions~\cite{2010MNRAS.401.1796A}. The public code \textsc{galacticus} is suitable for such a study~\cite{2012NewA...17..175B}. Such approaches can account for gravitational $N$-body-like dynamics, but will not address issues associated to the unique minicluster initial conditions, which can only be answered by field theory simulations of the axion sting network.

By treating the hierarchical structure formation as described by CDM on large scales, we computed the concentrations and formation times of MCHs. We used this to provide a simple estimate of the effect of tidal stripping on the ``seed'' miniclusters, and concluded it is likely a minor effect. Thus, miniclusters should be present as ``plums in a pudding'' in MCHs, a scenario we dubbed ``isolated miniclusters''. We also considered two other, less realistic, scenarios for mergers and their effects on observations. In the isolated minicluster case we concluded that the MCH mass function was irrelevant for microlensing constraints. 

In the case of the QCD axion we expect the individual miniclusters to have masses very roughly comparable to larger asteroids like Vesta or Pallas with radius of order 1 AU for $\delta=1$. After structure formation, these miniclusters find themselves gravitationally bound up into MCHs with typical masses up to that of Saturn and radii of approximately $10^4$ AU (0.1 light years).  We are left wondering if there are other consequences of MCHs that can be tested observationally.

We have not discussed the formation of miniclusters themselves, where the assembly history will be very different. Our mass function on scales smaller than $M_0$ is also subject to large uncertainties. The shape of the mass function for $M<M_0$ (describing fragmentation and formation of miniclusters) could be drastically different from our estimates. We have tried to quantify this uncertainty using different window functions and cut-off models. We predict a cut-off in the mass function at least at the axion Jeans scale, and probably not much below $M_0$ itself.

The largest uncertainties relate to the non-Gaussianity of the minicluster density field. The non-Gaussianity will also affect the mass function on small scales. We have treated the non-Gaussianity on scales of order $M_0$ as giving the distribution of minicluster concentrations, using a fit from simulations~\cite{Kolb:1995bu} giving the distribution of overdensities $\mathcal{F}(\delta>\delta_0)$. This predicts that miniclusters come in a very wide range of different sizes for a fixed initial mass. As we have assumed a particular form for $\mathcal{F}$, our results could rather be interpreted as a constraint on the integrated tails of this distribution with $\delta_0>\delta_{\rm lens}$, where $\delta_{\rm lens}$ is the critical density parameter for efficient microlensing.

Numerically verifying the distribution $\mathcal{F}$ for large $\delta$ is a key future project necessary, since our later results rest on an extrapolation of this function over a wide range.\footnote{The simulations of Wiebe, Redondo and Niemeyer~\cite{NiemeyerPATRAS} will be able to investigate this extrapolation. In the initial (no gravity) initial conditions some extrapolation seems justified to at least $\delta\gtrsim 100$. We thank Javier Redondo for discussion on this point.} Simulations of the minicluster field in the absence of gravity are not sufficient for this purpose: our results rely on the size distribution being preserved to the present day for bound objects of mass $M_0$, which take some time to assemble gravitationally. The numerical verification should be done by performing a large number of small-box simulations containing a mass of just a few $M_0$.

We computed the microlensing ``tube size'' and event rate for miniclusters. The distribution $\mathcal{F}$ was key to this computation, since the densest miniclusters in the tail of the distribution have the largest effective tube radius and contribute dominantly to the expected number of events. Using the recent HSC limit on the number of microlensing events $N_{\rm exp}\leq 3$ over event time scales between two minutes and seven hours we were able to place the first observational bound on miniclusters of $f_{\rm MC}<0.083 (m_a/100\,\mu {\rm eV})^{0.12}$ over the range relevant to the QCD axion. We also presented bounds for other axion-like particles.

In the event that observations are made which suggest microlensing events caused by DM, we would be faced with the pleasurable task of identifying the precise nature of these microlenses. The formation of miniclusters is some sense one of the more conservative scenarios for compact object formation, relying only on ordinary SSB leading to standard axion production. There are various other possible compact objects which could make up a significant fraction of the dark matter.  We have already pointed out that with enough statistics we could tell the difference between point like objects such as PBHs and extended density profiles such as axion miniclusters.  An interesting question is whether we could tell the difference between axion miniclusters and ``ultra compact mini-haloes'' that emerge as a result, for example, of features in the inflationary spectrum or a period of matter domination \cite{Aslanyan:2015hmi}. Just like in the case of miniclusters, there is significant theoretical uncertainty relating to the density profiles of such objects~\cite{2017arXiv171205421S}, and thus microlensing alone might not distinguish these types of compact object.

UCMHs can be composed of any type of DM, but they are typically thought of as being composed of WIMPs. Clearly, interactions distinguish axions and WIMPs, and could also be used to distinguish miniclusters and UCMHs. A clear consequence of the high UCMH density is in boosting the WIMP annihilation rate, causing UCMHs to appear as luminous gamma ray sources~\cite{2012PhRvD..85l5027B}. On the other hand, axion-photon conversion leads to miniclusters acting as radio sources, with a possible relation to fast radio bursts~\cite{2015JETPL.101....1T,2015PhRvD..91b3008I,2017arXiv170704827I}. Finally, on distinguishing different DM microlensig candidates, the early Universe physics responsible for each (minicluster, PBH, and UCMH) is in each case quite different, curvature perturbations for PBHs and UCMHs, isocurvature and SSB for miniclusters, and could lead to other indirect evidence supporting either case.

An $\mathcal{O}(1)$ uncertainty on the microlensing constraints comes from the minicluster density profile. We treated miniclusters as an NFW profile, which is close to a hard-sphere with mass located predominantly at a single radius. In the case of a self-similar density profile, miniclusters are somewhat denser, and we found tighter constraints on the density fraction. Again, simulations are necessary to confirm which density profile is the correct one.

An observational bound on $f_{\rm MC}$ can be seen two ways. Firstly, observationally it is good to know that $f_{\rm MC}$ is small, since in the absence of other information this allows direct detection constraints to be interpreted with more confidence. Secondly, it allows the possibility to exclude, or discover evidence for axions if theoretical predictions can be made.  The same simulations that will provide us with more information about the distribution of overdensities $\mathcal{F}$ could also narrow down predictions on the possible range of $f_{\rm MC}$, and investigate the minicluster density profiles \cite{NiemeyerPATRAS}. We hope that our preliminary investigations motivate the need for such further study.

If the eventual theoretical predictions are in violation of micolensing bounds, then microlensing constraints have excluded the QCD axion with late-time symmetry breaking. On the other hand, if theoretical predictions are within an order of magnitude or so below our constraints, then future microlensing surveys could expect to see evidence of a sub-dominant population of miniclusters in the galactic DM.  Upcoming observational efforts such as the Zwicky Transient Facility \cite{Bellm:2014pia} and eventually the Large Synoptic Survey Telescope \cite{Abell:2009aa} are expected to provide much more sensitivity to microlensing events.  We are excited to see what these telescopes observe and the implications of those observations for the nature of dark matter.


\section*{Acknowledgements}

The work of JQ and MF is supported by the UK STFC Grant ST/L000326/1.   MF is also funded by the European Research Council under the European Unions Horizon 2020 programme (ERC Grant Agreement no.648680 DARKHORIZONS). The work of DJEM is supported by a Royal Astronomical Society Postdoctoral Fellowship. We acknowledge useful discussions with Alberto Diez-Tejedor, Anne Green, Edward Hardy, Jens Niemeyer, Javier Redondo, and Paul Shellard. We also thank Torsten Bringmann, August Geelmuyden and Andrzej Jan Hryczuk for useful communications.

\appendix

\section{The Axion Relic Density}
\label{appendix:relic_density}

The following section uses known results concerning the axion relic density to find the range of axion masses where the minicluster scenario is possible. We reproduce and extend the treatment of the relic density in Ref.~\cite{2017JHEP...02..046H}, giving more details on the computation. 

\begin{figure}
\vspace{-0.2em}\includegraphics[width=\columnwidth]{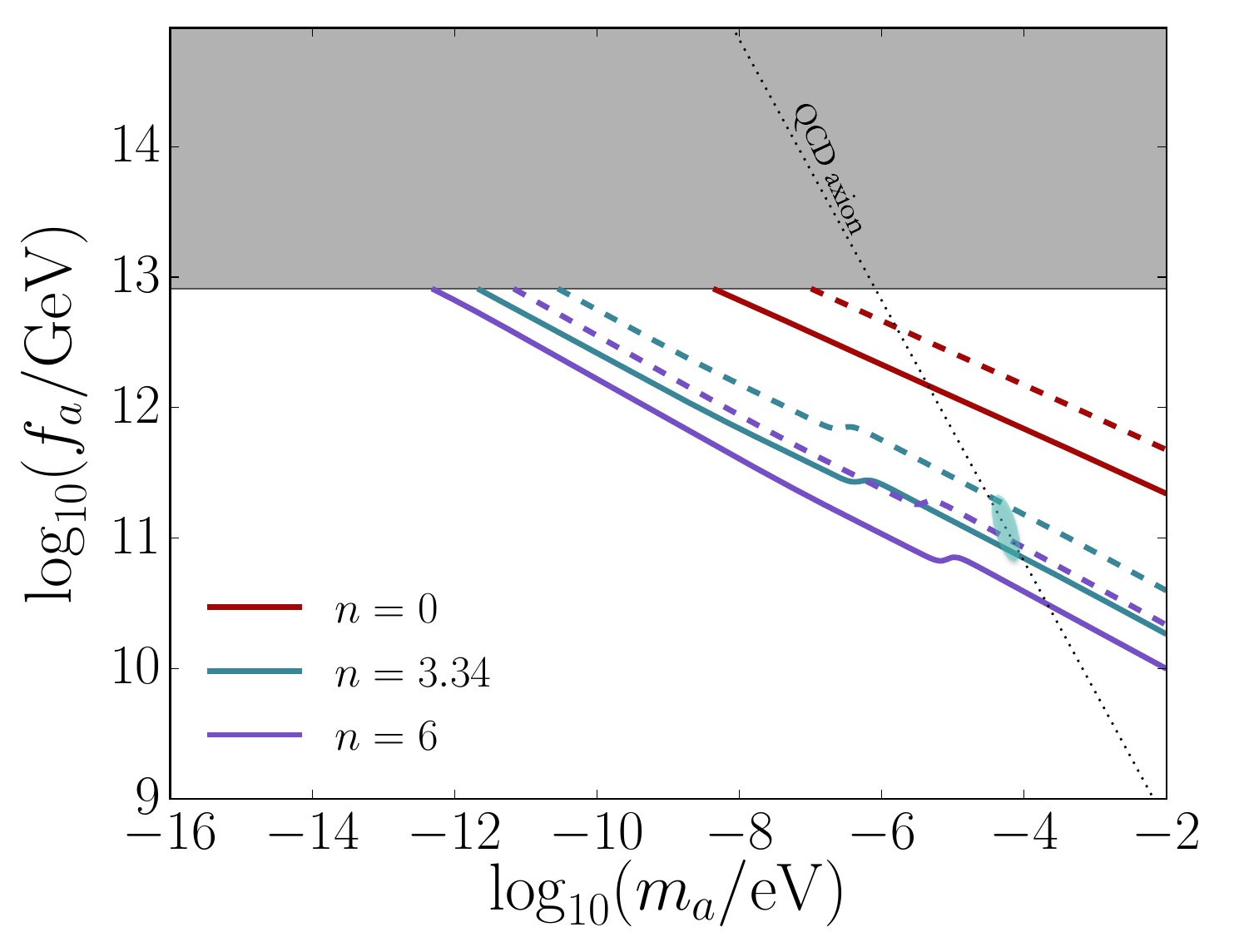}
\vspace{-2.5em}\caption{{\bf The Axion Relic Density}: Contours $\Omega_a h^2=0.12$ are shown for a variety of models. Solid (dashed) lines have $\alpha_{\rm dec}=2.48\,(1)$ and $c_{\rm an}=c_n\, (1)$ to account for uncertainty in the relic density from decay of topological defects and anharmonicities in the axion potential. The shaded area indicates $f_a\geq 8.2\times 10^{12}\text{ GeV}$, where inflationary constraints on the tensor-to-scalar ratio imply that miniclusters cannot be formed. The QCD axion is indicated by the ellipse giving $m_a\approx 100\,\mu\text{eV}$.}  
\label{fig:mass-lik}
\end{figure}

In the minicluster scenario the breaking of the PQ symmetry giving rise to axions must occur after any period of cosmic inflation (or indeed any other smoothing mechanism for the initial conditions). The axion relic density is produced by the coherent oscillations of the highly inhomogeneous classical axion condensate in a thermal background. In principle, the calculation of the relic density involves two computationally challenging pieces: the thermal evolution of the axion mass in finite temperature quantum field theory, and the classical evolution of the inhomogeneous condensate. We do not perform such calculations here, and in the literature results are only generally available for the QCD axion. However, both of these complexities can be parameterised into a much simpler computation, allowing to treat the more general ALP models, and account for uncertainties in the numerical results in the literature.

In order to avoid having to compute the inhomogeneous field evolution, we make the standard assumption that there are two sources of axion relic density: production of cold axion particles via the decay of topological defects, and vacuum misalignment production from the homogeneous axion condensate. This separation of axion populations is in fact artificial, the entire relic density being contained in coherent oscillations of the same classical field. It is a useful terminology, since the inhomogeneous calculation has the same parametric scaling as the homogeneous calculation, allowing a single ``fudge factor'' to be introduced and the parameter space can then be explored in the much simpler homogeneous case.

In order to account for the finite temperature effects, we parameterise the mass evolution as a power law. The power law index can be computed either using an instanton model, or a lattice computation, and depends on the particle content (which is of course fixed only in the case of QCD).

Of the total axion relic density, some fraction $f_{\rm MC}$ ends up bound in miniclusters. In principle this fraction can be determined by proper simulation of PQ symmetry breaking and axion production including the effects of gravity up to matter-radiation equality. Instead, we take $f_{\rm MC}$ as a phenomenological free parameter to be constrained by the data, and so our constraints on $f_{\rm MC}$ can be used to constrain models of PQ symmetry breaking.

The relic density of axions of mass $m_a$ is determined by the decay constant, $f_a$, which for the minicluster scenario must satisfy $f_a<8.2\times 10^{12}\text{ GeV}$. The bound applies in an inflationary scenario by imposing the observational constraint on the cosmic microwave background tensor-to-scalar ratio $r<0.07$~\cite{2016PhRvL.116c1302B}, which in turn constrains the Hubble scale during inflation. The constraint arises by imposing that the Gibbons-Hawking temperature, $T_{\rm GH}=H_I/2\pi$, must be higher than the PQ scale in order that the PQ symmetry remain unbroken during inflation, and we have assumed for simplicity that the phase transition occurs at $T=f_a$. In a non-inflationary scenario, if reheating is allowed to occur to temperatures $T_{\rm reh.}>8.2\times 10^{12}\text{ GeV}$, then the minicluster scenario could be extended to larger values of $f_a$. For definiteness, we consider only the inflationary scenario from now on. 

We must find the range of axion masses for which the minicluster scenario is relevant. To do this we impose the relic density constraint, $\Omega_a h^2=0.12$~\cite{planck_2015_params}, with $\Omega_a h^2$ computed in the minicluster scenario with the necessary requirement that $f_a<8.2\times 10^{12}\text{ GeV}$. 

Consider the homogeneous misalignment production of axions. The equation of motion for the homogeneous axion field, $\phi$, is:
\be
\ddot{\phi}+3H\dot{\phi}+V'(\phi)=0\, ,
\label{eqn:realignment}
\ee
where dots denote derivatives with respect to physical time $t$. The axion potential is $V(\phi)$, and prime denotes derivative with respect to $\phi$. The precise form of $V(\phi)$  is not important for the present treatment of the relic density. An additional parameter $N_{\rm DW}$ sets the periodicity of the axion field and then determines the number of domain walls. For the QCD axion, $N_{\rm DW}$ is determined by the color anomaly of the quarks carrying PQ charge, and is $N_{\rm DW}=1$ for the ``KSVZ'' axion~\cite{1979PhRvL..43..103K,1980NuPhB.166..493S}. For simplicity of presentation we set $N_{\rm DW}=1$, the $N_{\rm DW}>1$ case having quite different cosmology~\cite{2015PhRvD..91f5014K}.

Axion oscillations begin at temperature $T_{\rm osc}$ when the potential term dominates over the friction provided by the Hubble expansion. Thanks to the Kibble mechanism, the axion field remains homogeneous on scales up to the horizon size at this time. Once axion oscillations begin, the axion number density $n_a(T)=\rho_a(T)/m_a(T)$ becomes conserved (for slow $m_a$ variation) and the axion relic density at a later time when the temperature is $T_0$ is:
\begin{widetext}
\be
\rho_a^{\rm mis}=m_a(T_0)n(T_{\rm osc})\left(\frac{a(T_{\rm osc})}{a(T_0)}\right)^3 =\frac{1}{2} m_a(T_0)m_a(T_{\rm osc})f_a^2\theta_i^2\left(\frac{a(T_{\rm osc})}{a(T_0)}\right)^3\, , 
\label{eqn:mis_relic}
\ee
\end{widetext}
where $\theta_i=\phi_i/f_a\in [-\pi,\pi]$ is the ``initial misalignment angle''. The fraction of the critical density is given by $\Omega_a=\rho_a/(3H^2M_{pl}^2)$. The scale factor is related to the temperature by the condition of constant entropy:
\be
a(T)\propto g_{\star,{\rm S}}(T)^{-1/3}T^{-1}\, ,
\ee
with the proportionality normalised by fixing matter radiation equality at redshitf $z_{\rm eq}=3402$~\cite{planck_2015_params}.  We use the fit for the entropic degrees of freedom, $g_{\star,{\rm S}}(T)$, from Ref.~\cite{Wantz:2009it}.

In a harmonic potential, $T_{\rm osc}$ is given by:
\be
3H(T_{\rm osc})=m_a(T_{\rm osc}) \, . 
\label{eqn:tosc}
\ee
We allow temperature variation of the mass parameterised as:
\be
m_a(T)=m_{a,0}\left( \frac{T}{\mu}\right)^{-n}\, ,
\ee
for $T>\mu\propto\sqrt{m_{a,0}f_a}$ and $m_a(T<\mu)=m_{a,0}\equiv m_a$. Low-temperature variation of the mass occurs for axions acquiring their potential from a strongly coupled gauge theory such as QCD, and the index $n$ can be computed given the particle content. As a representative of QCD we take $n\approx 3.34$ from the ``interacting instanton liquid'' model~\cite{Wantz:2009it}, which is close to the results from lattice simulations ($n\approx 3.55\pm 0.30$~\cite{2016PhLB..752..175B,Borsanyi:2016ksw}) and the canonical dilute instanton gas ($n=4$~\cite{1981RvMP...53...43G}). For simplicity we neglect the case $\mu\neq\sqrt{m_af_a}$. The QCD axion has $\mu\approx \Lambda_{\rm QCD}\approx 2.5\sqrt{m_af_a}$. The very small effect of this factor on the relic density and $M_0$ is unimportant at our level of accuracy given the other associated uncertainties that we include below. The case $\mu\gg\sqrt{m_af_a}$ is approximated by $n=0$ and occurs for some string axions and ``accidental axions''~\cite{2014JHEP...06..037D,2016PhRvD..93b5027K}. 

Assuming radiation domination, the Hubble rate is given by the Friedmann equation:
\be
3H^2M_{pl}^2 = \frac{\pi^2}{30}g_{\star,{\rm R}}(T)T^4\, , \label{eqn:friedmann_temperature}
\ee
where $g_{\star,{\rm R}}(T)$ is the number of relativistic degrees of freedom at temperature $T$. The epoch when axion oscillations begin is found by solving Eqs.~\eqref{eqn:tosc} and \eqref{eqn:friedmann_temperature}, which we do numerically using the fit for $g_{\star,{\rm R}}(T)$ from Ref.~\cite{Wantz:2009it}, which includes all the standard model degrees of freedom and the QCD phase transition.

Anharmonicities in the potential lead to flattening away from the origin and delay the onset of oscillations for large $\theta$. The \emph{time} that oscillations begin grows as~\cite{1992PhRvD..45.3394L}
\be
t_{\rm osc} = m_a^{-1}(t_{\rm osc}) \ln [e/p(x)] \, ,
\label{eqn:anharmonic_time}
\ee
where $p(x)$ is a polynomial function of $x=\theta_i/\pi$ that is found from fitting to numerical solution of Eq.~\eqref{eqn:realignment} with a specific potential~\cite{1986PhRvD..33..889T}, and \emph{does not depend on $m_a$ or $f_a$} as each can be absorbed in an appropriate choice of units. For the potential $V(\theta)=m_a^2f_a^2(1-\cos\theta)$ we find that $p(x)=1-x^4$ gives a good fit to our own numerical solutions for axion-like particles with $T$-independent mass and $g_\star={\rm const.}$.

Using that $t\propto T^{-2}$ from the solution to the Friedmann equation, Eq.~\eqref{eqn:anharmonic_time} can be solved for the correction to $T_{\rm osc}(\theta_i)$ that is then substituted into Eq.~\eqref{eqn:mis_relic} to find the relic density. 

With the assumption of constant $g_\star$ during the epoch over which anharmonic corrections affect $T_{\rm osc}$, this solution can be found analytically, leading to an anharmonic correction to the relic density:
\be
\rho_a^{\rm mis.} \rightarrow f_{\rm an}(\theta_i)\rho_a^{\rm mis.} \, ,
\ee
with the anharmonic correction function $f_{\rm an}(\theta_i)=\{\ln [e/p(x)]\}^{q}$, and the power $q=3/2-n/(2n+4)$. With non-constant $g_\star$, $f_{\rm an}(\theta_i)$ cannot be found analytically, and furthermore it will in general depend on $m_a,f_a,n$, though this is not usually stated. 

In the minicluster scenario the vacuum misalignment relic density must be averaged over $\theta_i$, reflecting the fact that the current observable Universe is many times larger than the horizon size when axion oscillations begin. One must replace $\theta_i^2 f_{\rm an}(\theta_i)$ by:
\be
\langle \theta_i^2f_{\rm an}(\theta_i)\rangle = \frac{1}{2\pi}\int_{-\pi}^{\pi}\theta^2f_{\rm an}(\theta){\rm d}\theta \equiv c_{\rm an}\frac{\pi^2}{3}\, .
\ee
The number $c_{\rm an.}$ comes from the anharmonic corrections. Assuming one can treat $g_\star$ as constant over the period in which anharmonic corrections act, with $p(x)=(1-x^4)$ for the cosine potential and $n=(0,3.34,6)$ we find $c_{\rm an}=(2.7,2.1,2.0)$, which we term $c_n$. 

Now consider the population of axions from topological defect decay. For $N_{\rm DW}=1$, the population of axions produced by decay of the string-wall network at $T_{\rm osc}$ can be parameterised by writing the total relic density as
\be
\Omega_a h^2=(1+\alpha_{\rm dec})\Omega_a^{\rm mis} h^2 \, .
\ee
The parameter $\alpha_{\rm dec}$ is computed from numerical solution of the decay of the axion string-wall network~\cite{1985PhRvD..32.3172D,1987PhLB..195..361H,1994PhRvL..73.2954B,1996PhRvL..76.2203B,2012PhRvD..85j5020H}. The results can be expressed as a constant of proportionality, since the scalings are fixed by $T_{\rm osc}$, with only weak dependence from non-trivial $g_\star (T)$. The simulations of Ref.~\cite{2015PhRvD..91f5014K} find that $\alpha_{\rm dec.}=2.48$. 

For $N_{\rm DW}>1$ the relic density cannot be expressed in such a simple form. The domain wall network is long lived and the relic density of axions produced by its decay is too large (``overclosure'') unless some additional fine tuning on $CP$-violation is allowed~\cite{2015PhRvD..91f5014K}. We do not consider the $N_{\rm DW}>1$ case any further. It would be interesting to consider in futute whether miniclusters or similar dense DM objects are produced in this scenario.

Thus, our final expression for the relic density in the minicluster scenario is:
\begin{widetext}
\be
\Omega_a^{\rm total}=\frac{1}{6H_0^2M_{pl}^2}(1+\alpha_{\rm dec})\frac{c_{\rm an }\pi^2}{3}m_a(T_{\rm CMB})m_a(T_{\rm osc})f_a^2\left(\frac{a(T_{\rm osc})}{a(T_{\rm CMB})}\right)^3\, , \label{eqn:relic_analytic}
\ee
\end{widetext}
where $T_{\rm osc}$ is found from solving Eq.~\eqref{eqn:tosc} for oscillations in a harmonic potential, $T_{\rm CMB}=2.725\,{\rm K}$ is the CMB temperature, $a(T_{\rm CMB})=1$, and we vary the parameters $n$, $\alpha_{\rm dec}$ and $c_{\rm an.}$. We have not allowed for any additional entropy production or degrees of freedom beyond the standard model, which could suppress the relic density with an additional factor of $\gamma_{\rm ent}<1$~\cite{2004hep.th....9059F}. Allowing for $\gamma_{\rm ent}<1$ would raise the lower bound on $m_a$ for axion miniclusters.  

The results of our relic density computation are shown in Fig.~\ref{fig:mass-lik}. If axions are to provide the total DM relic denisty in the minicluster scenario we find a lower bound on the axion mass of $m_a\gtrsim 10^{-13}\text{ eV}$, which could be lowered slightly more for more extreme large values of $n$. For $n=0$ we find $m_a\gtrsim 10^{-8}\text{ eV}$. We find that the QCD axion is where dotted line $m_{a,{\rm QCD}}\approx 6.6\,\mu {\rm eV}(10^{12}\text{ GeV}/f_a)$ intersects $n\approx 3.34$. Thus, when the PQ symmetry is broken post-inflation, the QCD axion must have $m_a\approx 100\,\mu\text{eV}$, consistent with other estimates in the literature.

The relation between $(m_a,f_a)$ found from the relic density affects the $M_0(m_a)$ relationship via the weak dependence on $f_a$ for $n\neq 0$. For $n=0$, the relic density only affects this by imposing a lower limit on $n$. We have verified the accuracy of the misalignment contribution to Eq.~\eqref{eqn:relic_analytic} using a direct numerical solution of the temperature-dependent equations of motion following the method of Ref.~\cite{2016PhLB..752..175B}, confirming both the use of the estimate for $T_{\rm osc}$ and the computation of $c_{\rm an}$ for the single cosine and chiral axion potentials. The approximation only breaks down for oscillation temperatures near $\Lambda_{\rm QCD}$, where the effects of $g_\star$ on the dyamics are non-trivial. For $n=0$ using the analytic result does not affect $M_0$, and thus does not affect our microlensing results. For $n\neq 0$ the use of the analytic results has a smaller effect than the uncertainty in $\alpha_{\rm dec}$ already accounted for.  

The scenario discussed above does not apply to all possible axion models. For example, the relationship between $f_a$ and the scale of SSB can be drastically modified in the ``clockwork axion'' theories~\cite{2016PhRvD..93h5007K,2016arXiv161007962G}. Here a large hierarchy can be generated between $f_a$, which sets the periodicity of the axion field, and the symmetry breaking scale. For example, one could lower the symmetry breaking scale to $v_{\rm symm}\approx 1\text{ TeV}$ while keeping $f_a\approx 10^{16}\text{ GeV}$. Thus, lighter axions that require large $f_a>8.2\times 10^{12}\text{ GeV}$ could still give rise to miniclusters. In the clockwork scenario one must be careful that the additional axions involved in realising the clockwork do not produce dangerous relics themselves. The additional axions also significantly complicate the phase transition and subsequent production and decay of topological defects~\cite{2016JHEP...08..044H}. 

Geometric string theory (and supergravity) axions cannot give rise to miniclusters, as there is no notion of spontaneous PQ symmetry breaking in 3+1 dimensions that could produce the necessary large field fluctuations as initial conditions. Accidental axions, where the $U(1)$ symmetry is explicitly broken by Planck suppressed operators, and which also arise in string theory, \emph{can} undergo SSB and produce miniclusters.\footnote{We are grateful to Joseph Conlon, Matthew McCullough, Andreas Ringwald, and Fuminobu Takahashi for discussion on clockwork axions and string axions.} 

\section{Theoretical Uncertainties in the Mass Function and Analytic Results}
\label{appendix:modelling_mf}

There are four main theoretical uncertainties in our modelling of the MCH mass function:
\begin{itemize}
\item Non-Gaussianity of the minicluster density field.
\item The initial power spectrum, $P(k,\tau_0)$.
\item The filtering function, $W(kR)$.
\item The cut-off of the mass function.
\end{itemize}
\begin{figure}
\vspace{-0.2em}\includegraphics[width=\columnwidth]{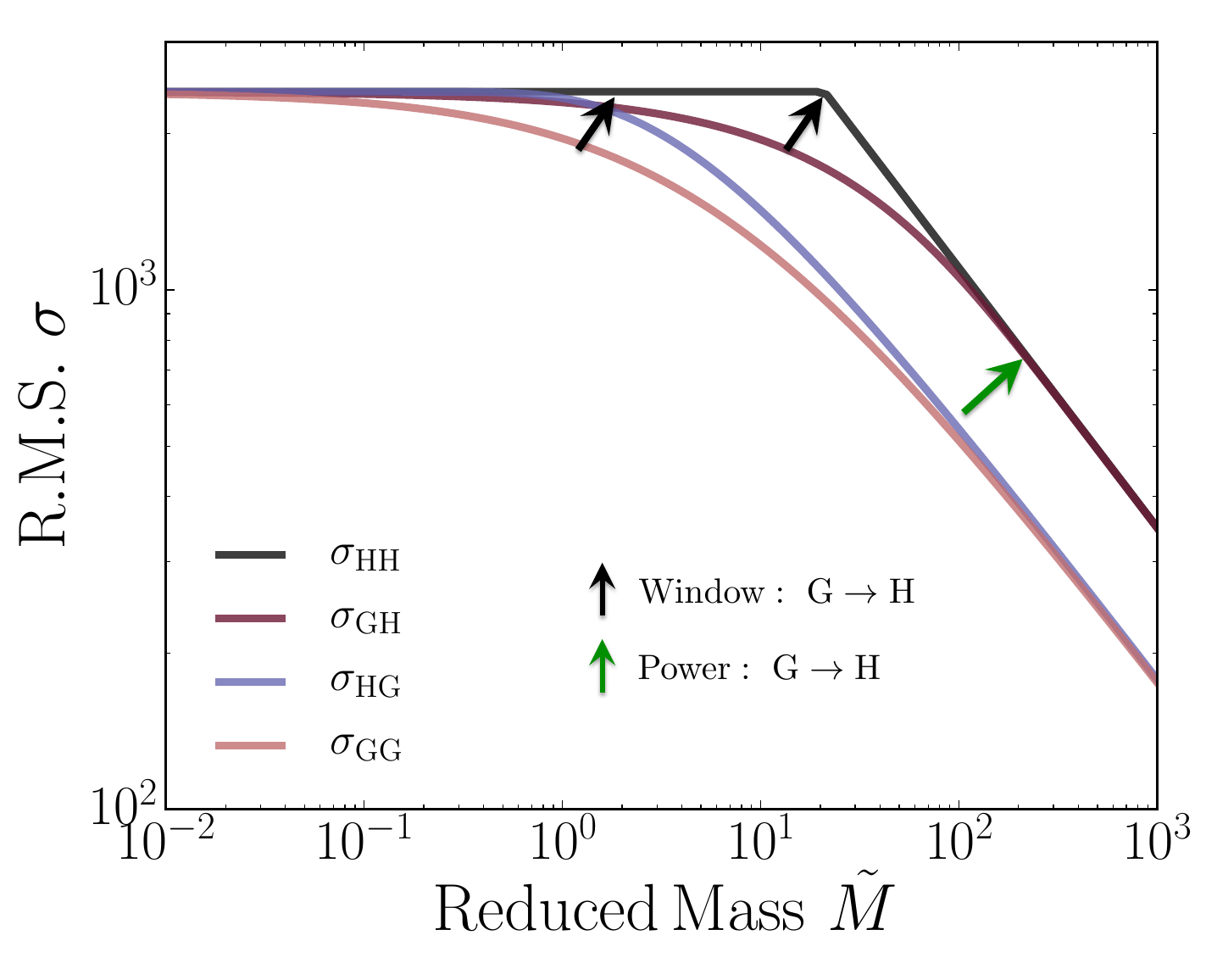}
\vspace{-2.5em}\caption{{\bf Theoretical Modelling of the R.M.S.}: The mass fluctuation, $\sigma$, is shown as a function of mass for four different combinations of initial power spectrum and window function. The Gaussian initial power cuts off power earlier than the step function. The Heaviside window function leads to a more pronounced flattening of $\sigma$ at low masses. The reduced mass $\tilde{M}=Mk_0^3/\bar{\rho}_a$.}  
\label{fig:sigma_exact}
\end{figure}

Non-Gaussianity is addressed in detail in Appendix~\ref{appendix:non-gauss}. In this Appendix we address the other three sources of uncertainty. We first compute the variance, $\sigma^2(M)$, at the initial time, $\tau_0$. We consider two simplified models for the initial power spectrum:
\begin{align}
P_{\rm G}(k) &= P_{0,{\rm G}} \exp \left[-\frac{1}{2}\left(\frac{k}{k_0}\right)^2\right] \, , \\
P_{\rm H}(k) &= P_{0,{\rm H}} \Theta (k_0-k) \, , \\
\end{align}
with ``G'' for ``Gaussian'' and ``H'' for ``Heaviside''. The Gaussian smoothing at $k_0$ to model the effects of the Kibble mechanism was used in Ref.~\cite{2007PhRvD..75d3511Z}. We note that the normalizations, $P_0$, of each power spectrum must be computed separately, and are given by:
\begin{align}
P_{0,{\rm H}}&=\frac{24}{5}\pi^2k_0^{-3} \, , \\
P_{0,{\rm G}}&= \frac{8\sqrt{2}}{5}\pi^{3/2}k_0^{-3} \, .
\end{align}

In addition to the power spectrum, we also consider two choices of window function (see e.g. Ref.~\cite{2007IJMPD..16..763Z}):
\begin{align}
W^2_{\rm G}(kR) &= e^{-k^2R^2}\, , \\
W^2_{\rm H}(kR) &= \Theta (1-kR)\, .
\end{align}

Since for all axion models considered we have $k_0<k_{J,{\rm eq}}$, the variance at $z=0$ is simply given by the variance at $z=z_{\rm eq}$ (i.e. computed with the initial power) divided by the CDM linear growth factor squared, $\sigma^2(z=0)=\sigma^2(z_{\rm eq})/D^2(z_{\rm eq})$. The CDM linear growth factor is given by the integral (assuming flatness, $\Omega_\Lambda=1-\Omega_m$):
\be
\tilde{D}(z)=\Omega_m\frac{5}{2}\frac{H(z)}{H_0}\int^\infty_z {\rm d}z' \left[\frac{H(z')}{(1+z')H_0}\right]^{-3} \, .
\ee
The growth factor is normalized to unity at $z=0$ such that $D(z)=\tilde{D}(z)/\tilde{D}(0)$.

For all combinations of initial power and window function the variance can be expressed analytically, in terms of error functions where necessary. We write the variance as $\sigma^2_{\rm XY}$, where X and Y take on either of the values G and H, with X labelling the window function and Y labelling the initial power spectrum. The closed-form expressions in terms of the dimensionless radius $\tR=Rk_0$ are:
\begin{align}
\sigma^2_{\rm GG}(\tR)&=\frac{P_{0,{\rm G}}k_0^3}{D^2(z_{\rm eq})(2\pi^2)}\frac{\sqrt{\pi/2}}{(1+2\tR^2)^{3/2}} \, , \\
\sigma^2_{\rm GH}(\tR)&=\frac{P_{0,{\rm H}}k_0^3}{D^2(z_{\rm eq})(2\pi^2)}\frac{1}{4\tR^3}\left[ -2\tR e^{-\tR^2}+\sqrt{\pi}\text{erf} (\tR)\right]\, , \\
\sigma^2_{\rm HG}(\tR)&=\frac{P_{0,{\rm G}}k_0^3}{D^2(z_{\rm eq})(2\pi^2)}\left\{ -\frac{e^{-1/(2\tR^2)}}{\tR}+\sqrt{\frac{\pi}{2}}\text{erf} [(\sqrt{2}\tR)^{-1}] \right\} \, , \label{eqn:analytic_sigHG}\\
\sigma^2_{\rm HH}(\tR)&=\frac{P_{0,{\rm H}}k_0^3}{D^2(z_{\rm eq})(2\pi^2)}\frac{1}{3}\left[ \frac{\Theta (\tR-1)}{\tR^3}+\Theta (1-\tR)\right] \, .
\end{align}

The mapping from radius to mass for the Gaussian window function is $M_{\rm G}=(2\pi)^{3/2}\bar{\rho}_a R^3$. For the Heaviside window function the mass is not well defined. We choose a mass-radius relation that gives the same variance at large $M$ as the Gaussian case, which is easily shown to be $M_{\rm H}=(4/3)M_{\rm G}$. We express the variance in terms of the dimensionless mass, $\tilde{M}=M/(\bar{\rho}_ak_0^{-3})$ (the dimensionless characteristic minicluster mass is $\tilde{M}_0\approx 130$). The variance $\sigma^2_{\rm HH}(\tilde{M})$ has a particularly simple closed form expression:
\be
\sigma_{\rm HH}(\tilde{M})=\sigma_0 \left[ \Theta(\tilde{M}-c_m)\left(\frac{\tilde{M}}{c_m}\right)^{-1/2}+\Theta (c_m-\tilde{M})\right] \, ,
\ee
with $c_m=(4/3)(2\pi)^{3/2}$ and $\sigma_0\approx 10^{3.4}$. In terms of $M_0$ defined in Eq.~\eqref{eqn:M0_def} this gives the cut-off in the variance at the physical mass scale $M_{\rm cut}=2^{3/2}\pi^{-5/2}M_0\approx M_0/6.2$.

The variances at $z=0$ as a function of $\tilde{M}$ for all four combinations of initial power spectrum and window function are plotted in Fig.~\ref{fig:sigma_exact}. Despite the different normalizations of the power, $P_0$, the variance has the same normalization in every case, as it should. We observe that the Gaussian initial power has a cut-off at lower $\tilde{M}$ than for the Heaviside initial power, and consequently the variance on larger mass scales is smaller for the Gaussian case. The effect of the window function is to give a flatter variance as $\tilde{M}\rightarrow 0$ for the Heaviside window compared to the Gaussian window, a fact which has important consequences for the cut-off in the mass function, which we now turn to.

The Press-Schechter mass function is given by:
\be
\frac{{\rm d}n}{{\rm d}\ln M} = \frac{1}{2}\frac{\bar{\rho}_a}{M}\left|\frac{{\rm d}\ln \sigma^2}{{\rm d}\ln M}\right|\sqrt{\frac{2}{\pi}}\frac{\delta_c}{\sigma}\exp \left[-\frac{1}{2}\left(\frac{\delta_c}{\sigma}\right)^2\right] \, .
\ee
When the logarithmic derivative of the variance goes to zero fast enough, this cuts-off the mass function at low $M$. This occurs for the Heaviside window function and was advocated by the authors of Ref.~\cite{2013MNRAS.433.1573S} to explain the downturn in the mass function seen in $N$-body simulations with truncated initial power spectra for warm DM. In our case, the flat variance with the Heaviside window also leads to a cut-off at low $M$ for the HMF using $\sigma_{\rm HH}$ and $\sigma_{\rm HG}$, with the cut-off for $\sigma_{\rm HH}$ being a step-function at $\tilde{M}=c_m$.

For the Gaussian window, the logarithmic derivative of the variance does not go to zero fast enough to cut-off the mass function. This is in conflict with simulations. We expect the Jeans scale to physically cut the power off in a simulation of the full axion field~\cite{2016PhRvD..94l3523V}. Even $N$-body simulations that simply use a truncated initial power spectrum see a downturn in the HMF after numerical artefacts have been removed~\cite{2016ApJ...818...89S,2017PhRvD..95h3512C}. 

Therefore, for the Gaussian window HMF we consider alternative cut-off procedures. The cut-off we adopt, as described in the main text, simply replaces $\delta_c$ with $\mathcal{G}(M)\delta_c$ in the Press-Schechter mass function. This is the approximation to the action of the Jeans scale considered in Refs.~\cite{Marsh:2013ywa,2015MNRAS.450..209B,Marsh:2016vgj}, which qualitatively agrees with the excursion set calculation using the same barrier of Ref.~\cite{2017MNRAS.465..941D}. We refer to this cut-off as the ``Jeans cut-off'': it is physically motivated, and qualitatively matches simulations. However, this cut-off requires one to know the axion mass, as well as $k_0$, and therefore it depends on the $M_0(m_a)$ relation. In the examples we use the fit for an $n=0$ ALP.
\begin{figure}
\vspace{-0.2em}\includegraphics[width=\columnwidth]{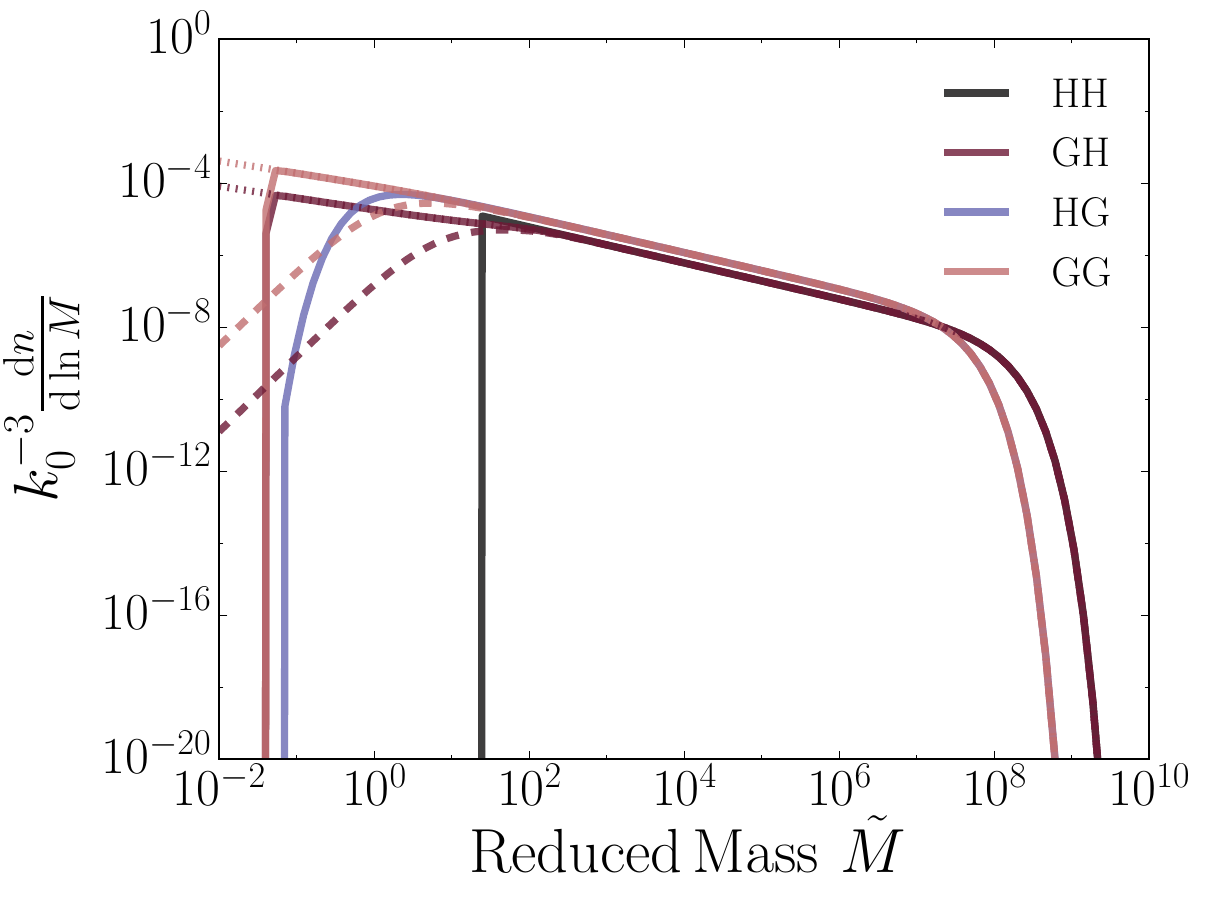}
\vspace{-2.5em}\caption{{\bf Theoretical Modelling of the Mass Function}: The minicluster mass function is shown as a function of mass for four different combinations of initial power spectrum and window function. In this case of the Gaussian window function, we also show the uncut (dotted), half-mode (dashed), and Jeans (solid) models for the low-mass cut-off. The reduced mass $\tilde{M}=Mk_0^3/\bar{\rho}_a$.}  
\label{fig:hmf_modelling}
\end{figure}

Another possibility is the fit to the cut-off seen in the simulations of Ref.~\cite{2016ApJ...818...89S} (see also Ref.~\cite{2017PhRvD..95h3512C}):
\be
\frac{{\rm d}n}{{\rm d}\ln M}\rightarrow \left[1+\left(\frac{2.4 M}{M_{1/2}}\right)^{-1.1} \right]^{-2.2}\frac{{\rm d}n}{{\rm d}\ln M} \, ,
\label{eqn:schive_cut}
\ee
where we have rescaled the cut-off to be given by the ``half-mode mass'', $M_{1/2}$, defined by the initial conditions cut at $k_0$~\cite{2016ApJ...818...89S}. The half-mode masses are:
\begin{align}
\tilde{M}_{1/2,{\rm H}}&= (2\pi)^{3/2} \, , \\
\tilde{M}_{1/2,{\rm G}}&\approx 0.2 (2\pi)^{3/2}\, .
\end{align}

The effects of all the modelling on the minicluster mass function are shown in Fig.~\ref{fig:hmf_modelling}, where we show all four combinations of initial power spectrum and window function, and for the Gaussian window we show the uncut, half-mode cut-off, and Jeans cut-offs. 

The largest difference between different cut-off models with the same initial power happens in the case of the Heaviside initial power. The Heaviside window gives a sharp cut-off at $\tilde{M}/c_m$ where the derivative of the variance goes exactly to zero. The half-mode cut-off comes in shortly after this, since we must identify the half-mode with $k_0$ in the absence of any other scale. The Jeans cut-off in this case gives considerably more low-mass halos, a number of orders of magnitude below. 
\begin{figure}[t!]
\includegraphics[width=\columnwidth]{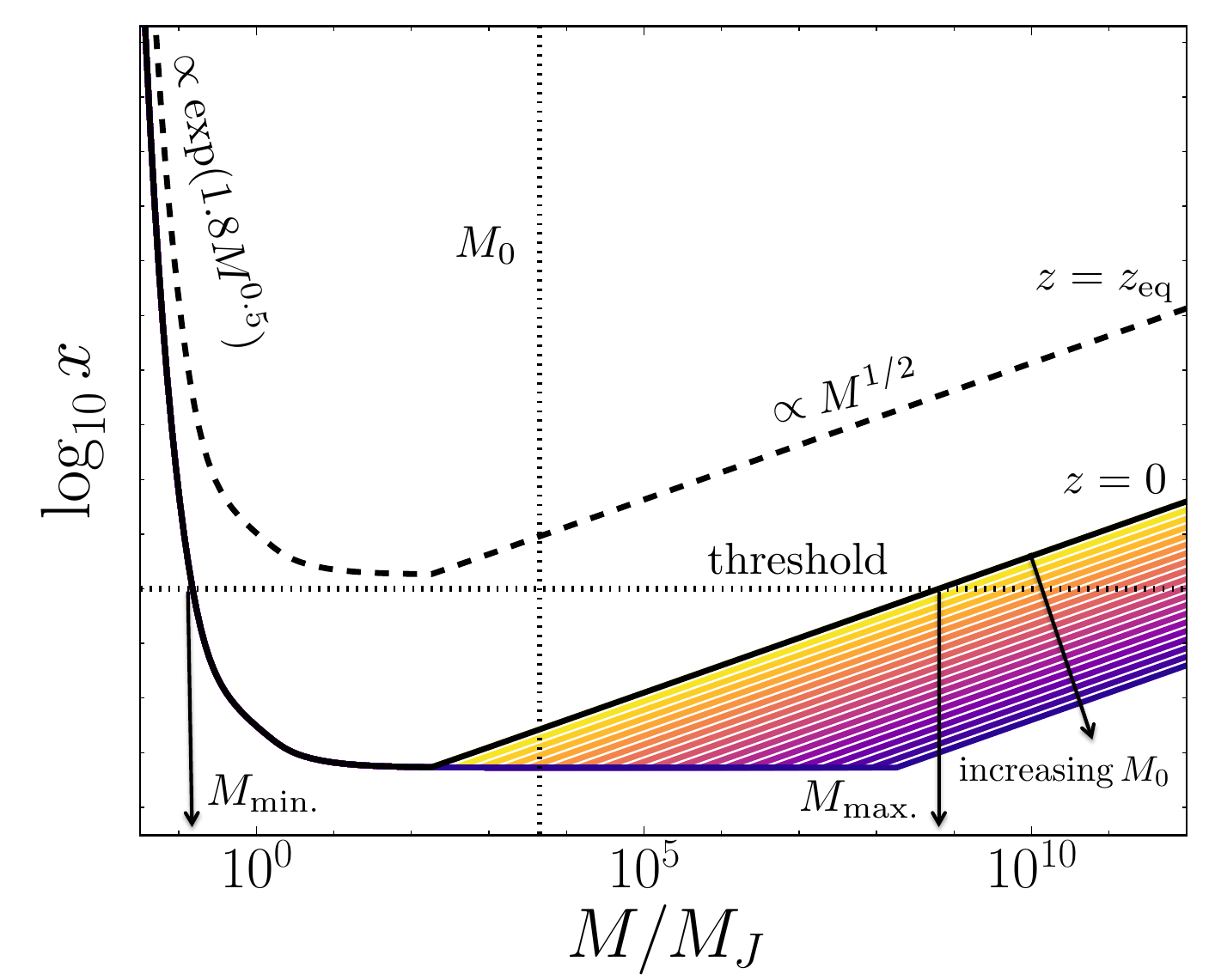}
\caption{{\bf Jeans Scale and Cut-offs in the Mass Function}: These are driven by the argument of the Press-Schechter Gaussian, $e^{-x^2/2}$ and the threshold for a $1\sigma$ cut-off is $x=1$. We plot $x$ as a function of $M/M_J$ with $M_J$ the Jeans mass. As the ratio $M_0/M_J$ increases from $1$ to $10^6$, the mass function remains centered near $M_0$ but continues to have support at $M_J\ll M_0$. The spread of minicluster masses increases from a narrow distribution near $M_0$ at $z_{\rm eq}$ to a much wider distribution today.}  
\label{fig:hmf_cut_plot}
\end{figure}

In the case of the Gaussian initial power, the different cut-offs make far less difference to the mass function. The Heaviside window leads to a cut-off around $\tilde{M}=1$, due to the shallower flattening of the variance. The half-mode is $\tilde{M}_{1/2,{\rm G}}\approx 3.1$, and the Jeans scale is $\tilde{M}_J\approx 0.3$, and thus the cut-off dependence occurs over less than one order of magnitude in $\tilde{M}$. For $n\neq 0$ we generically have $k_0<k_{J,{\rm eq}}$, and the Jeans mass can be much smaller than the characteristic minicluster mass. In this case the cut-off caused by the Kibble mechanism should dominate over the cut-off caused by the Jeans scale.

In the case of the barrier cut-off, the relation between $M_0(m_a,n)$ must be specified to derive an approximation to the cut-off, since the Jeans mass is given in terms of $m_a$ (see Fig.~\ref{fig:k0_plot}). Factoring out the linear growth, the argument of the Gaussian $e^{-x^2/2}$ is $x(M)=1.686\mathcal{G}(M)/\sigma_0(M)$, where $\mathcal{G}(M)$ is the fitting function of Ref.~\cite{Marsh:2016vgj}. The barrier $\mathcal{G}(M)$ depend on the Jeans mass, $M_J$, which itself depends on the axion mass as:
\begin{align}
M_J=&5.1\times 10^{-10}M_\odot \left(\frac{m_a}{10^{-10}\text{ eV}}\right)^{-3/2}\nonumber \\ & \left(\frac{\Omega_m}{0.32}\right)^{1/4} \left(\frac{h}{0.67}\right)^{1/2}\, .
\end{align}

Fig.~\ref{fig:hmf_cut_plot} shows $x(M/M_J,M_0/M_J)$ for $0<\mathcal{R}<10^6$. Cut-offs in the HMF occur at small and large masses, when $x(M)$ exceeds a particular threshold. The threshold for a $1\sigma$ cut-off is $x=1$. The scale $M_0$ is the largest scale to cross the threshold shortly after $z_{\rm eq}$, and thus sets the characteristic mass of miniclusters.

The Jeans mass depends only on the $T=0$ axion mass, and so the mass function cut off at low $M$ is always at approximately the same value for fixed $m_a$ regardless of the temperature dependence of the axion mass. However, as we saw in Fig.~\ref{fig:M0_mass}, as $n$ is increased and the axion mass switches on more sharply, the value $M_0$ of the characteristic minicluster mass is increased for fixed $m_a$. When $M_0/M_J$ is large, the spread of the minicluster mass function is increased. This demonstrates that the barrier cut-off has the largest difference in the low mass behaviour of the mass function for different values of $n$. Halos/miniclusters at the Jeans mass are formed by ``monolithic collapse'' and in simulations are shown to be composed of isolated ``axion stars''~\cite{Schive:2014dra}.

\section{Non-Gaussian Effects on the Mass Function}
\label{appendix:non-gauss}

We have already said something about the non-Gaussianity in the main text. We implictly assume that on large scales $k<k_0$ the fluctuations are close to Gaussian and that Press-Schechter can be applied to compute the MCH mass function for $M>M_0$. In the main analysis we account for non-Gaussianity in the distribution of density profiles on small scales at $M\sim M_0$ using the distribution $\mathcal{F}(\delta>\delta_0)$. 

In the following we discuss how non-Gaussianity in the white noise for $k_0<k_0$ can alter the mass function. In analogy to Eqs.~(\ref{eqn:2pt_corr}), (\ref{parsevalpowspec}) we consider the three-point correlation function of the overdensity:
\be
\xi_3(r_1,r_2)=\langle\delta({\bf x})\delta({\bf x}+{\bf r}_1)\delta({\bf x}+{\bf r}_2)\rangle \, ,
\ee
from which it can be shown that the bispectrum, $B(k_1,k_2)$ satisfies
\be
\int \frac{{\rm d}^3k_1}{(2\pi)^3}\frac{{\rm d}^3k_2}{(2\pi)^3}B(k_1,k_2)=\langle \delta({\bf x})^3\rangle \, .
\label{eqn:bispec_ident}
\ee
Another similar identity holds for the trispectrum, $T(k_1,k_2,k_3)$.

We model the axion initial field distribution as a random variable, $\theta$, with uniform distribution on $[-\pi,\pi]$. At early times when the axion interactions can be neglected (the potential has not switched on, $T\ll \mu$ and $n>0$), the uniform distribution is enough to allow us to calculate $\langle \delta({\bf x})^3\rangle=16/35$. We note that we cannot use the numerically derived distribution $\mathcal{F}(\delta>\delta_0)$ to compute the cumulants since it is not shown for negative $\delta$, and the Pearson distribution fit is not valid in this regime either. 

Next we assume, just like for the power spectrum, that the non-Gaussianity is constant on large scales but that the Kibble mechanism erases all correlations on small scales. Our ansatz for the bispectrum is thus:
\be
B(k_1,k_2)=B_0 \Theta (k_1-k_0)\Theta(k_2-k_0)\, .
\ee
Plugging this ansatz into Eq.~\eqref{eqn:bispec_ident} we can perform the integrals and compute the normalisation:
\be
B_0=\frac{576}{35}\pi^4k_0^{-6}\, .
\ee

A number of authors have considered the effect of a non-zero bispectrum on the mass function. For example, Ref.~\cite{2010ApJ...717..526M} perform an exact calculation using the path integral and excursion set theory. A simple formula to model non-Gaussian effects on the mass function is given by Ref.~\cite{2000ApJ...541...10M}:
\begin{widetext}
\be
\Pi_{\rm NL}(M)=\exp \left[ \frac{\delta_c^3\mathcal{S}_3}{6\sigma^2} \right]\left( \frac{1}{6}\frac{\delta_c}{\sqrt{1-\delta_c\mathcal{S}_3/3}}\frac{{\rm d}\mathcal{S}_3}{{\rm d}\ln \sigma}+\sqrt{1-\delta_c\mathcal{S}_3/3} \right)\, ,
\label{eqn:nl_hmf}
\ee
\end{widetext}
where $\Pi_{\rm NL}$ is the ratio of the non-Gaussian to Gaussian mass function. We do not use the related formula given by Ref.~\cite{2008JCAP...04..014L} since for large $\sigma^2$ on small scales as arises in our model it gives rise to the unphysical values $\Pi_{\rm NL}<0$.

The non-Gaussian corrections depend on the \emph{normalized skewness}:
\be
\mathcal{S}_3=\frac{\langle \delta^3\rangle}{[\langle\delta^2\rangle]^2}\equiv \frac{s_3(M)}{[\sigma^2(M)]^2}\, .
\ee
We are interested in the MCH mass dependence of this quantity. The denominator is simply the variance squared, computed under filtering by the appropriate window function. The bispectrum in the numerator must also be filtered to compute $s_3(M)$. The filtered bispectrum is
\be
s_3(R) = \int \frac{{\rm d}^3k_1}{(2\pi)^3}\frac{{\rm d}^3k_2}{(2\pi)^3}B(k_1,k_2)W^2(k_1R)W^2(k_2R)
\ee

In the following we work with the Gaussian filter to be explicit, but the results do not depend on this choice. Using the results for $P_0$ and $B_0$ we find
\begin{align}
\sigma^2(\tilde{R})&=\frac{12}{5}\int_1^\infty {\rm d}y y^2 e^{-y^2\tilde{R}} \, , \\
s_3(\tilde{R})&=\frac{144}{35}\left[ \int_1^\infty {\rm d}y y^2 e^{-y^2\tilde{R}} \right]^2 \, ,
\end{align}
and the normalized skewness is 
\be
\mathcal{S}_3=\frac{5}{7} \, .
\ee
It is important to note that the normalized skewness thus computed is \emph{independent of scale}.

Plugging this result into Eq.~\eqref{eqn:nl_hmf} we find that $\Pi_{\rm NL}$ diverges exponentially on large scales when $\sigma^2\rightarrow 0$. However, this is not a fundamental problem, since $\Pi_{\rm NL}$ multiplies the mass function, which is cut-off exponentially on large scales. On such large scales, the skewness-corrected non-linear mass function goes like:
\be
\frac{{\rm d} n}{{\rm d}\ln M}\propto \exp \left[\frac{\delta_c^2}{\sigma^2}\left(\frac{\delta_c\mathcal{S}_3}{6}-\frac{1}{2}  \right) \right] \approx \exp \left[-0.3\frac{\delta_c^2}{\sigma^2}\right]
\ee
and thus the mass function remains cut-off on large scales and gives sensible results in the presence of this type of non-Gaussianity. The change in the mass function cut-off leads to a small increase in the value of $M_{\rm max}$, the maximum MCH mass.

On intermediate and small scales $\Pi_{\rm NL}$ is approximately constant and equal to 0.77. Thus we find that the bispectrum correction to the mass function leads to a small $\mathcal{O}(20\%)$ suppression of the number density of intermediate mass MCHs. The number of large MCHs is increased by a much larger fraction, but their number density remains exponentially small.

Under our approximations, we can also compute the normalized excess kurtosis from the trispectrum:
\be
\mathcal{K}_4 = \frac{\langle\delta^4\rangle-3[\langle\delta^2\rangle]^2}{[\langle\delta^2\rangle]^3}=-\frac{15}{14}\, ,
\ee
which is also mass-independent. A negative excess kurtosis implies that the minicluster density field has a wider peak than a Gaussian. We are not aware of any study of the effect of kurtosis on the halo mass function.

We emphasise that the results presented here treat the axion field as a free random field and model the large scale effects of the non-Gaussianity due to white noise on the formation of MCHs. These results do not model the effects of the potential or the gradient energy, and only account for the purely random large scale effects. On large scales the gradients can be neglected and their effect on smoothing the perturbations on small scales via the Kibble mechanism is included in the ansatz for $B(k_1,k_2)$. On large scales the axion field values vary little, and the potential can be neglected. On smaller scales and at times $T\approx \mu$, the potential cannot be neglected and is important in establishing the exact form of the distribution $\mathcal{F}(\delta>\delta_0)$ and the formation of miniculsters and axion stars on scales $M\lesssim M_0$. Such effects can only be addressed by a combination of lattice field theory and $N$-simulations beyond the scope of the present work. For us, these effects fall under the umbrella of ``cut-off dependence''.

\section{Axion Stars}
\label{appendix:axion_stars}

We have treated minicluster density profiles as if axions are entirely cold. On small scales, however, scalar field dynamics becomes important~\cite{Schive:2014dra,Schive:2014hza,2015MNRAS.451.2479M,2017PhRvD..95d3541H,2016PhRvD..94d3513S,2017arXiv170505845M} and the final ingredient in any axion DM halo is the central solitonic core, or ``axion star''. The axion star is formed when gradient pressure in the axion field is sufficient to halt gravitational collapse, leading to a stable (in the non-relativistic limit) ground state solution (for axion star formation in halos, see the simulations of Ref.~\cite{Schive:2014hza}; for relativistic corrections and stability see Refs.~\cite{2016MPLA...3150090E,2017PhRvL.118a1301L,2017JCAP...03..055H,2017arXiv170505385E}; for interactions of axion stars with nuclear matter see Ref.~\cite{2016JHEP...12..127B}). We now briefly assess the role of axion stars in miniclusters, and whether they might modify any of our conclusions. We treat only the case where axion self-interactions can be neglected: the stable branch of dilute axion stars where collapse is driven by gravity.
\begin{figure}[t!]
\center
\includegraphics[width=1\columnwidth]{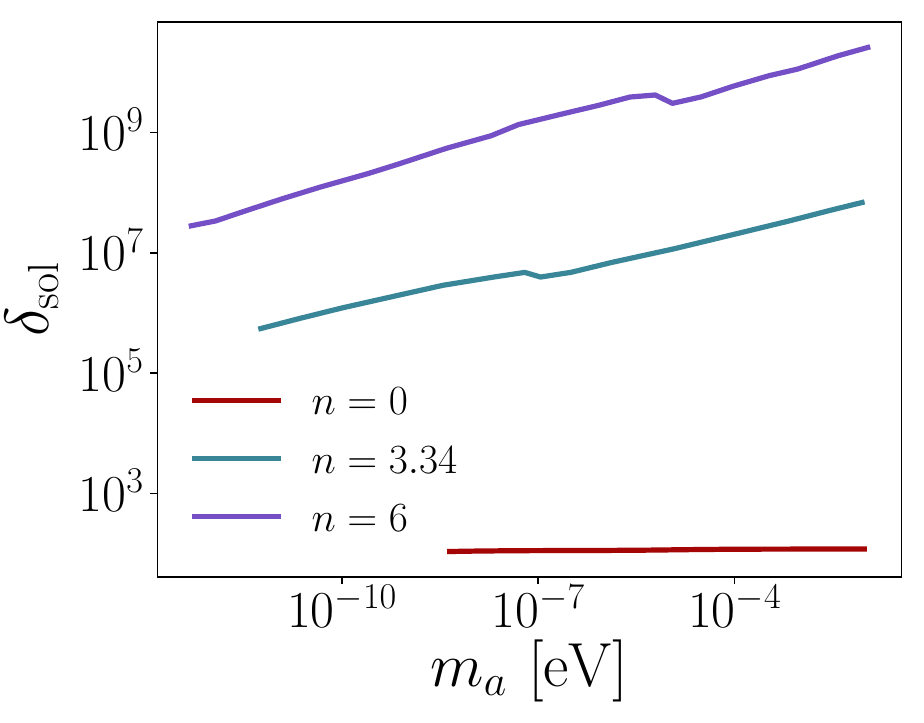}
\vspace{-0.1em}
  \caption{{\bf Axion Star Formation}: We find the critical value $\delta_{\rm sol}$ such that the minicluster radius is equal to the soliton radius. For $\delta>\delta_{\rm sol}$ miniclusters condense into axion stars.}
  \label{fig:soliton_test}
\end{figure}

The soliton is the ground state solution of the Schr\"{o}dinger-Poisson equations. For a given axion mass, the soliton solution is specified entirely by the soliton mass. The mass-radius relation is (e.g. Ref.~\cite{Schive:2014hza})
\begin{equation}
r_{\rm sol} \simeq 1.54 \times 10^{5} \left(\frac{M_{\odot}}{M_{\rm sol}}\right) \left(\frac{10^{-10}~\text{eV}}{m_{a}}\right)^2 \quad \text{cm}
\end{equation}
When an axion density perturbation becomes large enough, the axions inside will condense into the soliton solution under the influence of gravity. We estimate the value of $\delta$ for which a minicluster condenses to an axion star by equating $r_{\rm sol}$ to the hard-sphere minicluster radius, Eq.~\eqref{eqn:hard_sphere_mc}.

We set $M_{\rm sol}=M_0(m_a)$ and find $\delta_{\rm sol}$ such that $r_{\rm sol}(m_a)=r_{\rm mc}(\delta_{\rm sol},m_a)$. Using the numerical results for $M_0(m_a,n)$ we solve numerically for $\delta_{\rm sol}$ and plot the results for different $n$ in Fig.~\ref{fig:soliton_test}. For temperature-dependent axion masses with $n=3.34$ (the QCD axion) and $n=6$, we find that enormously large values of $\delta$ are required for soliton formation. For such axions, since $\delta_{\rm sol}>\delta_{\rm lens}$ giving the transition in the behaviour of the lensing tube parameter $\mathcal{R}$, soliton formation can be safely neglected in the lensing. The solitons are deep inside the point-like regime with $\mathcal{R}\approx 1$. This implies that numerical simulations of miniclusters for these values of $n$ will be free from large effects due to scalar field dynamics.

The situation is quite different for axion-like particles with a temperature-independent mass, $n=0$, where we find $\delta_{\rm sol}\approx 10^2$ approximately independent of axion mass. This smaller value of $\delta$ compared to $n\neq 0$ arises because miniclusters with $n=0$ are much lighter for fixed $m_a$. Such axions begin oscillating much earlier in cosmic history when the horizon is smaller, and thus the mass contained within the horizon is also much lower. The lighter miniclusters are closer to the soliton mass. 

Using the analytic result for $M_0(m_a)$ for $n=0$ (Eq.~\ref{eqn:m0_n0_approx}) we can derive the value of $\delta_{\rm sol}$ analytically. The axion mass dependence drops out of the ratio $r_{\rm mc}/r_{\rm sol}$ and we find $\delta_{\rm sol}\approx 120$ independent of axion mass. This value of $\delta_{\rm sol}$ is near the critical boundary for microlensing (Figs.~\ref{plotR} and \ref{fig:Rplot_self-similar}) and implies that our results for axion-like particles with $n=0$ could be altered by axion star formation. Miniclusters composed of such axions cannot become any denser without gaining mass, and this will reduce the expected number of microlensing events by effectively truncating the distribution $\mathcal{F}$ for $\delta\gtrsim 120$. Furthermore, this implies that numerical simulations of miniclusters with $n=0$ initial conditions cannot neglect scalar field dynamics.  

\bibliographystyle{h-physrev3.bst}
\bibliography{axion_review}

\end{document}